\documentclass[10pt,twocolumn,showpacs,preprintnumbers,amsmath,amssymb,aps,prb,superscriptaddress,longbibliography]{revtex4-2}
\usepackage{appendix}
\usepackage{bbm}
\usepackage{mathrsfs}
\usepackage{graphicx}
\usepackage{dcolumn}
\usepackage{bm}
\usepackage{braket}
\usepackage{amsmath}
\usepackage{amsfonts}
\usepackage[dvipsnames]{xcolor}
\usepackage[colorlinks=true,linkcolor=Blue,urlcolor=BlueViolet,citecolor=BlueViolet]{hyperref}
\usepackage{natbib}
\usepackage{multirow}
\usepackage{makecell}

\begin{document}

\title{Revealing superconducting chiral edge modes via resistance distributions}
\author{Linghao Huang}
\affiliation{State Key Laboratory of Surface Physics and Department of Physics, Fudan University, Shanghai 200433, China} 
\affiliation{Shanghai Research Center for Quantum Sciences, Shanghai 201315, China}
\author{Dongheng Qian}
\affiliation{State Key Laboratory of Surface Physics and Department of Physics, Fudan University, Shanghai 200433, China} 
\affiliation{Shanghai Research Center for Quantum Sciences, Shanghai 201315, China}
\author{Jing Wang}
\thanks{Contact author: wjingphys@fudan.edu.cn}
\affiliation{State Key Laboratory of Surface Physics and Department of Physics, Fudan University, Shanghai 200433, China}
\affiliation{Shanghai Research Center for Quantum Sciences, Shanghai 201315, China}
\affiliation{Institute for Nanoelectronic Devices and Quantum Computing, Fudan University, Shanghai 200433, China}
\affiliation{Hefei National Laboratory, Hefei 230088, China}

\begin{abstract}
  Inducing superconducting correlations in quantum anomalous Hall (QAH) states offers a promising route to realize topological superconductivity with chiral Majorana edge modes. However, the definitive identification of these modes is challenging. Here we propose detecting superconducting chiral edge modes via the probability distribution of the resistance, or equivalently the charge transmission of QAH-superconductor heterojunctions. Remarkably, the distribution for coherent edge exhibits distinct characteristics for different topological superconducting phases in sufficiently long junctions, and this difference remains robust against weak decoherence. These findings provide insights into transport phenomena beyond the clean limit and highlight the resistance distribution as a compelling signature for distinguishing topological superconducting phases.
\end{abstract}


\maketitle

\section{Introduction}
The superconducting chiral edge modes, which are unidirectional and dissipationless Bogoliubov edge states, could emerge along the boundaries of a two-dimensional chiral topological superconductor (TSC)~\cite{moore1991nonabelions,read2000paired,kitaev2006,fu2008superconducting,wilczek2009majorana,qi2010chiral,sau2010generic,alicea2010majorana,qi2011topological,elliott2015colloquium} and have potential applications in quantum information~\cite{nayak2008nonabelian,mong2014universal,clarke2014exotic,lian2018topological,hu2018fibonacci,Beenakker2019}. A promising approach to realizing these modes is to induce superconducting pair correlation in the chiral edge states of quantum Hall (QH) or quantum anomalous Hall (QAH) insulators via the proximity effect~\cite{Wan2015,Amet2016,Lee2017,Sahu2018,Matsuo2018,Seredinski2019,vignaud_2023,zhao2020interference,gul2022andreev,zhao2023loss,hatefipour2022induced,uday2024induced,sato2024,wang2024}. Distinctive transport phenomena have been observed in the QH-SC and QAH-SC junctions~\cite{zhao2020interference,gul2022andreev,hatefipour2022induced,zhao2023loss,uday2024induced} due to the charge-neutral nature of these edge modes~\cite{Hoppe2000,Giazotto2005,Akhmerov2007,Ostaay2011,fu2009probing,akhmerov2009electrically,chung2011conductance,wang2015chiral,strubi2011interferometric,lian2016edgestateinduced,lian2018quantum,wang2018multiple,Beconcini2018,lian2019distribution,he2019platform,manesco2022mechanisms,tang2022,schiller2023superconductivity,kurilovich2023disorderenabled,hu2024resistance,nava2024nonabeliana,bondarev2025}.
A significant goal is to achieve $p+ip$ TSC, which hosts a single chiral Majorana edge mode exhibiting quantized transport and interference signatures~\cite{fu2009probing,akhmerov2009electrically,chung2011conductance,wang2015chiral,katayama2025noisetocurrent}. However, its experimental confirmation remains elusive. At the same time, disorder and decoherence are inherent in realistic experimental devices~\cite{uday2024induced,zhao2023loss}, making it essential to study transport signatures beyond the idealized clean limit. In particular, understanding the resistance distribution under these effects is vital~\cite{lian2019distribution,kurilovich2023disorderenabled,hu2024resistance}, especially as such distributions can now be measured directly in experiments~\cite{ zhao2023loss}. Theoretically, an intriguing question is whether these distributions reveal novel behaviors due to the presence of exotic Majorana edge states. Experimentally, it is crucial to determine whether the distribution, beyond just its mean value, provides stronger evidence to distinguish between different topological phases in SC.

Here, we present a comprehensive analytical investigation of disorder-induced resistance distributions in QAH-SC junctions, supported by extensive numerical simulations. Using the transfer matrix formalism, we show that the quantum transport of edge modes under disorder can be intuitively described as random rotations on the Bloch sphere. This framework offers a more rigorous approach than that of Ref.~\cite{lian2019distribution}, which relied on simplifications from renormalization group analysis. We specifically consider QAH-SC heterojunctions commonly implemented in experiments as illustrated in Fig.~\ref{fig1}, where the QAH insulator has a Chern number $C=1$, hosting a single chiral electron edge mode~\cite{chang2013experimental,Checkelsky_2014,kou2014,mogi2015,deng2020quantum}, while the SC region is characterized by a Bogoliubov-de Gennes (BdG) Chern number $N=1$ or $N=2$, corresponding to transport mediated by one or two chiral Bogoliubov edge modes (CBEMs), respectively. The key results are summarized in Table~\ref{table}, which outlines the probability distributions of charge transmission [defined in Eq.~(\ref{transmission})] across different scenarios. Notably, for sufficiently long junctions, the distributions for $N=1$ and $N=2$ CBEMs exhibit distinct characteristics. Moreover, these differences remain robust against decoherence, suggesting that resistance distributions serve as a compelling signature to distinguish between different topological phases of SC.

\begin{table}[b]
\caption{Summary of the probability distributions of charge transmission fraction $T$ in two commonly implemented heterojunction configurations. $f(T)$: the probability density function of $T$;
$\mathcal{N}$: normal distribution; $\chi^2$: chi-squared distribution; and $U[-1,1]$: uniform distribution over $[-1,1]$. }
  \centering
  \renewcommand\arraystretch{1.4}
  \begin{tabular*}{3.4in}{c|c|c|c|@{\extracolsep{\fill}}c}
    \hline
    \hline
    Configuration & CBEM & Length & Cases & $f(T)$ \\
    \hline
    \multirow{3}{*}{QAH-SC-QAH} & \multirow{2}{*}{$N=2$} & Short & (i) & $\mathcal{N}$ or $\chi^2$ \\ 
    \cline{3-5}
    & & Long  & (ii) & $U[-1,1]$ \\
    \cline{2-5}
    & $N=1$ &  & (iii) & $\mathcal{N}$ \\ 
    \hline
    \multirow{4}{*}{QAH-SC} & \multirow{2}{*}{$N=2$} & Short & (i) & $\mathcal{N}$ or $\chi^2$ \\ 
    \cline{3-5}
    & & Long  & (ii) & $U[-1,1]$ \\
    \cline{2-5}
    & \multirow{2}{*}{$N=1$} & Short & (i)    & $\mathcal{N}$ or $\chi^2$ \\
    \cline{3-5}
    & & Long  & (iv)    & Generalized arcsine \quad  \\ 
    \hline
    \hline
  \end{tabular*}\label{table}
  \end{table}

The paper is organized as follows. We use transfer matrix formalism to solve the superconducting edge-mode transport problem of both $N=1$ and $N=2$ CBEM cases for different junction configurations in Sec.~\ref{edge}. Section~\ref{distribution} presents the derivation of the distribution function of the corresponding charge transmission fraction for each scenario. We use two-dimensional lattice models to numerically simulate this transport problem and verify our analytical results in Sec.~\ref{calculations}. In Sec.~\ref{decoherence}, the effect of decoherence and particle loss on the distribution is studied, and we find that the qualitative conclusion in Sec.~\ref{distribution} remain robust against weak decoherence. Finally, we conclude with some further discussions in Sec.~\ref{discussion}.

\section{Transport theory of superconducting edge modes}\label{edge}
\subsection{Edge transport with $N=2$ CBEMs}
\begin{figure*}[t]
  \begin{center}
  \includegraphics[width=5.0in,clip=true]{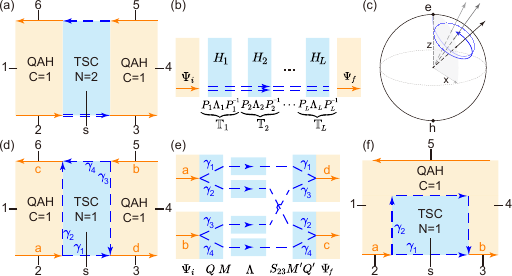}
  \end{center}
  \caption{(a) The QAH-SC-QAH junction with $N=2$ CBEM. The orange and dashed blue arrows represent the chiral electron edge mode of QAH state and the CBEM of TSC state, respectively. (b) Transfer matrix construction for the junction shown in (a). (c) Evolution of the state on the Bloch sphere. Both the axis and rotation angle exhibit weak fluctuations. The north (south) pole on the Bloch sphere represents a pure electron (hole) state. (d), (e) The QAH-SC-QAH junction with $N=1$ CBEM, and the corresponding transfer matrix construction. (f) The QAH-SC junction with $N=1$ CBEM.}
  \label{fig1}
\end{figure*}

By stacking a normal SC onto the middle region of a $C=1$ QAH insulator, one can experimentally implement a QAH-SC-QAH heterojunction in Fig.~\ref{fig1}(a). We first consider the middle region of the heterojunction 
to be in the same topological phase as the $C=1$ QAH state with broken charge $U(1)$ symmetry, namely a TSC with BdG Chern number $N=2$, which would be the case when the SC proximity gap is smaller than the insulating QAH bulk gap~\cite{qi2010chiral,wang2015chiral}. In Nambu basis $(\psi(x),~\psi^\dagger(x))^\text{T}$, the effective edge Hamiltonian is described as~\cite{huang2025}
\begin{equation}
  H(x)= v(x) k \tau_0 -\mu(x) \tau_z + \Delta(x) k \tau_x,
\end{equation}
where $k\equiv-i\partial_x$, $v(x)$ is the edge-mode velocity, $\mu(x)$ is the chemical potential, and $\Delta(x)k$ is the SC pairing amplitude, which is finite in the TSC regime and vanishes in the QAH regime. Here, $\tau_{x,y,z}$ denote the Pauli matrices in Nambu space and $\tau_0$ is the identity matrix.
Accordingly, the single chiral electron edge mode of the QAH state splits into a particle-hole conjugate pair of CBEM of the TSC state, also known as chiral Andreev edge modes~\cite{zhao2020interference}. In Fig.~\ref{fig1}(a), normal transmission and Andreev transmission~\cite{andreev1965thermal} refer to the processes in which an electron from the QAH region enters the SC and exits as an electron or a hole, with transmission coefficients denoted by $T_{\text{ee}}$ and $T_{\text{eh}}$, respectively. We define the charge transmission fraction~\cite{blonder1982transition} from lead 2 to 3 as, \begin{equation}\label{transmission}
T \equiv T_{\text{ee}}-T_{\text{eh}}. 
\end{equation}

The presence of static disorder induces spatial variations in the potential and pairing amplitude experienced by the edge modes. This effect is effectively described by random fluctuations of the parameters $(v,\mu,\Delta_0)$ in $H_{\text{edge}}$, making $T$ a random variable governed by a probability distribution—this forms the central focus of our study. Experimentally, this distribution can be observed by varying the magnetic field or the electron density \cite{lee1987universal,kurilovich2023disorderenabled, zhao2023loss}.
$T$ can be obtained via the transfer matrix method \cite{Ostaay2011,zhang2020disordered}:
We start from the eigenvalue equation for edge-mode transport: $H(x)\Psi(x)=\varepsilon\Psi(x)$, where $\varepsilon$ is the energy of the incident electron. By defining the following operators:
\begin{equation}
  \begin{aligned}
    & J(x) =\frac{\partial H}{\partial k}=
    \begin{pmatrix}
      v(x) & \Delta(x) \\
      \Delta(x) & v(x)
    \end{pmatrix}, \\
    & J^{\frac{1}{2}}(x) = 
    \begin{pmatrix}
      c_0(x) & c_x(x) \\
      c_x(x) & c_0(x)
    \end{pmatrix},~\text{where} \\
    & c_{0,x}(x) = \frac{\sqrt{v(x)+\Delta(x)}\pm\sqrt{v(x)-\Delta(x)}}{2},
  \end{aligned}
\end{equation}
which satisfies $\left[J^{\frac{1}{2}}(x)\right]^2 = J(x)$, the TSC effective edge Hamiltonian can be diagonalized as
\begin{equation}
  \begin{aligned}
  \tilde{H}(x)& =J^{-\frac{1}{2}}(x) H(x) J^{-\frac{1}{2}}(x) \\
  & = k \tau_0
  - \frac{\mu(x)}{\sqrt{[v(x)]^2-[\Delta(x)]^2}}\tau_z.
  \end{aligned}
\end{equation}
Here, $J^{-\frac{1}{2}}(x)$ denotes the inverse of $J^{\frac{1}{2}}(x)$. Transforming the eigenvalue equation to $\tilde{H}(x) \tilde{\Psi}(x) = \varepsilon \tilde{\Psi}(x)$ with $\tilde{\Psi}(x)=J^{\frac{1}{2}}(x)\Psi(x)$, one can straightforwardly solve it to obtain
\begin{equation}\label{transfer}
\begin{aligned}
    & \Psi_f = J^{-\frac{1}{2}}(x_f) e^{\int_{x_i}^{x_f} A(x) dx} J^{-\frac{1}{2}}(x_i)  \Psi_i = e^{\int_{x_i}^{x_f} A(x) dx} \Psi_i, 
    \\
     & A(x) = i \left[ \varepsilon J^{-1}(x) + \frac{1}{[v(x)]^2-[\mu(x)]^2} \right]   \mu(x) \tau_z,
     \end{aligned}
\end{equation}
where $x_{i(f)}$ are the initial (final) positions of the TSC region, $\Psi_f=\Psi(x_f)=(\psi_\text{e},\psi_\text{h})^{\text{T}}$ is the final state, and $\Psi_i=\Psi(x_i)=(1,0)^{\text{T}}$ is the incident state. In the first line we have used $J^{-\frac{1}{2}}(x_i-0^+)=J^{-\frac{1}{2}}(x_f+0^+)$ for the QAH regime, both proportional to $\tau_0$.
Disorder in the TSC region induces fluctuations in $v(x)$, $\mu(x)$, and $\Delta(x)$, which render the integral in Eq.~(\ref{transfer}) analytically intractable. For simplicity, we divide the SC region into $L$ segments labeled by the index $\ell=1,2,...,L$, where $L\propto\mathcal{L}/\xi_d$. Here $\mathcal{L}$ denotes the junction length and $\xi_d$ is the segment length. Within each segment, the parameters $(v_\ell,\mu_\ell,\Delta_\ell)$ remain constant in the local Hamiltonian $H_{\ell}$, as depicted in Fig.~\ref{fig1}(b). The integral in Eq.~(\ref{transfer}) is then converted into a sum, yielding the final state:
\begin{equation}
  \Psi_f = e^{A_L} \cdots e^{A_2} e^{A_1} \Psi_i = P_L^{-1} \Lambda_L P_L\cdots P_1^{-1} \Lambda_1 P_1 \Psi_i,
\end{equation}
In the second equation, the $A_\ell$ matrices are diagonalized piecewise. Denoting the transfer matrix as $\mathbb{T}_\ell = P_\ell^{-1} \Lambda_\ell P_\ell$, we obtain
\begin{equation}\label{final_state}
  \Psi_f = \mathbb{T}_L \cdots \mathbb{T}_2 \mathbb{T}_1 \Psi_i \equiv \mathbb{T} \Psi_i,
\end{equation}
where $\Lambda_\ell$ represents the propagation matrix of each segment and $P_\ell$ transforms the Nambu basis into the eigenbasis of $H_\ell$.

The explicit expressions for $P_\ell$ and $\Lambda_\ell$ are
\begin{widetext}
\begin{equation}\label{PL}
\begin{aligned}
  P_\ell & = 
\begin{pmatrix}
 \frac{\sqrt{(v_\ell^2-\Delta_\ell^2)\mu_\ell^2 + \Delta_\ell^2 \varepsilon^2}-\mu_\ell  \sqrt{v_\ell^2-\Delta_\ell^2}}{\sqrt{\Delta_\ell ^2 \varepsilon ^2+\left(\sqrt{(v_\ell^2-\Delta_\ell^2)\mu_\ell^2 + \Delta_\ell^2 \varepsilon^2}-\mu_\ell  \sqrt{v_\ell^2-\Delta_\ell^2}\right)^2}} & 
 -\frac{\sqrt{(v_\ell^2-\Delta_\ell^2)\mu_\ell^2 + \Delta_\ell^2 \varepsilon^2}+\mu_\ell \sqrt{v_\ell^2-\Delta_\ell^2}}{\sqrt{\Delta_\ell ^2 \varepsilon ^2+\left(\sqrt{(v_\ell^2-\Delta_\ell^2)\mu_\ell^2 + \Delta_\ell^2 \varepsilon^2}+\mu_\ell  \sqrt{v_\ell^2-\Delta_\ell^2}\right)^2}} \\
 \frac{\Delta_\ell \varepsilon}{\sqrt{\Delta_\ell^2 \varepsilon ^2+\left(\sqrt{(v_\ell^2-\Delta_\ell^2)\mu_\ell^2 + \Delta_\ell^2 \varepsilon^2}-\mu_\ell  \sqrt{v_\ell^2-\Delta_\ell^2}\right)^2}} & 
 \frac{\Delta_\ell \varepsilon }{\sqrt{\Delta_\ell^2 \varepsilon^2+\left(\sqrt{(v_\ell^2-\Delta_\ell^2)\mu_\ell^2 + \Delta_\ell^2 \varepsilon^2}+\mu_\ell \sqrt{v_\ell^2-\Delta_\ell^2}\right)^2}}
\end{pmatrix}, \\
 \Lambda_\ell & = 
\begin{pmatrix}
  e^{i \frac{v_\ell\varepsilon - \sqrt{(v_\ell^2-\Delta_\ell^2)\mu_\ell^2 + \Delta_\ell^2 \varepsilon^2}}{v_\ell^2-\Delta_\ell^2} L_\ell} & 0 \\
  0 & e^{i \frac{v_\ell\varepsilon + \sqrt{(v_\ell^2-\Delta_\ell^2)\mu_\ell^2 + \Delta_\ell^2 \varepsilon^2}}{v_\ell^2-\Delta_\ell^2} L_\ell}
\end{pmatrix},
\end{aligned}
\end{equation}
where $L_\ell$ is the length of the $\ell$th TSC section.
\end{widetext}

Applying a gauge transformation to make $\psi_\text{e}$ real, the wavefunction can be parametrized as $(\cos(\theta/2),~e^{i\phi}\sin(\theta/2))^\text{T}$, where $\theta$ and $\phi$ are the polar and azimuthal angles on the Bloch sphere, respectively. In this representation, $\Psi_i$ lies at the north pole. The $2\times2$ unitary matrix $\mathbb{T}_\ell$ can be parametrized as $ N_\ell e^{i \alpha_\ell (\mathbf{n}_\ell \cdot \bm{\sigma})/2}$, where $\boldsymbol{\sigma}=(\sigma_x, \sigma_y, \sigma_z)$. This indicates that $\mathbb{T}_\ell$ corresponds to a rotation on the Bloch sphere by an angle $\alpha_\ell$ around the axis $\mathbf{n}_\ell$, as shown in Fig.~\ref{fig1}(c). From Eq.~(\ref{PL}), we find 
\begin{equation}\label{rotation}
\begin{aligned}
N_\ell &= e^{\frac{i L_\ell v_\ell\varepsilon}{v_\ell^2-\Delta_\ell^2}}, \quad \alpha_\ell=\frac{2 L_\ell \Delta_\ell \varepsilon \sqrt{1+s_\ell}}{v_\ell^2-\Delta_\ell^2}~\text{mod}~2\pi, \\
\mathbf{n}_\ell &=\left(\frac{1}{\sqrt{1+s_\ell}},~0,~\frac{1}{\sqrt{1+s_\ell^{-1}}}\right)^{\text{T}},
\end{aligned}
\end{equation}
where $s_\ell\equiv\frac{\mu_\ell^2}{\varepsilon^2} \left(\frac{v_\ell^2}{\Delta_\ell^2} -1\right)$. The rotation axis always lies in the $xoz$ plane. When the parameters of the effective Hamiltonian fluctuate slightly due to disorder, both $\alpha_\ell$ and $\mathbf{n}_\ell$ exhibit small variations, with standard deviations satisfying $\delta \alpha_\ell \ll 2\pi$ and $|\delta \mathbf{n}_\ell| \ll 1$, particularly when $s_\ell$ is large \cite{sm}. Hence, the total transport process can be regarded as the evolution of a point at the north pole under successive rotations.
The cumulative effect of these random rotations can be represented by a single rotation by an angle $\alpha$ around an axis $\mathbf{n}\equiv(n_x, n_y, n_z)$, such that $\mathbb{T}\propto e^{i \alpha (\mathbf{n}\cdot\boldsymbol{\sigma})/2}$. The final state then takes the form $\Psi_f = ( \cos{(\alpha/2)} +i n_z  \sin{(\alpha/2)},  i(n_x+i n_y)  \sin{(\alpha/2)})^{\text{T}}$, and the charge transmission fraction is given by
\begin{equation}\label{twoMajorana}
  T=|\psi_{\text{e}}|^2-|\psi_{\text{h}}|^2= n_z^2 + (1-n_z^2) \cos{\alpha}.
\end{equation}

Although the resulting rotation is generally complicated, one simple limit can be identified:
as $\varepsilon \to 0$, $s_\ell \to \infty$ and $\mathbf{n}_\ell$ aligns with the $z$ axis. Since the initial state lies at the north pole, the final state also remains at the north pole, leading to no Andreev conversion and $T=1$. This behavior arises from the Pauli exclusion principle when forming the Cooper pair \cite{fisher1994cooperpair,beri2009quantum,Ostaay2011,zhang2020disordered}. Furthermore, this Bloch sphere picture shows that $T$ depends on the incident energy $\varepsilon$.
Larger incident energy may increase the probability of Andreev conversion. In the following, we therefore restrict to finite $\varepsilon$ to ensure Andreev conversion.

\subsection{Edge transport with $N=1$ CBEM}
Next, we consider the case where the middle SC region of the heterojunction has a BdG Chern number $N=1$, hosting a single self-conjugate CBEM, also known as chiral Majorana edge mode~\cite{qi2010chiral,wang2015chiral}. As shown in Fig.~\ref{fig1}(d), two incoherent incident charged modes, originating from different sources and labeled $a$ and $b$, decompose into four Majorana edge modes, labeled $\gamma_1$ through $\gamma_4$. Among them, $\gamma_2$ and $\gamma_3$ undergo braiding, after which all four modes recombine into two outgoing charged modes, labeled $c$ and $d$. This process, depicted in Fig.~\ref{fig1}(e), is described by
\begin{equation}\label{N1QTQ}
  \Psi_f=Q' M' S_{23} \Lambda M Q \Psi_i.
\end{equation}
Here, $\Psi_i$ is a four-dimensional vector representing the initial state. The transformation sequence, proceeding from right to left, is as follows: $Q$ accounts for propagation near the corners upon entering the SC region. $M$ transforms the Nambu basis into the Majorana basis:
\begin{equation}
M = \frac{1}{\sqrt{2}}\begin{pmatrix}
  1 & 1 & 0 & 0 \\ 
  -i & i & 0 & 0 \\ 
  0 & 0 & 1 & 1 \\ 
  0 & 0 & -i & i
\end{pmatrix},
\end{equation}
under the basis $(\psi^\dagger,\psi,\psi'^\dagger,\psi')^\text{T}$.
$\Lambda$ describes propagation within the SC region in the Majorana basis. $S_{23}$ captures the braiding of $\gamma_2$ and $\gamma_3$:
\begin{equation}
  S_{23} =
\begin{pmatrix}
  1 & 0 & 0 & 0 \\ 
  0 & 0 & 1 & 0 \\ 
  0 & -1 & 0 & 0 \\ 
  0 & 0 & 0 & 1
\end{pmatrix}.
\end{equation}
$M'=M^{-1}$ rotates the Majorana basis back to the Nambu basis. $Q'$ describes propagation near the corners before exiting the SC region. The matrix $\Lambda$ is diagonal, whose diagonal elements are all phase factors, reflecting the well-separated and decoupled nature of the Majorana modes away from the SC region's corners. However, near these corners, the Majorana modes overlap, leading to non-trivial coupling. This coupling takes the form $\eta\tau_z$ in the Nambu basis, where $\eta$ is a disorder-dependent coupling coefficient \cite{lian2018quantum,lian2019distribution}.
This coupling term corresponds to the chemical potential term in $H_{\text{edge}}$. Consequently, the effect of random coupling is captured by $Q = \text{diag}(\mathbb{T}^a, \mathbb{T}^b)$ and $Q'=\text{diag}(\mathbb{T}^c, \mathbb{T}^d)$, where $\mathbb{T}^{\zeta}=\begin{pmatrix}\zeta_{11} & \zeta_{12} \\ \zeta_{21} & \zeta_{22}\end{pmatrix}$ represents the total transfer matrix at corner $\zeta=a,b,c,d$, structured similarly to the transfer matrix product in Eq.~(\ref{final_state}).

As shown in Fig.~\ref{fig1}(d), the two incoming electrons are phase-incoherent. The charge transmission from lead 2 to lead 3, denoted by $T$, is obtained by setting $\Psi_i=(1,0,0,0)^{\text{T}}$ and evaluating the first two components of $\Psi_f$. Similarly, the charge transmission from lead 5 to lead 3, denoted by $R$, is obtained by setting $\Psi_i=(0,0,1,0)^{\text{T}}$ and evaluating the first two components of $\Psi_f$. The final result is:
\begin{equation}
  \begin{aligned}
  T & =\frac{1}{4} |a_{11}+a_{21}|^2 \left(|d_{11}+d_{12}|^2-|d_{21}+d_{22}|^2\right), \\ 
  R & =\frac{1}{4} |b_{11}+b_{21}|^2 \left(|d_{11}-d_{12}|^2-|d_{21}-d_{22}|^2\right).
  \end{aligned}
\end{equation}

Then we examine an alternative junction configuration~\cite{zhao2020interference,gul2022andreev,hatefipour2022induced,zhao2023loss,uday2024induced}, depicted in Fig.~\ref{fig1}(f). This setup, known as a QAH-SC junction, is realized by stacking a normal SC onto the edge region of a $C=1$ QAH insulator. When the SC region has a BdG Chern number $N=2$, the charge transmission fraction $T$ from lead 2 to 3 in the QAH-SC junction matches that of the QAH-SC-QAH junction in Fig.~\ref{fig1}(a). Therefore, we focus on the SC region that has $N=1$. In this configuration, the incident charged mode $a$ splits into two Majorana modes, $\gamma_1$ and $\gamma_2$, within the SC region, which subsequently recombine and emerge as the outgoing charged mode $b$. A key distinction between this setup and the QAH-SC-QAH junction in Fig.~\ref{fig1}(d) is that in the QAH-SC junction, the Majorana modes originate from the same source, preserving their coherence. This coherence leads to different transport behaviors of $T$, depending on the junction length. Specifically, the final state can be expressed as
\begin{equation}
\Psi_f = \mathbb{T}^{b} M' \Lambda M \mathbb{T}^{a} \Psi_i.
\end{equation} 
Here, fluctuations near the corners are negligible compared to those along the edges of the SC region. This allows us to approximate the total transfer matrix as $\overline{\mathbb{T}}^{-1} M' \Lambda M \overline{\mathbb{T}}$, where $\overline{\mathbb{T}}$ represents the average transfer matrix at the corner. As a result, the entire process can be interpreted as a random rotation by an angle $\alpha$ around a fixed axis $\mathbf{n}$, with $\alpha$ and $\mathbf{n}$ determined by $\Lambda$ and $M\overline{\mathbb{T}}$, respectively. The charge transmission fraction from lead 2 to 3 is the same as Eq.~(\ref{twoMajorana}).

\section{Distribution of charge transmission fraction}\label{distribution}
The distribution of $T$ exhibits distinct yet general behaviors depending on the number of CBEM and the length of SC region, as analyzed in the following cases and summarized in Table~\ref{table}.

\subsection{Case (i)}
For a relatively short QAH-SC-QAH junction ($L$ is small but $L>2\ell_d/a$, where $\ell_d$ is the decay length of the CBEM wavefuntion, and $a$ is the lattice constant) with $N=2$ CBEMs, as shown in Fig.~\ref{fig1}(a), $T$ approximately follows either a normal distribution ($\mathcal{N}$), or a chi-squared distribution ($\chi^2$).

In this regime, fluctuations in the rotation axis are negligible, allowing us to approximate $\mathbf{n} \approx \overline{\mathbf{n}}_\ell$.
In this case all $P_\ell$s become identical, denoted as $P_1$, so that $\Psi_f= P_1^{-1} \Lambda_L P_1 \cdots P_1^{-1} \Lambda_1 P_1 \Psi_i=P_1^{-1} \Lambda_L \cdots \Lambda_1 P_1 \Psi_i$.
As a result, the state undergoes sequential rotations around a nearly fixed axis, with each segment contributing a random rotation angle. The total rotation angle is $\alpha=\sum_{\ell=1}^{L} \alpha_\ell$.
To exclude other effects such as crossed Andreev reflection and elastic cotunneling \cite{galambos2022crossed,tang2022,kurilovich2023criticality,schiller2023superconductivity}, which would complicate the transport processes of interest, we require $L>2\ell_d/a$, which is typically at least of order 10.
By the central limit theorem, after multiple segments, $\alpha$ follows a normal distribution with mean $\overline{\alpha}=L\overline{\alpha}_\ell$ and standard deviation $\delta\alpha = \sqrt{L}\delta\alpha_\ell$, irrespective of the specific distribution of the individual rotation angles $\alpha_\ell$.
Because of the cosine term in Eq. (\ref{twoMajorana}), $T(\alpha)=A+(1-A) \cos{\alpha'}$, where $A=n_z^2,~\alpha' = \alpha~\text{mod}~2\pi$. The normal distribution of $\alpha$ induces a wrapped normal distribution \cite{mardia2010directional} of $\alpha'$, with probability density function
\begin{equation}\label{wrapped}
  f_\text{w}(\alpha')=\frac{1}{\sqrt{2\pi}\delta\alpha} \sum_{k=-\infty}^{+\infty} \exp\left[-\frac{(\alpha'-\overline{\alpha'}-2\pi k)^2}{2(\delta\alpha)^2}\right],
\end{equation}
where $\overline{\alpha'}= \overline{\alpha}~\text{mod}~2\pi$. The probability density function of $T$ is then
\begin{equation}\label{wrappedT}
  f(T) = \frac{f_\text{w}\left(\arccos{\left(\frac{T - A}{1-A}\right)}\right) + f_\text{w}\left(-\arccos{\left(\frac{T - A}{1-A}\right)}\right)}{(1-A) \sqrt{1 - \left(\frac{T - A}{1-A}\right)^2}},
\end{equation}
for $T \in [2A-1, 1]$ and zero otherwise.
Since $\delta\alpha_\ell$ and $L$ are small, only one term in Eq.~(\ref{wrapped}) contributes significantly, so $\alpha'$ is approximately distributed as $N\left(\overline{\alpha'},(\delta\alpha)^2\right)$. When $\overline{\alpha'}$ is not close to 0 or $\pi$, expanding $\alpha'$ around $\overline{\alpha'}$ to first order, $\cos{\alpha'}\approx\cos{\overline{\alpha'}}-\sin{\overline{\alpha'}}\Delta\alpha'$, gives $T\approx A+(1-A)\cos{\overline{\alpha'}}-(1-A)\sin{\overline{\alpha'}}\Delta\alpha'$, where $\Delta\alpha'\sim N\left(0,(\delta\alpha)^2\right)$. Thus Eq.~(\ref{wrappedT}) reduces to
\begin{equation}
  f(T) \approx \frac{1}{\sqrt{2\pi} \sigma_T} e^{-\frac{(T - E)^2}{2\sigma_T^2}},
\end{equation}
where $\sigma_T = (1 - A) \delta\alpha \sin\overline{\alpha'}$ and $E=A+(1 - A) \cos{\overline{\alpha'}}$. Therefore, $T$ follows the normal distribution $N(E,\sigma_T^2)$.
When $\overline{\alpha'}\approx0$ or $\pi$, the first-order term vanishes and $\alpha'$ must be expanded to the second order: $\cos{\alpha'}\approx\pm1\mp(\Delta\alpha')^2/2$. Here $X=(\Delta\alpha')^2$ follows a non-central chi-squared distribution with $k=1$, whose probability density function is: $f(X)= e^{-\frac{X}{2(\delta\alpha)^2}}/(\delta\alpha \sqrt{2 \pi X})$. Thus the probability density function of $T\approx A+(1-A)(\pm1\mp X/2)$ is:
\begin{equation}
  f(T)=
  \begin{cases}
    \frac{1}{\sqrt{\pi (1-T)}\sigma_T} e^{-\frac{1-T}{\sigma_T^2}}, & \overline{\alpha'}=0 \\
    \frac{1}{\sqrt{\pi (T-(2A-1))}\sigma_T} e^{-\frac{T-(2A-1)}{\sigma_T^2}}, & \overline{\alpha'}=\pi \\ 
  \end{cases},
\end{equation}
where $\sigma_T=\sqrt{1-A}\delta\alpha$. This implies that $T$ follows a chi-squared distribution.
As the junction length varies, the distribution of $T$ oscillates between normal and chi-squared distributions. This behavior extends beyond the results in Ref.~\cite{kurilovich2023disorderenabled}, where only a normal distribution was proposed.

\subsection{Case (ii)}
For a long QAH-SC-QAH junction with (sufficiently large $L$ and) $N=2$ CBEM, as shown in Fig.~\ref{fig1}(a), $T$ approximates a continous uniform distribution over $[-1,1]$, denoted as $U[-1,1]$.

As $L$ increases, the fluctuations of the rotation axis become significant, causing the state to spread across the entire Bloch sphere, as illustrated in Fig.~\ref{fig1}(c).
The uniform distribution of transmission can be explained using the theorem of It\^{o} and Kawada \cite{kawada1940probability}: Consider a stochastic process on a group $G$: $S_L = R_1 R_2 \cdots R_L$, where $R_i$ are independent random variables distributed according to a single probability measure $\mu$ on $G$. The probability measure of $S_L$, denoted $\mu^L$, is the $L$-fold convolution of $\mu$. If $\mu$ is aperiodic, then the sequence $\mu^L$ converges in distribution to the normalized Haar measure on $G$.

Here we focus on the group $G=SU(2)$. The fluctuation region of $\mathbf{n}_\ell$ defines the probability measure $\mu$, whose support is not contained in a coset of any proper closed subgroup of $G$, ensuring aperiodicity. A sequence of such random rotations thus forms a stochastic process on $G$. According to the theorem, after sufficiently many rotations, the probability measure converges to the normalized Haar measure of $SU(2)$, implying that $\Psi_f$ becomes equidistributed on the Bloch sphere. Consequently, the $z$ component of $\Psi_f$ follows $U[-1,1]$. Parametrizing the wavefunction as $\Psi_f=(\cos(\theta/2),e^{i\phi}\sin(\theta/2))^\text{T}$, we obtain $T=\cos{\theta}=z \sim U[-1,1]$, i.e., a continuous uniform distribution.

\subsection{Case (iii)}
For the QAH-SC-QAH junction with $N=1$ CBEM shown in Fig.~\ref{fig1}(d), $T$ consistently follows a normal distribution, regardless of the junction length.

\begin{figure*}[t]
  \begin{center}
  \includegraphics[width=5.0in,clip=true]{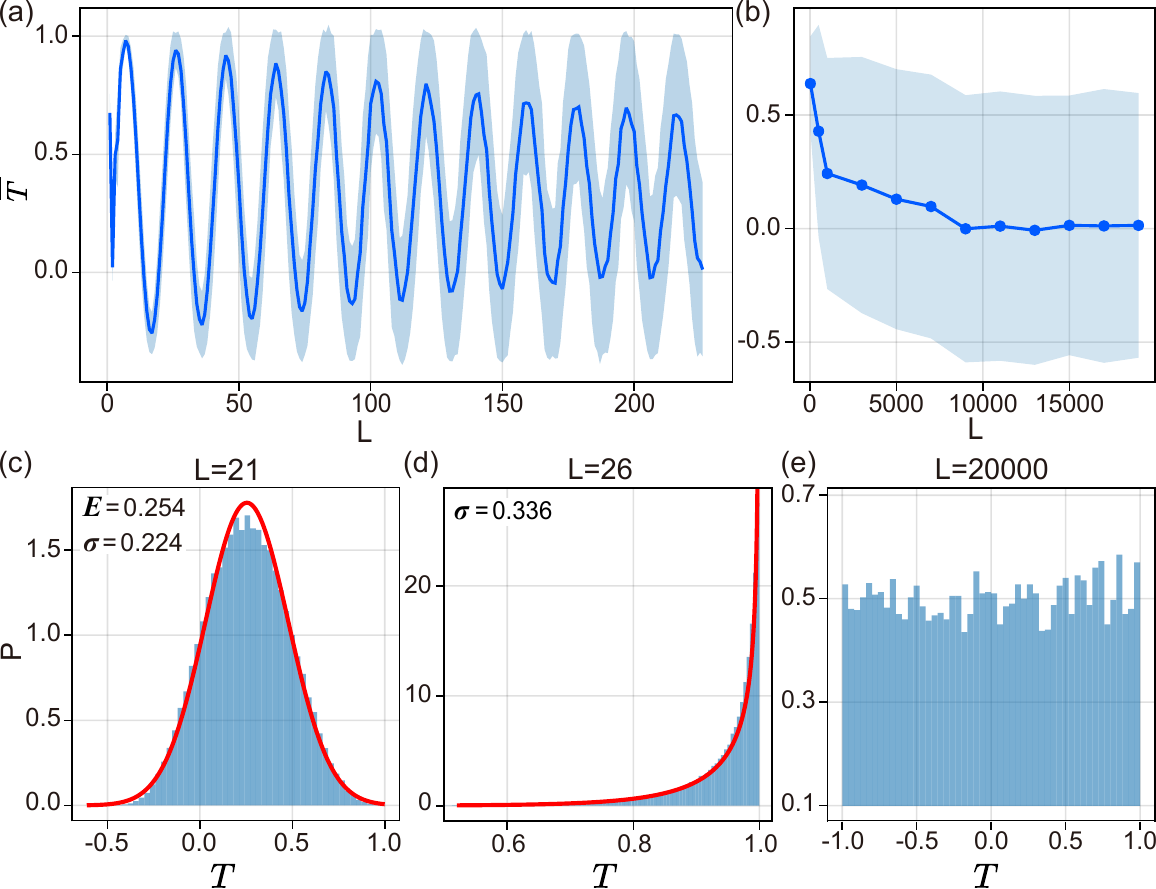}
  \end{center}
  \caption{Mean and distribution of the charge transmission fraction $T$ for the junction in Fig.~\ref{fig1}(a). (a), (b) $\overline{T}$ vs $L$, where solid lines and shaded region indicate the mean and standard deviation, respectively. (c-e) The distributions of $T$ for different $L$, with histograms representing numerical results and red lines showing fitted distributions. (c) The distribution is fitted with a normal distribution $f(T)=\frac{1}{\sigma  \sqrt{2 \pi}}e^{-\frac{(T-E)^2}{2 \sigma ^2}}$. (d) The distribution is fitted with a chi-squared distribution $f(T)=\frac{1}{\sigma\sqrt{\pi(1-T)}}e^{-\frac{1-T}{\sigma ^2}}$. (e) The mean and standard deviation of $T$ are $8.54\times10^{-3}$ and 0.582, respectively, which match those of a uniform distribution $U[-1,1]$.}
  \label{fig2}
\end{figure*}

In this case, the total transfer matrix at corner $\zeta^a$ can also be parametrized as $\mathbb{T}^a=e^{i \alpha (\mathbf{n}\cdot\boldsymbol{\sigma})/2}$, thus:
\begin{equation}
  |a_{11}\pm a_{21}|^2 = 1 \pm n_x n_y -\mp n_x n_z \cos{\alpha} \mp n_y \sin{\alpha}.
\end{equation}
Because the edge-mode wavefunction has a finite decay length, $\eta\neq0$ only near the corner. As a result, the effective $L$ is small, resembling case (i). Expanding around $\overline{a}$ to first order, $|a_{11}+a_{21}|^2 \sim \overline{a}+\Delta a$ is normally distributed with small variance. Similar results hold for $|a_{11}-a_{21}|^2$, $|b_{11}\pm b_{21}|^2$, and related terms. Consequently, $|a_{11}+a_{21}|^2 |d_{11}+d_{12}|^2 \sim \overline{a} \overline{d} + \overline{a}\Delta d + \overline{d}\Delta a$ is also normally distributed to first order. Therefore, both $T$ and $R$ follow normal distributions with small variance.
The above analysis reveals that the transmission probability is independent of $L$, which results from the incoherence of incident electrons, making the phase accumulation in $\Lambda$ in Eq. (\ref{N1QTQ}) have no effect on the final result.

\subsection{Case (iv)}
For a long QAH-SC junction (sufficiently large $L$) with $N=1$ CBEM, as shown in Fig.~\ref{fig1}(f), $T$ approximately follows a generalized arcsine distribution.

As discussed in case (i), $\delta \alpha \propto \sqrt{L}$, which becomes significant for large $L$. Therefore, $\cos{\alpha}$ will converge to the standard arcsine distribution~\cite{feller1991introduction}.
This can be seen by expanding Eq.~(\ref{wrapped}) in a Fourier series:
\begin{equation}
    f_\text{w}(\alpha') = \frac{1}{2\pi}\sum_{n=-\infty}^\infty e^{in(\alpha' - \overline{\alpha'}) - \frac{(\delta \alpha)^2 n^2}{2}}.
\end{equation}
When $\delta\alpha\rightarrow\infty$, only the $n=0$ term contributes, and $f_\text{w}(\alpha')$ reduces to a constant $1/(2\pi)$, which implies that $\alpha'$ is uniformly distributed as $U[0,2\pi]$. Then Eq.~(\ref{wrappedT}) approaches
\begin{equation}
  f(T)=\begin{cases}
  \frac{1}{\pi\sqrt{(1-A)^2-(T-A)^2}}\ , & (T\in [2A-1,1])\\
  0\ ,                                   & (T\notin [2A-1,1])
  \end{cases}
  \label{arcsine}
\end{equation}
Thus, $T$ follows a shifted and rescaled variant of the standard arcsine distribution, characterized by a U-shaped profile that diverges near the endpoints at $T=1$ and $T=2A-1$.

\begin{figure*}[t]
  \begin{center}
  \includegraphics[width=5.0in,clip=true]{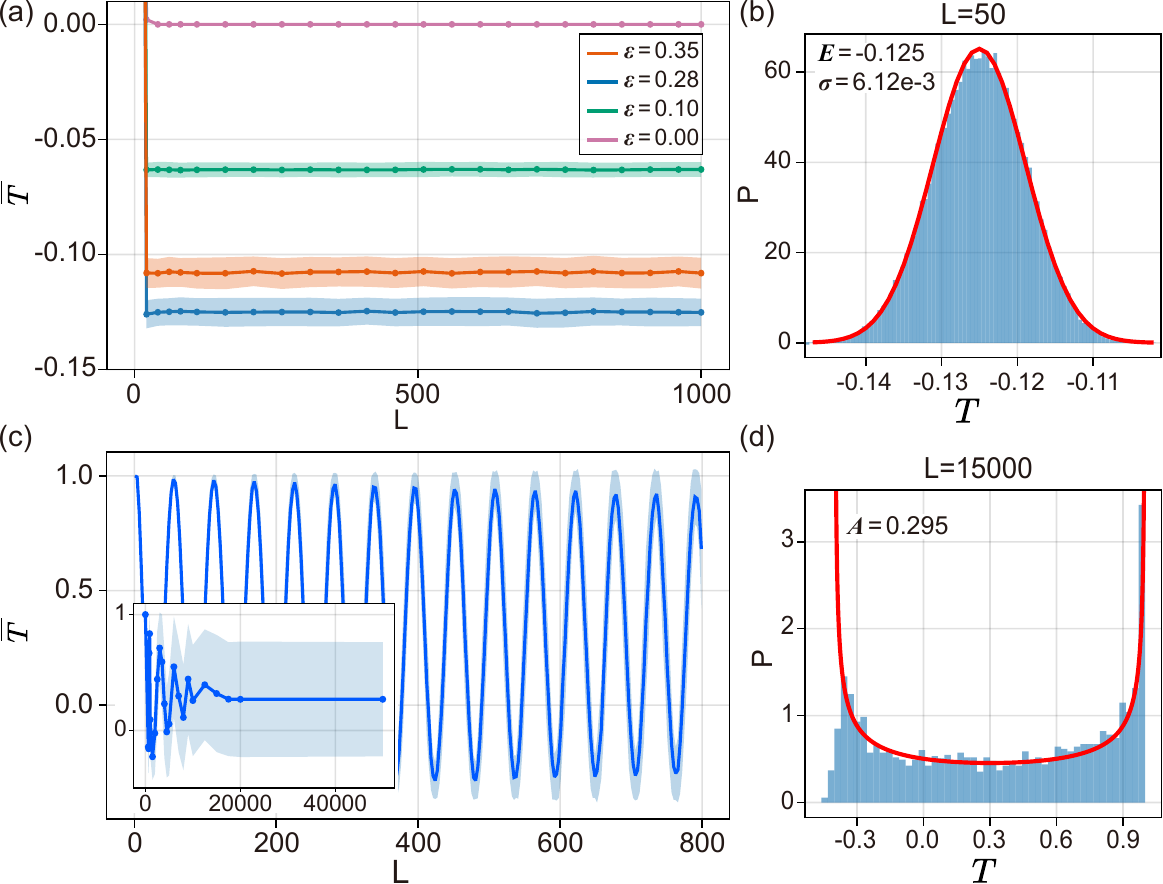}
  \end{center}
  \caption{(a), (b) Numerical results of $T$ for junction in Fig.~\ref{fig1}(d). (a) $\overline{T}$ vs $L$, where $\varepsilon$ denotes the incident electron energy. The solid lines and shading indicate the mean and standard deviation, respectively. (b) Distribution of $T$ for $\varepsilon=0.28$, which follows a normal distribution depicted as a red line. (c), (d) Numerical results for junction of Fig.~\ref{fig1}(f). (c) $\overline{T}$ vs $L$. (d) Distribution of $T$, with the red line showing the fitting distribution as a generalized arcsine distribution in Eq.~(\ref{arcsine}).}
  \label{fig3}
\end{figure*}

Finally, we make the following remarks:

(a) When the SC region of the junction is topologically trivial ($N=0$), two CBEMs appear along the SC-QAH interface. The transport properties in this case closely resemble those of the $N=2$ junction~\cite{sm}.

(b) There is no universal threshold for a ``sufficiently large'' $L$, as it depends on disorder, incident energy, and other parameters of the system. For smaller incident energies, a longer junction is required to reach convergence in the distribution of $T$. In contrast, a stronger disorder leads to convergence at shorter lengths.

(c) The explanation of case (iii) is based on the central limit theorem, which assumes a sufficiently large number of random rotations. Deviations may occur when the edge-mode decay length is short or the disorder correlation length is long, both of which reduce the effective number of independent rotations, potentially altering the distribution of $T$.

\section{Numerical simulations}\label{calculations}
To validate the analytical results presented above, we numerically compute $T$ in each case. We implemented two-dimensional tight-binding Hamiltonians of both QAH and SC for the configurations in Figs.~\ref{fig1} (a),~\ref{fig1}(d), and~\ref{fig1}(f) on a rectangular geometry. The disorder is introduced through a uniformly distributed onsite potential and spatially uncorrelated. We set $a\equiv1$. The charge transmission $T$ is then calculated using the recursive Green's function method~\cite{thouless1981conductivity,sancho1984quick,sancho1985highly,mackinnon1985calculation,datta2017lessons,sun2009quantum}. Additional numerical details are provided in Ref.~\cite{sm}.

The results for the junction configuration of Fig.~\ref{fig1}(a) are presented in Fig.~\ref{fig2}. Here, $\overline{T}$ denotes the mean charge transmission fraction from lead 2 to 3. When disorder is introduced into the SC region, $\overline{T}$ exhibits oscillations with a decaying amplitude as the junction length $L$ increases, eventually converging to zero. Meanwhile, the fluctuations in $T$ increase, as shown in Figs.~\ref{fig2}(a) and~\ref{fig2}(b). For $L\gtrsim2\ell_d$, $T$ follows a normal or chi-squared distribution, as illustrated in Fig.~\ref{fig2}(c) and Fig.~\ref{fig2}(d), respectively, in accordance with case (i). For sufficiently large $L$, in this parameter setup $L\gtrsim10^4$, the distribution of $T$ approaches a continuous uniform distribution $U[-1,1]$, as shown in Fig.~\ref{fig2}(e), corresponding to case (ii). 

For the junction configuration of Fig.~\ref{fig1}(d), the results are presented in Figs.~\ref{fig3}(a) and~\ref{fig3}(b). When $L\gtrsim2\ell_d$, both the mean and fluctuations of $T$ remain constant as $L$ varies, with $\overline{T}$ aligning with its value in the clean limit. Furthermore, the distribution of $T$ follows a normal distribution, 
consistent with case (iii), confirming that this distribution arises from the random coupling of Majorana modes near the junction corners. 
The behavior of $R$, the charge transmission probability from lead 3 to lead 5, exhibits a similar behavior~\cite{sm}.
Next, the results for the QAH-SC junction of Fig.~\ref{fig1}(f) are shown in Figs.~\ref{fig3}(c) and~\ref{fig3}(d). While $\overline{T}$ 
exhibits oscillations with decaying amplitude—similar to the $N=2$ case—it does not necessarily converge to zero, depending on the incident electron energy $\varepsilon$. Furthermore, for sufficiently large $L$, in this parameter setup $L\gtrsim2\times10^4$, the distribution transitions to a generalized arcsine form, as predicted in case (iv). The left peak is broadened due to the neglected fluctuations of the rotation axis in the analytical analysis, which introduce slight variations in the lower bound of the distribution.

\section{Effect of decoherence and particle loss }\label{decoherence}
Two additional factors, decoherence and particle loss, play a significant role in shaping experimental observations beyond the effects of disorder~\cite{zhao2023loss}.
Decoherence refers to the loss of quantum phase coherence of the state $\Psi(x)$ during propagation, arising from two sources: (1) inelastic scattering with other gapless degrees of freedom, which can be induced by magnetic vortices, incomplete SC proximity, or thermal fluctuations \cite{hu2024resistance}; and (2) temperature-induced dephasing.
Loss occurs when particles tunnel into magnetic vortices and escape through the vortex line to the substrate~\cite{kurilovich2023disorderenabled,tang2022, michelsen2023supercurrentenabled}.
In the following, we discuss the effects of decoherence and loss separately.

\subsection{Decoherence effect}
We first consider decoherence induced by inelastic scattering. Phenomenologically, this effect can be incorporated by introducing floating leads \cite{buttiker1988symmetry} at the boundaries of the TSC region, which serve as decoherence sources. In the simplest case, with only one inelastic scattering center along the edge, we introduce a parameter $p$ to describe the probability that edge modes enter a floating lead.
For a QAH-SC-QAH junction or a QAH-SC junction with $N=2$ CBEMs, the effective charge transmission fraction from lead 2 to lead 3 is given by the generalized Landauer-B\"uttiker formula \cite{sm}:
\begin{equation}\label{N2onelead}
  T^{\text{eff}} = (1-p) T^{(l,r)} + p T^{(l,f)} T^{(f,r)}, 
\end{equation}
where $T^{(l,r)}$ denotes the coherent charge transmission fraction from lead 2 to lead 3, $T^{(l,f)}$ is the fraction from lead 2 to the floating lead $f$, and $T^{(f,r)}$ is the fraction from lead $f$ to lead 3. For a QAH-SC junction with $N=1$ CBEM, the situation is slightly different because the two Majorana modes propagate along distinct edges of the TSC. For simplicity, we consider one floating lead on each edge, assuming the same decoherence probability $p$ for both. The effective charge transmission fraction is then given by
\begin{equation}\label{N1onelead}
  T^{\text{eff}} = p\left[T^{(l,u)} T^{(u,r)}+T^{(l,d)} T^{(d,r)}\right]+(1-p) T^{(l,r)},
\end{equation}
where $T^{(l,u)}$ and $T^{(l,d)}$ are the transmission fractions from lead 2 to the upper and lower floating leads, respectively, $T^{(u,r)}$ and $T^{(d,r)}$ are the corresponding fractions from the floating leads to lead 3, and $T^{(l,r)}$ is the coherent transmission fraction from lead 2 directly to lead 3.

\begin{figure*}[t]
  \begin{center}
  \includegraphics[width=5.0in,clip=true]{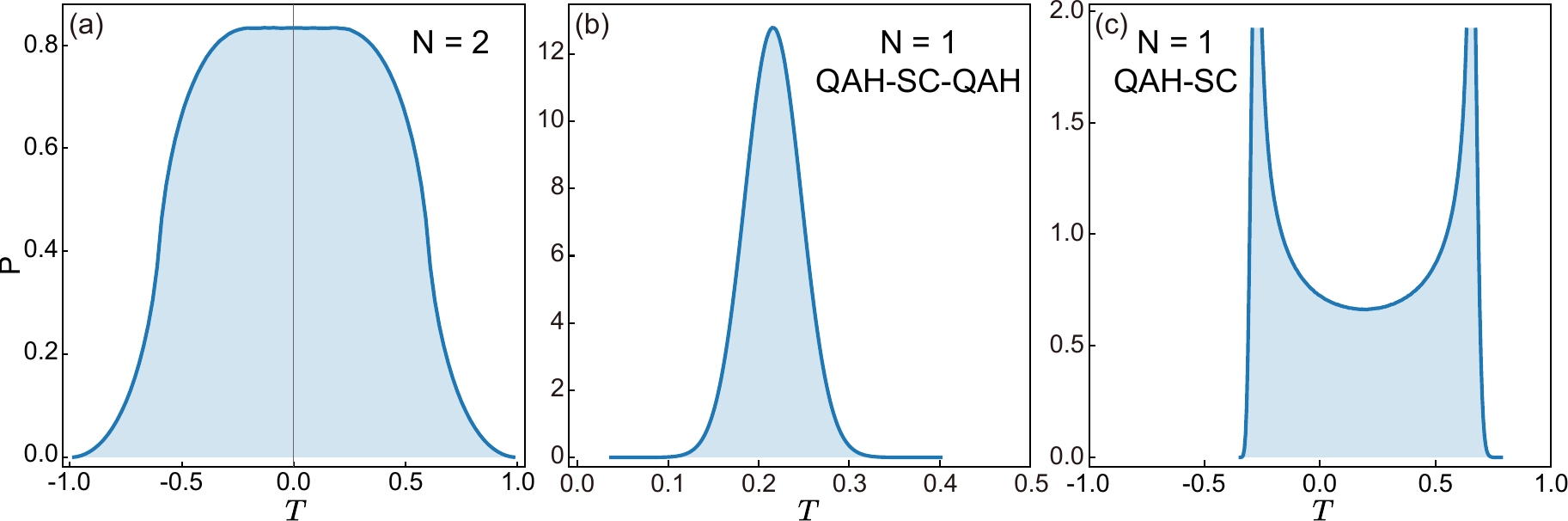}
  \end{center}
  \caption{(a), (b), (c) Decoherence effect in scenarios Figs.~\ref{fig1}(a),~\ref{fig1}(d) and~\ref{fig1}(f), respectively. The distribution is calculated from Monte Carlo sampling~\cite{sm}, and we set $p=0.4$. We set the distribution of $T$ without decoherence follows $U[-1,1]$, $\mathcal{N}(0.3,0.05^2)$ and arcsin distribution with $A=0.6$ for Figs.~\ref{fig1}(a),~\ref{fig1}(d) and~\ref{fig1}(f), respectively.}
  \label{fig4}
\end{figure*}

When the decoherence effect is weak, the resulting effective charge transmission fraction for different scenarios in Figs.~\ref{fig1}(a),~\ref{fig1}(d) and~\ref{fig1}(f) is shown in Fig.~\ref{fig4}, where we focus on the long junction case, i.e., each charge transmission fraction converges to the form of its large $L$ limit. The distribution for $N=2$ CBEMs remains distinct from that for $N=1$ CBEM, indicating that the characteristic behavior of the distribution in the coherent process persists under sufficiently weak decoherence. As the decoherence effect intensifies, characterized by an increase in $p$ and the occurrence of multiple scattering events, the distribution of the scenario in Fig.~\ref{fig1}(a) will become logarithmic-divergent peaked at $T=0$, while the scenarios in Fig.~\ref{fig1}(d) and~\ref{fig1}(f) become bell-shape peaked at finite $T$~\cite{sm}.

If multiple inelastic scattering centers exist, additional floating leads need to be introduced, which effectively divides the TSC region into several segments and leads to a more complicated multi-terminal formula. We consider the simple case when $p=1$, which means that edge modes fully enter every floating lead and completely lose coherence. Then Eq.~(\ref{N2onelead}) reduces to \cite{hu2024resistance}
\begin{equation}\label{N2Nlead}
  T^{\text{eff}} = T_{l,1} T_{1,2} \cdots T_{m,r},
\end{equation}
when there are $m$ leads at the edge, where the subscripts $1$ to $m$ represent virtual floating leads. For QAH-SC junction with $N=1$ CBEM, the effective transmission fraction is instead given by the sum of two paths:
\begin{equation}\label{Nlead2}
  T^{\text{eff}} = T^u_{l,1} T^u_{1,2} \cdots T^u_{m,r} + T^d_{l,1} T^d_{1,2} \cdots T^d_{m,r}.
\end{equation}
When more floating leads are introduced, the distribution of $T^{\text{eff}}$ will become increasingly sharply peaked, and its peak position will shift toward zero \cite{sm}.

We now briefly address the decoherence induced by temperature. Unlike the discrete nature of inelastic scattering centers, the temperature effect can be modeled as a continuous decoherence source by attaching floating leads to every site along the edge while keeping the parameter $p$ very small. Intuitively, in the Bloch-sphere picture, decoherence transforms a pure state located on the Bloch surface into a mixed state inside the sphere. In the case of temperature-induced decoherence, this corresponds to the point gradually shrinking toward the center of the Bloch sphere as the edge state propagates. Consequently, the overall effect is a narrowing of the distribution of $T$, without changing its qualitative form. Since our objective is to use the distribution to distinguish different edge scenarios, temperature-induced decoherence does not qualitatively affect the results, provided the effect remains weak, i.e., low temperature.

\subsection{Particle loss}
The phenomenological treatment of particle loss is straightforward. Since this process corresponds to particles that escape from the device, the parameters $q$ (with values between 0 and 1) can be introduced to describe the reduction of particle number: $\Psi_f=q_m \cdots q_2 q_1 \Psi^0_f$, where $m$ is the number of vortices and $\Psi^0_f$ denotes the state in the absence of particle loss. This factor merely suppresses the amplitude of $T$, which means that particle loss compresses the distribution of $T$ without altering its overall form, similar to the effect of temperature-induced decoherence. Therefore, only inelastic scattering modifies the functional form of the distribution and exerts a qualitative influence, whereas temperature and particle loss have only quantitative effects.

\section{Discussion and conclusion}\label{discussion}
Finally, we discuss the relation between experimentally measurable quantities and the charge transmission fraction introduced above. We restrict the discussion to the linear-response regime, corresponding to a small applied bias. Within this approximation, the charge transmission function can be defined as~\cite{datta1995electronic}
\begin{equation}
\mathcal{T}(\varepsilon)= \int T(\xi)\left[-\frac{\partial f_\text{FD}(\xi-\varepsilon)}{\partial \xi}\right] \mathrm{d}\xi,
\end{equation}
where $f_\text{FD}$ is the Fermi-Dirac distribution, $\varepsilon$ is the incident energy, and we have explicitly written the energy dependence of $T$ defined in previous sections. Roughly the kernel $-\frac{\partial f_\text{FD}(\xi-\varepsilon)}{\partial \xi}=\left(4k_B \tau \cosh^2 \frac{\xi-\varepsilon}{2k_B \tau}\right)^{-1}$ effectively averages $T(\xi)$ over an energy window of order $\varepsilon\pm k_B\tau$, where $\tau$ denotes the temperature. Taking the QAH-SC junction shown in Fig.~\ref{fig1}(f) as an example, the current is from lead 1 to the grounded SC, namely $I_1=-I_s=I$, and voltage leads are on electrodes 2, 3, 4, and 5. Under linear-response approximation, the experimentally measurable  downstream resistance \cite{Lee2017,gul2022andreev,uday2024induced}, a nonlocal differential resistance, is given by the generalized Landauer-B\"uttiker formula~\cite{entin-wohlman2008conductance}
\begin{equation}\label{RD}
R_d \equiv R_{3s} = \frac{V_3-V_s}{I} = \frac{h}{e^2}\frac{\mathcal{T}}{1-\mathcal{T}},
\end{equation}
where $h$ is the Planck constant and $e$ is the electrical
charge. Thus $\mathcal{T}$ can be directly extracted from the measured $R_d$. Other nonlocal resistances in different scenarios are provided in Ref.~\cite{sm}.
In general, the distribution of $\mathcal{T}$ can differ from that of $T$; the mapping between them—given by the energy convolution with the kernel $-\partial f_\text{FD}/\partial \xi$—is analytically complicated and depends on the detailed energy dependence of $T(\xi)$ and on the temperature $\tau$.
However, if the smearing window $k_B\tau$ is much smaller than the superconducting gap, or $T(\xi)$ is not sensitive to $\xi$ (by, e.g., decoherence effects), then $T(\xi)$ varies only weakly and thus the distribution of $\mathcal{T}(\varepsilon)$ approaches that of $T(\varepsilon)$. Under this situation, thermal smearing effect is negligible; therefore the distinct behaviors of $\mathcal{T}$ for the $N=1$ and $N=2$ cases remain observable.

In conclusion, our theory shows that the probability distribution of charge transmission $T$ provides a signature distinguished between the different topological phases of the SC region. As summarized in Table~\ref{table}, for sufficiently long junctions, the distributions for $N = 1 $ and $N = 2$ CBEMs exhibit distinct characteristics. Moreover, these differences remain robust against weak decoherence. Given the stark differences between these distributions, they provide stronger evidence for phase differentiation, offering extra insight than mean alone, which may be insufficient to distinguish between topological phases of the SC region~\cite{uday2024induced}.

\begin{acknowledgments}
We thank Yuanbo Zhang for enlightening discussions. This work is supported by the National Key Research Program of China under Grant No. 2025YFA1411400, the Natural Science Foundation of China through Grants No.~12350404 and No.~12174066, the Innovation Program for Quantum Science and Technology through Grant No.~2021ZD0302600, the Science and Technology Commission of Shanghai Municipality under Grants No.~23JC1400600, No.~24LZ1400100 and No.~2019SHZDZX01, and is sponsored by the ``Shuguang Program'' supported by the Shanghai Education Development Foundation and Shanghai Municipal Education Commission.
\end{acknowledgments}


\begin{thebibliography}{83}%
\makeatletter
\providecommand \@ifxundefined [1]{%
 \@ifx{#1\undefined}
}%
\providecommand \@ifnum [1]{%
 \ifnum #1\expandafter \@firstoftwo
 \else \expandafter \@secondoftwo
 \fi
}%
\providecommand \@ifx [1]{%
 \ifx #1\expandafter \@firstoftwo
 \else \expandafter \@secondoftwo
 \fi
}%
\providecommand \natexlab [1]{#1}%
\providecommand \enquote  [1]{``#1''}%
\providecommand \bibnamefont  [1]{#1}%
\providecommand \bibfnamefont [1]{#1}%
\providecommand \citenamefont [1]{#1}%
\providecommand \href@noop [0]{\@secondoftwo}%
\providecommand \href [0]{\begingroup \@sanitize@url \@href}%
\providecommand \@href[1]{\@@startlink{#1}\@@href}%
\providecommand \@@href[1]{\endgroup#1\@@endlink}%
\providecommand \@sanitize@url [0]{\catcode `\\12\catcode `\$12\catcode
  `\&12\catcode `\#12\catcode `\^12\catcode `\_12\catcode `\%12\relax}%
\providecommand \@@startlink[1]{}%
\providecommand \@@endlink[0]{}%
\providecommand \url  [0]{\begingroup\@sanitize@url \@url }%
\providecommand \@url [1]{\endgroup\@href {#1}{\urlprefix }}%
\providecommand \urlprefix  [0]{URL }%
\providecommand \Eprint [0]{\href }%
\providecommand \doibase [0]{https://doi.org/}%
\providecommand \selectlanguage [0]{\@gobble}%
\providecommand \bibinfo  [0]{\@secondoftwo}%
\providecommand \bibfield  [0]{\@secondoftwo}%
\providecommand \translation [1]{[#1]}%
\providecommand \BibitemOpen [0]{}%
\providecommand \bibitemStop [0]{}%
\providecommand \bibitemNoStop [0]{.\EOS\space}%
\providecommand \EOS [0]{\spacefactor3000\relax}%
\providecommand \BibitemShut  [1]{\csname bibitem#1\endcsname}%
\let\auto@bib@innerbib\@empty
\bibitem [{\citenamefont {Moore}\ and\ \citenamefont
  {Read}(1991)}]{moore1991nonabelions}%
  \BibitemOpen
  \bibfield  {author} {\bibinfo {author} {\bibfnamefont {G.}~\bibnamefont
  {Moore}}\ and\ \bibinfo {author} {\bibfnamefont {N.}~\bibnamefont {Read}},\
  }\bibfield  {title} {\bibinfo {title} {Nonabelions in the fractional quantum
  hall effect},\ }\href {https://doi.org/10.1016/0550-3213(91)90407-O}
  {\bibfield  {journal} {\bibinfo  {journal} {Nucl. Phys. B}\ }\textbf
  {\bibinfo {volume} {360}},\ \bibinfo {pages} {362} (\bibinfo {year}
  {1991})}\BibitemShut {NoStop}%
\bibitem [{\citenamefont {Read}\ and\ \citenamefont
  {Green}(2000)}]{read2000paired}%
  \BibitemOpen
  \bibfield  {author} {\bibinfo {author} {\bibfnamefont {N.}~\bibnamefont
  {Read}}\ and\ \bibinfo {author} {\bibfnamefont {D.}~\bibnamefont {Green}},\
  }\bibfield  {title} {\bibinfo {title} {Paired states of fermions in two
  dimensions with breaking of parity and time-reversal symmetries and the
  fractional quantum hall effect},\ }\href
  {https://doi.org/10.1103/PhysRevB.61.10267} {\bibfield  {journal} {\bibinfo
  {journal} {Phys. Rev. B}\ }\textbf {\bibinfo {volume} {61}},\ \bibinfo
  {pages} {10267} (\bibinfo {year} {2000})}\BibitemShut {NoStop}%
\bibitem [{\citenamefont {Kitaev}(2006)}]{kitaev2006}%
  \BibitemOpen
  \bibfield  {author} {\bibinfo {author} {\bibfnamefont {A.}~\bibnamefont
  {Kitaev}},\ }\bibfield  {title} {\bibinfo {title} {Anyons in an exactly
  solved model and beyond},\ }\href
  {https://doi.org/http://doi.org/10.1016/j.aop.2005.10.005} {\bibfield
  {journal} {\bibinfo  {journal} {Ann. Phys.}\ }\textbf {\bibinfo {volume}
  {321}},\ \bibinfo {pages} {2} (\bibinfo {year} {2006})}\BibitemShut {NoStop}%
\bibitem [{\citenamefont {Fu}\ and\ \citenamefont
  {Kane}(2008)}]{fu2008superconducting}%
  \BibitemOpen
  \bibfield  {author} {\bibinfo {author} {\bibfnamefont {L.}~\bibnamefont
  {Fu}}\ and\ \bibinfo {author} {\bibfnamefont {C.~L.}\ \bibnamefont {Kane}},\
  }\bibfield  {title} {\bibinfo {title} {Superconducting proximity effect and
  majorana fermions at the surface of a topological insulator},\ }\href
  {https://doi.org/10.1103/PhysRevLett.100.096407} {\bibfield  {journal}
  {\bibinfo  {journal} {Phys. Rev. Lett.}\ }\textbf {\bibinfo {volume} {100}},\
  \bibinfo {pages} {096407} (\bibinfo {year} {2008})}\BibitemShut {NoStop}%
\bibitem [{\citenamefont {Wilczek}(2009)}]{wilczek2009majorana}%
  \BibitemOpen
  \bibfield  {author} {\bibinfo {author} {\bibfnamefont {F.}~\bibnamefont
  {Wilczek}},\ }\bibfield  {title} {\bibinfo {title} {Majorana returns},\
  }\href {https://doi.org/10.1038/nphys1380} {\bibfield  {journal} {\bibinfo
  {journal} {Nat. Phys.}\ }\textbf {\bibinfo {volume} {5}},\ \bibinfo {pages}
  {614} (\bibinfo {year} {2009})}\BibitemShut {NoStop}%
\bibitem [{\citenamefont {Qi}\ \emph {et~al.}(2010)\citenamefont {Qi},
  \citenamefont {Hughes},\ and\ \citenamefont {Zhang}}]{qi2010chiral}%
  \BibitemOpen
  \bibfield  {author} {\bibinfo {author} {\bibfnamefont {X.-L.}\ \bibnamefont
  {Qi}}, \bibinfo {author} {\bibfnamefont {T.~L.}\ \bibnamefont {Hughes}},\
  and\ \bibinfo {author} {\bibfnamefont {S.-C.}\ \bibnamefont {Zhang}},\
  }\bibfield  {title} {\bibinfo {title} {Chiral topological superconductor from
  the quantum hall state},\ }\href {https://doi.org/10.1103/PhysRevB.82.184516}
  {\bibfield  {journal} {\bibinfo  {journal} {Phys. Rev. B}\ }\textbf {\bibinfo
  {volume} {82}},\ \bibinfo {pages} {184516} (\bibinfo {year}
  {2010})}\BibitemShut {NoStop}%
\bibitem [{\citenamefont {Sau}\ \emph {et~al.}(2010)\citenamefont {Sau},
  \citenamefont {Lutchyn}, \citenamefont {Tewari},\ and\ \citenamefont
  {Das~Sarma}}]{sau2010generic}%
  \BibitemOpen
  \bibfield  {author} {\bibinfo {author} {\bibfnamefont {J.~D.}\ \bibnamefont
  {Sau}}, \bibinfo {author} {\bibfnamefont {R.~M.}\ \bibnamefont {Lutchyn}},
  \bibinfo {author} {\bibfnamefont {S.}~\bibnamefont {Tewari}},\ and\ \bibinfo
  {author} {\bibfnamefont {S.}~\bibnamefont {Das~Sarma}},\ }\bibfield  {title}
  {\bibinfo {title} {Generic new platform for topological quantum computation
  using semiconductor heterostructures},\ }\href
  {https://doi.org/10.1103/PhysRevLett.104.040502} {\bibfield  {journal}
  {\bibinfo  {journal} {Phys. Rev. Lett.}\ }\textbf {\bibinfo {volume} {104}},\
  \bibinfo {pages} {040502} (\bibinfo {year} {2010})}\BibitemShut {NoStop}%
\bibitem [{\citenamefont {Alicea}(2010)}]{alicea2010majorana}%
  \BibitemOpen
  \bibfield  {author} {\bibinfo {author} {\bibfnamefont {J.}~\bibnamefont
  {Alicea}},\ }\bibfield  {title} {\bibinfo {title} {Majorana fermions in a
  tunable semiconductor device},\ }\href
  {https://doi.org/10.1103/PhysRevB.81.125318} {\bibfield  {journal} {\bibinfo
  {journal} {Phys. Rev. B}\ }\textbf {\bibinfo {volume} {81}},\ \bibinfo
  {pages} {125318} (\bibinfo {year} {2010})}\BibitemShut {NoStop}%
\bibitem [{\citenamefont {Qi}\ and\ \citenamefont
  {Zhang}(2011)}]{qi2011topological}%
  \BibitemOpen
  \bibfield  {author} {\bibinfo {author} {\bibfnamefont {X.-L.}\ \bibnamefont
  {Qi}}\ and\ \bibinfo {author} {\bibfnamefont {S.-C.}\ \bibnamefont {Zhang}},\
  }\bibfield  {title} {\bibinfo {title} {Topological insulators and
  superconductors},\ }\href {https://doi.org/10.1103/RevModPhys.83.1057}
  {\bibfield  {journal} {\bibinfo  {journal} {Rev. Mod. Phys.}\ }\textbf
  {\bibinfo {volume} {83}},\ \bibinfo {pages} {1057} (\bibinfo {year}
  {2011})}\BibitemShut {NoStop}%
\bibitem [{\citenamefont {Elliott}\ and\ \citenamefont
  {Franz}(2015)}]{elliott2015colloquium}%
  \BibitemOpen
  \bibfield  {author} {\bibinfo {author} {\bibfnamefont {S.~R.}\ \bibnamefont
  {Elliott}}\ and\ \bibinfo {author} {\bibfnamefont {M.}~\bibnamefont
  {Franz}},\ }\bibfield  {title} {\bibinfo {title} {Colloquium: Majorana
  fermions in nuclear, particle, and solid-state physics},\ }\href
  {https://doi.org/10.1103/RevModPhys.87.137} {\bibfield  {journal} {\bibinfo
  {journal} {Rev. Mod. Phys.}\ }\textbf {\bibinfo {volume} {87}},\ \bibinfo
  {pages} {137} (\bibinfo {year} {2015})}\BibitemShut {NoStop}%
\bibitem [{\citenamefont {Nayak}\ \emph {et~al.}(2008)\citenamefont {Nayak},
  \citenamefont {Simon}, \citenamefont {Stern}, \citenamefont {Freedman},\ and\
  \citenamefont {Das~Sarma}}]{nayak2008nonabelian}%
  \BibitemOpen
  \bibfield  {author} {\bibinfo {author} {\bibfnamefont {C.}~\bibnamefont
  {Nayak}}, \bibinfo {author} {\bibfnamefont {S.~H.}\ \bibnamefont {Simon}},
  \bibinfo {author} {\bibfnamefont {A.}~\bibnamefont {Stern}}, \bibinfo
  {author} {\bibfnamefont {M.}~\bibnamefont {Freedman}},\ and\ \bibinfo
  {author} {\bibfnamefont {S.}~\bibnamefont {Das~Sarma}},\ }\bibfield  {title}
  {\bibinfo {title} {Non-abelian anyons and topological quantum computation},\
  }\href {https://doi.org/10.1103/RevModPhys.80.1083} {\bibfield  {journal}
  {\bibinfo  {journal} {Rev. Mod. Phys.}\ }\textbf {\bibinfo {volume} {80}},\
  \bibinfo {pages} {1083} (\bibinfo {year} {2008})}\BibitemShut {NoStop}%
\bibitem [{\citenamefont {Mong}\ \emph {et~al.}(2014)\citenamefont {Mong},
  \citenamefont {Clarke}, \citenamefont {Alicea}, \citenamefont {Lindner},
  \citenamefont {Fendley}, \citenamefont {Nayak}, \citenamefont {Oreg},
  \citenamefont {Stern}, \citenamefont {Berg}, \citenamefont {Shtengel},\ and\
  \citenamefont {Fisher}}]{mong2014universal}%
  \BibitemOpen
  \bibfield  {author} {\bibinfo {author} {\bibfnamefont {R.~S.~K.}\
  \bibnamefont {Mong}}, \bibinfo {author} {\bibfnamefont {D.~J.}\ \bibnamefont
  {Clarke}}, \bibinfo {author} {\bibfnamefont {J.}~\bibnamefont {Alicea}},
  \bibinfo {author} {\bibfnamefont {N.~H.}\ \bibnamefont {Lindner}}, \bibinfo
  {author} {\bibfnamefont {P.}~\bibnamefont {Fendley}}, \bibinfo {author}
  {\bibfnamefont {C.}~\bibnamefont {Nayak}}, \bibinfo {author} {\bibfnamefont
  {Y.}~\bibnamefont {Oreg}}, \bibinfo {author} {\bibfnamefont {A.}~\bibnamefont
  {Stern}}, \bibinfo {author} {\bibfnamefont {E.}~\bibnamefont {Berg}},
  \bibinfo {author} {\bibfnamefont {K.}~\bibnamefont {Shtengel}},\ and\
  \bibinfo {author} {\bibfnamefont {M.~P.~A.}\ \bibnamefont {Fisher}},\
  }\bibfield  {title} {\bibinfo {title} {Universal topological quantum
  computation from a superconductor-abelian quantum hall heterostructure},\
  }\href {https://doi.org/10.1103/PhysRevX.4.011036} {\bibfield  {journal}
  {\bibinfo  {journal} {Phys. Rev. X}\ }\textbf {\bibinfo {volume} {4}},\
  \bibinfo {pages} {011036} (\bibinfo {year} {2014})}\BibitemShut {NoStop}%
\bibitem [{\citenamefont {Clarke}\ \emph {et~al.}(2014)\citenamefont {Clarke},
  \citenamefont {Alicea},\ and\ \citenamefont {Shtengel}}]{clarke2014exotic}%
  \BibitemOpen
  \bibfield  {author} {\bibinfo {author} {\bibfnamefont {D.~J.}\ \bibnamefont
  {Clarke}}, \bibinfo {author} {\bibfnamefont {J.}~\bibnamefont {Alicea}},\
  and\ \bibinfo {author} {\bibfnamefont {K.}~\bibnamefont {Shtengel}},\
  }\bibfield  {title} {\bibinfo {title} {Exotic circuit elements from
  zero-modes in hybrid superconductor--quantum-hall systems},\ }\href
  {https://doi.org/10.1038/nphys3114} {\bibfield  {journal} {\bibinfo
  {journal} {Nat. Phys.}\ }\textbf {\bibinfo {volume} {10}},\ \bibinfo {pages}
  {877} (\bibinfo {year} {2014})}\BibitemShut {NoStop}%
\bibitem [{\citenamefont {Lian}\ \emph
  {et~al.}(2018{\natexlab{a}})\citenamefont {Lian}, \citenamefont {Sun},
  \citenamefont {Vaezi}, \citenamefont {Qi},\ and\ \citenamefont
  {Zhang}}]{lian2018topological}%
  \BibitemOpen
  \bibfield  {author} {\bibinfo {author} {\bibfnamefont {B.}~\bibnamefont
  {Lian}}, \bibinfo {author} {\bibfnamefont {X.-Q.}\ \bibnamefont {Sun}},
  \bibinfo {author} {\bibfnamefont {A.}~\bibnamefont {Vaezi}}, \bibinfo
  {author} {\bibfnamefont {X.-L.}\ \bibnamefont {Qi}},\ and\ \bibinfo {author}
  {\bibfnamefont {S.-C.}\ \bibnamefont {Zhang}},\ }\bibfield  {title} {\bibinfo
  {title} {Topological quantum computation based on chiral majorana fermions},\
  }\href {https://doi.org/10.1073/pnas.1810003115} {\bibfield  {journal}
  {\bibinfo  {journal} {Proc. Natl. Acad. Sci. U.S.A.}\ }\textbf {\bibinfo
  {volume} {115}},\ \bibinfo {pages} {10938} (\bibinfo {year}
  {2018}{\natexlab{a}})}\BibitemShut {NoStop}%
\bibitem [{\citenamefont {Hu}\ and\ \citenamefont
  {Kane}(2018)}]{hu2018fibonacci}%
  \BibitemOpen
  \bibfield  {author} {\bibinfo {author} {\bibfnamefont {Y.}~\bibnamefont
  {Hu}}\ and\ \bibinfo {author} {\bibfnamefont {C.~L.}\ \bibnamefont {Kane}},\
  }\bibfield  {title} {\bibinfo {title} {Fibonacci topological
  superconductor},\ }\href {https://doi.org/10.1103/PhysRevLett.120.066801}
  {\bibfield  {journal} {\bibinfo  {journal} {Phys. Rev. Lett.}\ }\textbf
  {\bibinfo {volume} {120}},\ \bibinfo {pages} {066801} (\bibinfo {year}
  {2018})}\BibitemShut {NoStop}%
\bibitem [{\citenamefont {Beenakker}\ \emph {et~al.}(2019)\citenamefont
  {Beenakker}, \citenamefont {Baireuther}, \citenamefont {Herasymenko},
  \citenamefont {Adagideli}, \citenamefont {Wang},\ and\ \citenamefont
  {Akhmerov}}]{Beenakker2019}%
  \BibitemOpen
  \bibfield  {author} {\bibinfo {author} {\bibfnamefont {C.~W.~J.}\
  \bibnamefont {Beenakker}}, \bibinfo {author} {\bibfnamefont {P.}~\bibnamefont
  {Baireuther}}, \bibinfo {author} {\bibfnamefont {Y.}~\bibnamefont
  {Herasymenko}}, \bibinfo {author} {\bibfnamefont {I.}~\bibnamefont
  {Adagideli}}, \bibinfo {author} {\bibfnamefont {L.}~\bibnamefont {Wang}},\
  and\ \bibinfo {author} {\bibfnamefont {A.~R.}\ \bibnamefont {Akhmerov}},\
  }\bibfield  {title} {\bibinfo {title} {Deterministic creation and braiding of
  chiral edge vortices},\ }\href
  {https://doi.org/10.1103/PhysRevLett.122.146803} {\bibfield  {journal}
  {\bibinfo  {journal} {Phys. Rev. Lett.}\ }\textbf {\bibinfo {volume} {122}},\
  \bibinfo {pages} {146803} (\bibinfo {year} {2019})}\BibitemShut {NoStop}%
\bibitem [{\citenamefont {Wan}\ \emph {et~al.}(2015)\citenamefont {Wan},
  \citenamefont {Kazakov}, \citenamefont {Manfra}, \citenamefont {Pfeiffer},
  \citenamefont {West},\ and\ \citenamefont {Rokhinson}}]{Wan2015}%
  \BibitemOpen
  \bibfield  {author} {\bibinfo {author} {\bibfnamefont {Z.}~\bibnamefont
  {Wan}}, \bibinfo {author} {\bibfnamefont {A.}~\bibnamefont {Kazakov}},
  \bibinfo {author} {\bibfnamefont {M.~J.}\ \bibnamefont {Manfra}}, \bibinfo
  {author} {\bibfnamefont {L.~N.}\ \bibnamefont {Pfeiffer}}, \bibinfo {author}
  {\bibfnamefont {K.~W.}\ \bibnamefont {West}},\ and\ \bibinfo {author}
  {\bibfnamefont {L.~P.}\ \bibnamefont {Rokhinson}},\ }\bibfield  {title}
  {\bibinfo {title} {Induced superconductivity in high-mobility two-dimensional
  electron gas in gallium arsenide heterostructures},\ }\href
  {https://doi.org/10.1038/ncomms8426} {\bibfield  {journal} {\bibinfo
  {journal} {Nat. Commun.}\ }\textbf {\bibinfo {volume} {6}},\ \bibinfo {pages}
  {7426} (\bibinfo {year} {2015})}\BibitemShut {NoStop}%
\bibitem [{\citenamefont {Amet}\ \emph {et~al.}(2016)\citenamefont {Amet},
  \citenamefont {Ke}, \citenamefont {Borzenets}, \citenamefont {Wang},
  \citenamefont {Watanabe}, \citenamefont {Taniguchi}, \citenamefont {Deacon},
  \citenamefont {Yamamoto}, \citenamefont {Bomze}, \citenamefont {Tarucha},\
  and\ \citenamefont {Finkelstein}}]{Amet2016}%
  \BibitemOpen
  \bibfield  {author} {\bibinfo {author} {\bibfnamefont {F.}~\bibnamefont
  {Amet}}, \bibinfo {author} {\bibfnamefont {C.~T.}\ \bibnamefont {Ke}},
  \bibinfo {author} {\bibfnamefont {I.~V.}\ \bibnamefont {Borzenets}}, \bibinfo
  {author} {\bibfnamefont {J.}~\bibnamefont {Wang}}, \bibinfo {author}
  {\bibfnamefont {K.}~\bibnamefont {Watanabe}}, \bibinfo {author}
  {\bibfnamefont {T.}~\bibnamefont {Taniguchi}}, \bibinfo {author}
  {\bibfnamefont {R.~S.}\ \bibnamefont {Deacon}}, \bibinfo {author}
  {\bibfnamefont {M.}~\bibnamefont {Yamamoto}}, \bibinfo {author}
  {\bibfnamefont {Y.}~\bibnamefont {Bomze}}, \bibinfo {author} {\bibfnamefont
  {S.}~\bibnamefont {Tarucha}},\ and\ \bibinfo {author} {\bibfnamefont
  {G.}~\bibnamefont {Finkelstein}},\ }\bibfield  {title} {\bibinfo {title}
  {Supercurrent in the quantum hall regime},\ }\href
  {https://doi.org/10.1126/science.aad6203} {\bibfield  {journal} {\bibinfo
  {journal} {Science}\ }\textbf {\bibinfo {volume} {352}},\ \bibinfo {pages}
  {966} (\bibinfo {year} {2016})}\BibitemShut {NoStop}%
\bibitem [{\citenamefont {Lee}\ \emph {et~al.}(2017)\citenamefont {Lee},
  \citenamefont {Huang}, \citenamefont {Efetov}, \citenamefont {Wei},
  \citenamefont {Hart}, \citenamefont {Taniguchi}, \citenamefont {Watanabe},
  \citenamefont {Yacoby},\ and\ \citenamefont {Kim}}]{Lee2017}%
  \BibitemOpen
  \bibfield  {author} {\bibinfo {author} {\bibfnamefont {G.-H.}\ \bibnamefont
  {Lee}}, \bibinfo {author} {\bibfnamefont {K.-F.}\ \bibnamefont {Huang}},
  \bibinfo {author} {\bibfnamefont {D.~K.}\ \bibnamefont {Efetov}}, \bibinfo
  {author} {\bibfnamefont {D.~S.}\ \bibnamefont {Wei}}, \bibinfo {author}
  {\bibfnamefont {S.}~\bibnamefont {Hart}}, \bibinfo {author} {\bibfnamefont
  {T.}~\bibnamefont {Taniguchi}}, \bibinfo {author} {\bibfnamefont
  {K.}~\bibnamefont {Watanabe}}, \bibinfo {author} {\bibfnamefont
  {A.}~\bibnamefont {Yacoby}},\ and\ \bibinfo {author} {\bibfnamefont
  {P.}~\bibnamefont {Kim}},\ }\bibfield  {title} {\bibinfo {title} {Inducing
  superconducting correlation in quantum hall edge states},\ }\href
  {https://doi.org/10.1038/nphys4084} {\bibfield  {journal} {\bibinfo
  {journal} {Nat. Phys.}\ }\textbf {\bibinfo {volume} {13}},\ \bibinfo {pages}
  {693} (\bibinfo {year} {2017})}\BibitemShut {NoStop}%
\bibitem [{\citenamefont {Sahu}\ \emph {et~al.}(2018)\citenamefont {Sahu},
  \citenamefont {Liu}, \citenamefont {Paul}, \citenamefont {Das}, \citenamefont
  {Raychaudhuri}, \citenamefont {Jain},\ and\ \citenamefont {Das}}]{Sahu2018}%
  \BibitemOpen
  \bibfield  {author} {\bibinfo {author} {\bibfnamefont {M.~R.}\ \bibnamefont
  {Sahu}}, \bibinfo {author} {\bibfnamefont {X.}~\bibnamefont {Liu}}, \bibinfo
  {author} {\bibfnamefont {A.~K.}\ \bibnamefont {Paul}}, \bibinfo {author}
  {\bibfnamefont {S.}~\bibnamefont {Das}}, \bibinfo {author} {\bibfnamefont
  {P.}~\bibnamefont {Raychaudhuri}}, \bibinfo {author} {\bibfnamefont {J.~K.}\
  \bibnamefont {Jain}},\ and\ \bibinfo {author} {\bibfnamefont
  {A.}~\bibnamefont {Das}},\ }\bibfield  {title} {\bibinfo {title}
  {Inter-landau-level andreev reflection at the dirac point in a graphene
  quantum hall state coupled to a ${\mathrm{nbse}}_{2}$ superconductor},\
  }\href {https://doi.org/10.1103/PhysRevLett.121.086809} {\bibfield  {journal}
  {\bibinfo  {journal} {Phys. Rev. Lett.}\ }\textbf {\bibinfo {volume} {121}},\
  \bibinfo {pages} {086809} (\bibinfo {year} {2018})}\BibitemShut {NoStop}%
\bibitem [{\citenamefont {Matsuo}\ \emph {et~al.}(2018)\citenamefont {Matsuo},
  \citenamefont {Ueda}, \citenamefont {Baba}, \citenamefont {Kamata},
  \citenamefont {Tateno}, \citenamefont {Shabani}, \citenamefont {Palmstrøm},\
  and\ \citenamefont {Tarucha}}]{Matsuo2018}%
  \BibitemOpen
  \bibfield  {author} {\bibinfo {author} {\bibfnamefont {S.}~\bibnamefont
  {Matsuo}}, \bibinfo {author} {\bibfnamefont {K.}~\bibnamefont {Ueda}},
  \bibinfo {author} {\bibfnamefont {S.}~\bibnamefont {Baba}}, \bibinfo {author}
  {\bibfnamefont {H.}~\bibnamefont {Kamata}}, \bibinfo {author} {\bibfnamefont
  {M.}~\bibnamefont {Tateno}}, \bibinfo {author} {\bibfnamefont
  {J.}~\bibnamefont {Shabani}}, \bibinfo {author} {\bibfnamefont {C.~J.}\
  \bibnamefont {Palmstrøm}},\ and\ \bibinfo {author} {\bibfnamefont
  {S.}~\bibnamefont {Tarucha}},\ }\bibfield  {title} {\bibinfo {title}
  {Equal-spin andreev reflection on junctions of spin-resolved quantum hall
  bulk state and spin-singlet superconductor},\ }\href
  {https://doi.org/10.1038/s41598-018-21707-0} {\bibfield  {journal} {\bibinfo
  {journal} {Sci. Rep.}\ }\textbf {\bibinfo {volume} {8}},\ \bibinfo {pages}
  {3454} (\bibinfo {year} {2018})}\BibitemShut {NoStop}%
\bibitem [{\citenamefont {Seredinski}\ \emph {et~al.}(2019)\citenamefont
  {Seredinski}, \citenamefont {Draelos}, \citenamefont {Arnault}, \citenamefont
  {Wei}, \citenamefont {Li}, \citenamefont {Fleming}, \citenamefont {Watanabe},
  \citenamefont {Taniguchi}, \citenamefont {Amet},\ and\ \citenamefont
  {Finkelstein}}]{Seredinski2019}%
  \BibitemOpen
  \bibfield  {author} {\bibinfo {author} {\bibfnamefont {A.}~\bibnamefont
  {Seredinski}}, \bibinfo {author} {\bibfnamefont {A.~W.}\ \bibnamefont
  {Draelos}}, \bibinfo {author} {\bibfnamefont {E.~G.}\ \bibnamefont
  {Arnault}}, \bibinfo {author} {\bibfnamefont {M.-T.}\ \bibnamefont {Wei}},
  \bibinfo {author} {\bibfnamefont {H.}~\bibnamefont {Li}}, \bibinfo {author}
  {\bibfnamefont {T.}~\bibnamefont {Fleming}}, \bibinfo {author} {\bibfnamefont
  {K.}~\bibnamefont {Watanabe}}, \bibinfo {author} {\bibfnamefont
  {T.}~\bibnamefont {Taniguchi}}, \bibinfo {author} {\bibfnamefont
  {F.}~\bibnamefont {Amet}},\ and\ \bibinfo {author} {\bibfnamefont
  {G.}~\bibnamefont {Finkelstein}},\ }\bibfield  {title} {\bibinfo {title}
  {Quantum hall-based superconducting interference device},\ }\href
  {https://doi.org/10.1126/sciadv.aaw8693} {\bibfield  {journal} {\bibinfo
  {journal} {Sci. Adv.}\ }\textbf {\bibinfo {volume} {5}},\ \bibinfo {pages}
  {eaaw8693} (\bibinfo {year} {2019})}\BibitemShut {NoStop}%
\bibitem [{\citenamefont {Vignaud}\ \emph {et~al.}(2023)\citenamefont
  {Vignaud}, \citenamefont {Perconte}, \citenamefont {Yang}, \citenamefont
  {Kousar}, \citenamefont {Wagner}, \citenamefont {Gay}, \citenamefont
  {Watanabe}, \citenamefont {Taniguchi}, \citenamefont {Courtois},
  \citenamefont {Han}, \citenamefont {Sellier},\ and\ \citenamefont
  {Sacépé}}]{vignaud_2023}%
  \BibitemOpen
  \bibfield  {author} {\bibinfo {author} {\bibfnamefont {H.}~\bibnamefont
  {Vignaud}}, \bibinfo {author} {\bibfnamefont {D.}~\bibnamefont {Perconte}},
  \bibinfo {author} {\bibfnamefont {W.}~\bibnamefont {Yang}}, \bibinfo {author}
  {\bibfnamefont {B.}~\bibnamefont {Kousar}}, \bibinfo {author} {\bibfnamefont
  {E.}~\bibnamefont {Wagner}}, \bibinfo {author} {\bibfnamefont
  {F.}~\bibnamefont {Gay}}, \bibinfo {author} {\bibfnamefont {K.}~\bibnamefont
  {Watanabe}}, \bibinfo {author} {\bibfnamefont {T.}~\bibnamefont {Taniguchi}},
  \bibinfo {author} {\bibfnamefont {H.}~\bibnamefont {Courtois}}, \bibinfo
  {author} {\bibfnamefont {Z.}~\bibnamefont {Han}}, \bibinfo {author}
  {\bibfnamefont {H.}~\bibnamefont {Sellier}},\ and\ \bibinfo {author}
  {\bibfnamefont {B.}~\bibnamefont {Sacépé}},\ }\bibfield  {title} {\bibinfo
  {title} {Evidence for chiral supercurrent in quantum hall josephson
  junctions},\ }\href {https://doi.org/10.1038/s41586-023-06764-4} {\bibfield
  {journal} {\bibinfo  {journal} {Nature}\ }\textbf {\bibinfo {volume} {624}},\
  \bibinfo {pages} {545} (\bibinfo {year} {2023})}\BibitemShut {NoStop}%
\bibitem [{\citenamefont {Zhao}\ \emph {et~al.}(2020)\citenamefont {Zhao},
  \citenamefont {Arnault}, \citenamefont {Bondarev}, \citenamefont
  {Seredinski}, \citenamefont {Larson}, \citenamefont {Draelos}, \citenamefont
  {Li}, \citenamefont {Watanabe}, \citenamefont {Taniguchi}, \citenamefont
  {Amet}, \citenamefont {Baranger},\ and\ \citenamefont
  {Finkelstein}}]{zhao2020interference}%
  \BibitemOpen
  \bibfield  {author} {\bibinfo {author} {\bibfnamefont {L.}~\bibnamefont
  {Zhao}}, \bibinfo {author} {\bibfnamefont {E.~G.}\ \bibnamefont {Arnault}},
  \bibinfo {author} {\bibfnamefont {A.}~\bibnamefont {Bondarev}}, \bibinfo
  {author} {\bibfnamefont {A.}~\bibnamefont {Seredinski}}, \bibinfo {author}
  {\bibfnamefont {T.~F.~Q.}\ \bibnamefont {Larson}}, \bibinfo {author}
  {\bibfnamefont {A.~W.}\ \bibnamefont {Draelos}}, \bibinfo {author}
  {\bibfnamefont {H.}~\bibnamefont {Li}}, \bibinfo {author} {\bibfnamefont
  {K.}~\bibnamefont {Watanabe}}, \bibinfo {author} {\bibfnamefont
  {T.}~\bibnamefont {Taniguchi}}, \bibinfo {author} {\bibfnamefont
  {F.}~\bibnamefont {Amet}}, \bibinfo {author} {\bibfnamefont {H.~U.}\
  \bibnamefont {Baranger}},\ and\ \bibinfo {author} {\bibfnamefont
  {G.}~\bibnamefont {Finkelstein}},\ }\bibfield  {title} {\bibinfo {title}
  {Interference of chiral andreev edge states},\ }\href
  {https://doi.org/10.1038/s41567-020-0898-5} {\bibfield  {journal} {\bibinfo
  {journal} {Nature Phys.}\ }\textbf {\bibinfo {volume} {16}},\ \bibinfo
  {pages} {862} (\bibinfo {year} {2020})}\BibitemShut {NoStop}%
\bibitem [{\citenamefont {G{\"u}l}\ \emph {et~al.}(2022)\citenamefont
  {G{\"u}l}, \citenamefont {Ronen}, \citenamefont {Lee}, \citenamefont
  {Shapourian}, \citenamefont {Zauberman}, \citenamefont {Lee}, \citenamefont
  {Watanabe}, \citenamefont {Taniguchi}, \citenamefont {Vishwanath},
  \citenamefont {Yacoby},\ and\ \citenamefont {Kim}}]{gul2022andreev}%
  \BibitemOpen
  \bibfield  {author} {\bibinfo {author} {\bibfnamefont {{\"O}.}~\bibnamefont
  {G{\"u}l}}, \bibinfo {author} {\bibfnamefont {Y.}~\bibnamefont {Ronen}},
  \bibinfo {author} {\bibfnamefont {S.~Y.}\ \bibnamefont {Lee}}, \bibinfo
  {author} {\bibfnamefont {H.}~\bibnamefont {Shapourian}}, \bibinfo {author}
  {\bibfnamefont {J.}~\bibnamefont {Zauberman}}, \bibinfo {author}
  {\bibfnamefont {Y.~H.}\ \bibnamefont {Lee}}, \bibinfo {author} {\bibfnamefont
  {K.}~\bibnamefont {Watanabe}}, \bibinfo {author} {\bibfnamefont
  {T.}~\bibnamefont {Taniguchi}}, \bibinfo {author} {\bibfnamefont
  {A.}~\bibnamefont {Vishwanath}}, \bibinfo {author} {\bibfnamefont
  {A.}~\bibnamefont {Yacoby}},\ and\ \bibinfo {author} {\bibfnamefont
  {P.}~\bibnamefont {Kim}},\ }\bibfield  {title} {\bibinfo {title} {Andreev
  reflection in the fractional quantum hall state},\ }\href
  {https://doi.org/10.1103/PhysRevX.12.021057} {\bibfield  {journal} {\bibinfo
  {journal} {Phys. Rev. X}\ }\textbf {\bibinfo {volume} {12}},\ \bibinfo
  {pages} {021057} (\bibinfo {year} {2022})}\BibitemShut {NoStop}%
\bibitem [{\citenamefont {Zhao}\ \emph {et~al.}(2023)\citenamefont {Zhao},
  \citenamefont {Iftikhar}, \citenamefont {Larson}, \citenamefont {Arnault},
  \citenamefont {Watanabe}, \citenamefont {Taniguchi}, \citenamefont {Amet},\
  and\ \citenamefont {Finkelstein}}]{zhao2023loss}%
  \BibitemOpen
  \bibfield  {author} {\bibinfo {author} {\bibfnamefont {L.}~\bibnamefont
  {Zhao}}, \bibinfo {author} {\bibfnamefont {Z.}~\bibnamefont {Iftikhar}},
  \bibinfo {author} {\bibfnamefont {T.~F.~Q.}\ \bibnamefont {Larson}}, \bibinfo
  {author} {\bibfnamefont {E.~G.}\ \bibnamefont {Arnault}}, \bibinfo {author}
  {\bibfnamefont {K.}~\bibnamefont {Watanabe}}, \bibinfo {author}
  {\bibfnamefont {T.}~\bibnamefont {Taniguchi}}, \bibinfo {author}
  {\bibfnamefont {F.}~\bibnamefont {Amet}},\ and\ \bibinfo {author}
  {\bibfnamefont {G.}~\bibnamefont {Finkelstein}},\ }\bibfield  {title}
  {\bibinfo {title} {Loss and decoherence at the quantum hall-superconductor
  interface},\ }\href {https://doi.org/10.1103/PhysRevLett.131.176604}
  {\bibfield  {journal} {\bibinfo  {journal} {Phys. Rev. Lett.}\ }\textbf
  {\bibinfo {volume} {131}},\ \bibinfo {pages} {176604} (\bibinfo {year}
  {2023})}\BibitemShut {NoStop}%
\bibitem [{\citenamefont {Hatefipour}\ \emph {et~al.}(2022)\citenamefont
  {Hatefipour}, \citenamefont {Cuozzo}, \citenamefont {Kanter}, \citenamefont
  {Strickland}, \citenamefont {Allemang}, \citenamefont {Lu}, \citenamefont
  {Rossi},\ and\ \citenamefont {Shabani}}]{hatefipour2022induced}%
  \BibitemOpen
  \bibfield  {author} {\bibinfo {author} {\bibfnamefont {M.}~\bibnamefont
  {Hatefipour}}, \bibinfo {author} {\bibfnamefont {J.~J.}\ \bibnamefont
  {Cuozzo}}, \bibinfo {author} {\bibfnamefont {J.}~\bibnamefont {Kanter}},
  \bibinfo {author} {\bibfnamefont {W.~M.}\ \bibnamefont {Strickland}},
  \bibinfo {author} {\bibfnamefont {C.~R.}\ \bibnamefont {Allemang}}, \bibinfo
  {author} {\bibfnamefont {T.-M.}\ \bibnamefont {Lu}}, \bibinfo {author}
  {\bibfnamefont {E.}~\bibnamefont {Rossi}},\ and\ \bibinfo {author}
  {\bibfnamefont {J.}~\bibnamefont {Shabani}},\ }\bibfield  {title} {\bibinfo
  {title} {Induced superconducting pairing in integer quantum hall edge
  states},\ }\href {https://doi.org/10.1021/acs.nanolett.2c01413} {\bibfield
  {journal} {\bibinfo  {journal} {Nano Lett.}\ }\textbf {\bibinfo {volume}
  {22}},\ \bibinfo {pages} {6173} (\bibinfo {year} {2022})}\BibitemShut
  {NoStop}%
\bibitem [{\citenamefont {Uday}\ \emph {et~al.}(2024)\citenamefont {Uday},
  \citenamefont {Lippertz}, \citenamefont {Moors}, \citenamefont {Legg},
  \citenamefont {Joris}, \citenamefont {Bliesener}, \citenamefont {Pereira},
  \citenamefont {Taskin},\ and\ \citenamefont {Ando}}]{uday2024induced}%
  \BibitemOpen
  \bibfield  {author} {\bibinfo {author} {\bibfnamefont {A.}~\bibnamefont
  {Uday}}, \bibinfo {author} {\bibfnamefont {G.}~\bibnamefont {Lippertz}},
  \bibinfo {author} {\bibfnamefont {K.}~\bibnamefont {Moors}}, \bibinfo
  {author} {\bibfnamefont {H.~F.}\ \bibnamefont {Legg}}, \bibinfo {author}
  {\bibfnamefont {R.}~\bibnamefont {Joris}}, \bibinfo {author} {\bibfnamefont
  {A.}~\bibnamefont {Bliesener}}, \bibinfo {author} {\bibfnamefont {L.~M.~C.}\
  \bibnamefont {Pereira}}, \bibinfo {author} {\bibfnamefont {A.~A.}\
  \bibnamefont {Taskin}},\ and\ \bibinfo {author} {\bibfnamefont
  {Y.}~\bibnamefont {Ando}},\ }\bibfield  {title} {\bibinfo {title} {Induced
  superconducting correlations in a quantum anomalous hall insulator},\ }\href
  {https://doi.org/10.1038/s41567-024-02574-1} {\bibfield  {journal} {\bibinfo
  {journal} {Nat. Phys.}\ }\textbf {\bibinfo {volume} {20}},\ \bibinfo {pages}
  {1589} (\bibinfo {year} {2024})}\BibitemShut {NoStop}%
\bibitem [{\citenamefont {Sato}\ \emph {et~al.}(2024)\citenamefont {Sato},
  \citenamefont {Nagahama}, \citenamefont {Belopolski}, \citenamefont
  {Yoshimi}, \citenamefont {Kawamura}, \citenamefont {Tsukazaki}, \citenamefont
  {Kanazawa}, \citenamefont {Takahashi}, \citenamefont {Kawasaki},\ and\
  \citenamefont {Tokura}}]{sato2024}%
  \BibitemOpen
  \bibfield  {author} {\bibinfo {author} {\bibfnamefont {Y.}~\bibnamefont
  {Sato}}, \bibinfo {author} {\bibfnamefont {S.}~\bibnamefont {Nagahama}},
  \bibinfo {author} {\bibfnamefont {I.}~\bibnamefont {Belopolski}}, \bibinfo
  {author} {\bibfnamefont {R.}~\bibnamefont {Yoshimi}}, \bibinfo {author}
  {\bibfnamefont {M.}~\bibnamefont {Kawamura}}, \bibinfo {author}
  {\bibfnamefont {A.}~\bibnamefont {Tsukazaki}}, \bibinfo {author}
  {\bibfnamefont {N.}~\bibnamefont {Kanazawa}}, \bibinfo {author}
  {\bibfnamefont {K.~S.}\ \bibnamefont {Takahashi}}, \bibinfo {author}
  {\bibfnamefont {M.}~\bibnamefont {Kawasaki}},\ and\ \bibinfo {author}
  {\bibfnamefont {Y.}~\bibnamefont {Tokura}},\ }\bibfield  {title} {\bibinfo
  {title} {Molecular beam epitaxy of superconducting
  $\mathrm{FeS}{\mathrm{e}}_{x}\mathrm{T}{\mathrm{e}}_{1\ensuremath{-}x}$ thin
  films interfaced with magnetic topological insulators},\ }\href
  {https://doi.org/10.1103/PhysRevMaterials.8.L041801} {\bibfield  {journal}
  {\bibinfo  {journal} {Phys. Rev. Mater.}\ }\textbf {\bibinfo {volume} {8}},\
  \bibinfo {pages} {L041801} (\bibinfo {year} {2024})}\BibitemShut {NoStop}%
\bibitem [{\citenamefont {Wang}\ and\ \citenamefont {Liu}(2024)}]{wang2024}%
  \BibitemOpen
  \bibfield  {author} {\bibinfo {author} {\bibfnamefont {J.}~\bibnamefont
  {Wang}}\ and\ \bibinfo {author} {\bibfnamefont {Z.}~\bibnamefont {Liu}},\
  }\bibfield  {title} {\bibinfo {title} {A way to cross the andreev bridge},\
  }\href {https://doi.org/10.1038/s41567-024-02575-0} {\bibfield  {journal}
  {\bibinfo  {journal} {Nat. Phys.}\ }\textbf {\bibinfo {volume} {20}},\
  \bibinfo {pages} {1525–1526} (\bibinfo {year} {2024})}\BibitemShut
  {NoStop}%
\bibitem [{\citenamefont {Hoppe}\ \emph {et~al.}(2000)\citenamefont {Hoppe},
  \citenamefont {Z\"ulicke},\ and\ \citenamefont {Sch\"on}}]{Hoppe2000}%
  \BibitemOpen
  \bibfield  {author} {\bibinfo {author} {\bibfnamefont {H.}~\bibnamefont
  {Hoppe}}, \bibinfo {author} {\bibfnamefont {U.}~\bibnamefont {Z\"ulicke}},\
  and\ \bibinfo {author} {\bibfnamefont {G.}~\bibnamefont {Sch\"on}},\
  }\bibfield  {title} {\bibinfo {title} {Andreev reflection in strong magnetic
  fields},\ }\href {https://doi.org/10.1103/PhysRevLett.84.1804} {\bibfield
  {journal} {\bibinfo  {journal} {Phys. Rev. Lett.}\ }\textbf {\bibinfo
  {volume} {84}},\ \bibinfo {pages} {1804} (\bibinfo {year}
  {2000})}\BibitemShut {NoStop}%
\bibitem [{\citenamefont {Giazotto}\ \emph {et~al.}(2005)\citenamefont
  {Giazotto}, \citenamefont {Governale}, \citenamefont {Z\"ulicke},\ and\
  \citenamefont {Beltram}}]{Giazotto2005}%
  \BibitemOpen
  \bibfield  {author} {\bibinfo {author} {\bibfnamefont {F.}~\bibnamefont
  {Giazotto}}, \bibinfo {author} {\bibfnamefont {M.}~\bibnamefont {Governale}},
  \bibinfo {author} {\bibfnamefont {U.}~\bibnamefont {Z\"ulicke}},\ and\
  \bibinfo {author} {\bibfnamefont {F.}~\bibnamefont {Beltram}},\ }\bibfield
  {title} {\bibinfo {title} {Andreev reflection and cyclotron motion at
  superconductor---normal-metal interfaces},\ }\href
  {https://doi.org/10.1103/PhysRevB.72.054518} {\bibfield  {journal} {\bibinfo
  {journal} {Phys. Rev. B}\ }\textbf {\bibinfo {volume} {72}},\ \bibinfo
  {pages} {054518} (\bibinfo {year} {2005})}\BibitemShut {NoStop}%
\bibitem [{\citenamefont {Akhmerov}\ and\ \citenamefont
  {Beenakker}(2007)}]{Akhmerov2007}%
  \BibitemOpen
  \bibfield  {author} {\bibinfo {author} {\bibfnamefont {A.~R.}\ \bibnamefont
  {Akhmerov}}\ and\ \bibinfo {author} {\bibfnamefont {C.~W.~J.}\ \bibnamefont
  {Beenakker}},\ }\bibfield  {title} {\bibinfo {title} {Detection of valley
  polarization in graphene by a superconducting contact},\ }\href
  {https://doi.org/10.1103/PhysRevLett.98.157003} {\bibfield  {journal}
  {\bibinfo  {journal} {Phys. Rev. Lett.}\ }\textbf {\bibinfo {volume} {98}},\
  \bibinfo {pages} {157003} (\bibinfo {year} {2007})}\BibitemShut {NoStop}%
\bibitem [{\citenamefont {van Ostaay}\ \emph {et~al.}(2011)\citenamefont {van
  Ostaay}, \citenamefont {Akhmerov},\ and\ \citenamefont
  {Beenakker}}]{Ostaay2011}%
  \BibitemOpen
  \bibfield  {author} {\bibinfo {author} {\bibfnamefont {J.~A.~M.}\
  \bibnamefont {van Ostaay}}, \bibinfo {author} {\bibfnamefont {A.~R.}\
  \bibnamefont {Akhmerov}},\ and\ \bibinfo {author} {\bibfnamefont {C.~W.~J.}\
  \bibnamefont {Beenakker}},\ }\bibfield  {title} {\bibinfo {title}
  {Spin-triplet supercurrent carried by quantum hall edge states through a
  josephson junction},\ }\href {https://doi.org/10.1103/PhysRevB.83.195441}
  {\bibfield  {journal} {\bibinfo  {journal} {Phys. Rev. B}\ }\textbf {\bibinfo
  {volume} {83}},\ \bibinfo {pages} {195441} (\bibinfo {year}
  {2011})}\BibitemShut {NoStop}%
\bibitem [{\citenamefont {Fu}\ and\ \citenamefont
  {Kane}(2009)}]{fu2009probing}%
  \BibitemOpen
  \bibfield  {author} {\bibinfo {author} {\bibfnamefont {L.}~\bibnamefont
  {Fu}}\ and\ \bibinfo {author} {\bibfnamefont {C.~L.}\ \bibnamefont {Kane}},\
  }\bibfield  {title} {\bibinfo {title} {Probing neutral majorana fermion edge
  modes with charge transport},\ }\href
  {https://doi.org/10.1103/PhysRevLett.102.216403} {\bibfield  {journal}
  {\bibinfo  {journal} {Phys. Rev. Lett.}\ }\textbf {\bibinfo {volume} {102}},\
  \bibinfo {pages} {216403} (\bibinfo {year} {2009})}\BibitemShut {NoStop}%
\bibitem [{\citenamefont {Akhmerov}\ \emph {et~al.}(2009)\citenamefont
  {Akhmerov}, \citenamefont {Nilsson},\ and\ \citenamefont
  {Beenakker}}]{akhmerov2009electrically}%
  \BibitemOpen
  \bibfield  {author} {\bibinfo {author} {\bibfnamefont {A.~R.}\ \bibnamefont
  {Akhmerov}}, \bibinfo {author} {\bibfnamefont {J.}~\bibnamefont {Nilsson}},\
  and\ \bibinfo {author} {\bibfnamefont {C.~W.~J.}\ \bibnamefont {Beenakker}},\
  }\bibfield  {title} {\bibinfo {title} {Electrically detected interferometry
  of majorana fermions in a topological insulator},\ }\href
  {https://doi.org/10.1103/PhysRevLett.102.216404} {\bibfield  {journal}
  {\bibinfo  {journal} {Phys. Rev. Lett.}\ }\textbf {\bibinfo {volume} {102}},\
  \bibinfo {pages} {216404} (\bibinfo {year} {2009})}\BibitemShut {NoStop}%
\bibitem [{\citenamefont {Chung}\ \emph {et~al.}(2011)\citenamefont {Chung},
  \citenamefont {Qi}, \citenamefont {Maciejko},\ and\ \citenamefont
  {Zhang}}]{chung2011conductance}%
  \BibitemOpen
  \bibfield  {author} {\bibinfo {author} {\bibfnamefont {S.~B.}\ \bibnamefont
  {Chung}}, \bibinfo {author} {\bibfnamefont {X.-L.}\ \bibnamefont {Qi}},
  \bibinfo {author} {\bibfnamefont {J.}~\bibnamefont {Maciejko}},\ and\
  \bibinfo {author} {\bibfnamefont {S.-C.}\ \bibnamefont {Zhang}},\ }\bibfield
  {title} {\bibinfo {title} {Conductance and noise signatures of majorana
  backscattering},\ }\href {https://doi.org/10.1103/PhysRevB.83.100512}
  {\bibfield  {journal} {\bibinfo  {journal} {Phys. Rev. B}\ }\textbf {\bibinfo
  {volume} {83}},\ \bibinfo {pages} {100512} (\bibinfo {year}
  {2011})}\BibitemShut {NoStop}%
\bibitem [{\citenamefont {Wang}\ \emph {et~al.}(2015)\citenamefont {Wang},
  \citenamefont {Zhou}, \citenamefont {Lian},\ and\ \citenamefont
  {Zhang}}]{wang2015chiral}%
  \BibitemOpen
  \bibfield  {author} {\bibinfo {author} {\bibfnamefont {J.}~\bibnamefont
  {Wang}}, \bibinfo {author} {\bibfnamefont {Q.}~\bibnamefont {Zhou}}, \bibinfo
  {author} {\bibfnamefont {B.}~\bibnamefont {Lian}},\ and\ \bibinfo {author}
  {\bibfnamefont {S.-C.}\ \bibnamefont {Zhang}},\ }\bibfield  {title} {\bibinfo
  {title} {Chiral topological superconductor and half-integer conductance
  plateau from quantum anomalous hall plateau transition},\ }\href
  {https://doi.org/10.1103/PhysRevB.92.064520} {\bibfield  {journal} {\bibinfo
  {journal} {Phys. Rev. B}\ }\textbf {\bibinfo {volume} {92}},\ \bibinfo
  {pages} {064520} (\bibinfo {year} {2015})}\BibitemShut {NoStop}%
\bibitem [{\citenamefont {Str{\"u}bi}\ \emph {et~al.}(2011)\citenamefont
  {Str{\"u}bi}, \citenamefont {Belzig}, \citenamefont {Choi},\ and\
  \citenamefont {Bruder}}]{strubi2011interferometric}%
  \BibitemOpen
  \bibfield  {author} {\bibinfo {author} {\bibfnamefont {G.}~\bibnamefont
  {Str{\"u}bi}}, \bibinfo {author} {\bibfnamefont {W.}~\bibnamefont {Belzig}},
  \bibinfo {author} {\bibfnamefont {M.-S.}\ \bibnamefont {Choi}},\ and\
  \bibinfo {author} {\bibfnamefont {C.}~\bibnamefont {Bruder}},\ }\bibfield
  {title} {\bibinfo {title} {Interferometric and noise signatures of majorana
  fermion edge states in transport experiments},\ }\href
  {https://doi.org/10.1103/PhysRevLett.107.136403} {\bibfield  {journal}
  {\bibinfo  {journal} {Phys. Rev. Lett.}\ }\textbf {\bibinfo {volume} {107}},\
  \bibinfo {pages} {136403} (\bibinfo {year} {2011})}\BibitemShut {NoStop}%
\bibitem [{\citenamefont {Lian}\ \emph {et~al.}(2016)\citenamefont {Lian},
  \citenamefont {Wang},\ and\ \citenamefont
  {Zhang}}]{lian2016edgestateinduced}%
  \BibitemOpen
  \bibfield  {author} {\bibinfo {author} {\bibfnamefont {B.}~\bibnamefont
  {Lian}}, \bibinfo {author} {\bibfnamefont {J.}~\bibnamefont {Wang}},\ and\
  \bibinfo {author} {\bibfnamefont {S.-C.}\ \bibnamefont {Zhang}},\ }\bibfield
  {title} {\bibinfo {title} {Edge-state-induced andreev oscillation in quantum
  anomalous hall insulator-superconductor junctions},\ }\href
  {https://doi.org/10.1103/PhysRevB.93.161401} {\bibfield  {journal} {\bibinfo
  {journal} {Phys. Rev. B}\ }\textbf {\bibinfo {volume} {93}},\ \bibinfo
  {pages} {161401} (\bibinfo {year} {2016})}\BibitemShut {NoStop}%
\bibitem [{\citenamefont {Lian}\ \emph
  {et~al.}(2018{\natexlab{b}})\citenamefont {Lian}, \citenamefont {Wang},
  \citenamefont {Sun}, \citenamefont {Vaezi},\ and\ \citenamefont
  {Zhang}}]{lian2018quantum}%
  \BibitemOpen
  \bibfield  {author} {\bibinfo {author} {\bibfnamefont {B.}~\bibnamefont
  {Lian}}, \bibinfo {author} {\bibfnamefont {J.}~\bibnamefont {Wang}}, \bibinfo
  {author} {\bibfnamefont {X.-Q.}\ \bibnamefont {Sun}}, \bibinfo {author}
  {\bibfnamefont {A.}~\bibnamefont {Vaezi}},\ and\ \bibinfo {author}
  {\bibfnamefont {S.-C.}\ \bibnamefont {Zhang}},\ }\bibfield  {title} {\bibinfo
  {title} {Quantum phase transition of chiral majorana fermions in the presence
  of disorder},\ }\href {https://doi.org/10.1103/PhysRevB.97.125408} {\bibfield
   {journal} {\bibinfo  {journal} {Phys. Rev. B}\ }\textbf {\bibinfo {volume}
  {97}},\ \bibinfo {pages} {125408} (\bibinfo {year}
  {2018}{\natexlab{b}})}\BibitemShut {NoStop}%
\bibitem [{\citenamefont {Wang}\ and\ \citenamefont
  {Lian}(2018)}]{wang2018multiple}%
  \BibitemOpen
  \bibfield  {author} {\bibinfo {author} {\bibfnamefont {J.}~\bibnamefont
  {Wang}}\ and\ \bibinfo {author} {\bibfnamefont {B.}~\bibnamefont {Lian}},\
  }\bibfield  {title} {\bibinfo {title} {Multiple chiral majorana fermion modes
  and quantum transport},\ }\href
  {https://doi.org/10.1103/PhysRevLett.121.256801} {\bibfield  {journal}
  {\bibinfo  {journal} {Phys. Rev. Lett.}\ }\textbf {\bibinfo {volume} {121}},\
  \bibinfo {pages} {256801} (\bibinfo {year} {2018})}\BibitemShut {NoStop}%
\bibitem [{\citenamefont {Beconcini}\ \emph {et~al.}(2018)\citenamefont
  {Beconcini}, \citenamefont {Polini},\ and\ \citenamefont
  {Taddei}}]{Beconcini2018}%
  \BibitemOpen
  \bibfield  {author} {\bibinfo {author} {\bibfnamefont {M.}~\bibnamefont
  {Beconcini}}, \bibinfo {author} {\bibfnamefont {M.}~\bibnamefont {Polini}},\
  and\ \bibinfo {author} {\bibfnamefont {F.}~\bibnamefont {Taddei}},\
  }\bibfield  {title} {\bibinfo {title} {Nonlocal superconducting correlations
  in graphene in the quantum hall regime},\ }\href
  {https://doi.org/10.1103/PhysRevB.97.201403} {\bibfield  {journal} {\bibinfo
  {journal} {Phys. Rev. B}\ }\textbf {\bibinfo {volume} {97}},\ \bibinfo
  {pages} {201403} (\bibinfo {year} {2018})}\BibitemShut {NoStop}%
\bibitem [{\citenamefont {Lian}\ and\ \citenamefont
  {Wang}(2019)}]{lian2019distribution}%
  \BibitemOpen
  \bibfield  {author} {\bibinfo {author} {\bibfnamefont {B.}~\bibnamefont
  {Lian}}\ and\ \bibinfo {author} {\bibfnamefont {J.}~\bibnamefont {Wang}},\
  }\bibfield  {title} {\bibinfo {title} {Distribution of conductances in chiral
  topological superconductor junctions},\ }\href
  {https://doi.org/10.1103/PhysRevB.99.041404} {\bibfield  {journal} {\bibinfo
  {journal} {Phys. Rev. B}\ }\textbf {\bibinfo {volume} {99}},\ \bibinfo
  {pages} {041404} (\bibinfo {year} {2019})}\BibitemShut {NoStop}%
\bibitem [{\citenamefont {He}\ \emph {et~al.}(2019)\citenamefont {He},
  \citenamefont {Liang}, \citenamefont {Tanaka},\ and\ \citenamefont
  {Nagaosa}}]{he2019platform}%
  \BibitemOpen
  \bibfield  {author} {\bibinfo {author} {\bibfnamefont {J.~J.}\ \bibnamefont
  {He}}, \bibinfo {author} {\bibfnamefont {T.}~\bibnamefont {Liang}}, \bibinfo
  {author} {\bibfnamefont {Y.}~\bibnamefont {Tanaka}},\ and\ \bibinfo {author}
  {\bibfnamefont {N.}~\bibnamefont {Nagaosa}},\ }\bibfield  {title} {\bibinfo
  {title} {Platform of chiral majorana edge modes and its quantum transport
  phenomena},\ }\href {https://doi.org/10.1038/s42005-019-0250-5} {\bibfield
  {journal} {\bibinfo  {journal} {Commun. Phys.}\ }\textbf {\bibinfo {volume}
  {2}},\ \bibinfo {pages} {149} (\bibinfo {year} {2019})}\BibitemShut {NoStop}%
\bibitem [{\citenamefont {Manesco}\ \emph {et~al.}(2022)\citenamefont
  {Manesco}, \citenamefont {Fl{\'o}r}, \citenamefont {Liu},\ and\ \citenamefont
  {Akhmerov}}]{manesco2022mechanisms}%
  \BibitemOpen
  \bibfield  {author} {\bibinfo {author} {\bibfnamefont {A.}~\bibnamefont
  {Manesco}}, \bibinfo {author} {\bibfnamefont {I.~M.}\ \bibnamefont
  {Fl{\'o}r}}, \bibinfo {author} {\bibfnamefont {C.-X.}\ \bibnamefont {Liu}},\
  and\ \bibinfo {author} {\bibfnamefont {A.}~\bibnamefont {Akhmerov}},\
  }\bibfield  {title} {\bibinfo {title} {Mechanisms of andreev reflection in
  quantum hall graphene},\ }\href
  {https://doi.org/10.21468/SciPostPhysCore.5.3.045} {\bibfield  {journal}
  {\bibinfo  {journal} {SciPost Phys. Core}\ }\textbf {\bibinfo {volume} {5}},\
  \bibinfo {pages} {045} (\bibinfo {year} {2022})}\BibitemShut {NoStop}%
\bibitem [{\citenamefont {Tang}\ \emph {et~al.}(2022)\citenamefont {Tang},
  \citenamefont {Knapp},\ and\ \citenamefont {Alicea}}]{tang2022}%
  \BibitemOpen
  \bibfield  {author} {\bibinfo {author} {\bibfnamefont {Y.}~\bibnamefont
  {Tang}}, \bibinfo {author} {\bibfnamefont {C.}~\bibnamefont {Knapp}},\ and\
  \bibinfo {author} {\bibfnamefont {J.}~\bibnamefont {Alicea}},\ }\bibfield
  {title} {\bibinfo {title} {Vortex-enabled andreev processes in quantum
  hall--superconductor hybrids},\ }\href
  {https://doi.org/10.1103/PhysRevB.106.245411} {\bibfield  {journal} {\bibinfo
   {journal} {Phys. Rev. B}\ }\textbf {\bibinfo {volume} {106}},\ \bibinfo
  {pages} {245411} (\bibinfo {year} {2022})}\BibitemShut {NoStop}%
\bibitem [{\citenamefont {Schiller}\ \emph {et~al.}(2023)\citenamefont
  {Schiller}, \citenamefont {Katzir}, \citenamefont {Stern}, \citenamefont
  {Berg}, \citenamefont {Lindner},\ and\ \citenamefont
  {Oreg}}]{schiller2023superconductivity}%
  \BibitemOpen
  \bibfield  {author} {\bibinfo {author} {\bibfnamefont {N.}~\bibnamefont
  {Schiller}}, \bibinfo {author} {\bibfnamefont {B.~A.}\ \bibnamefont
  {Katzir}}, \bibinfo {author} {\bibfnamefont {A.}~\bibnamefont {Stern}},
  \bibinfo {author} {\bibfnamefont {E.}~\bibnamefont {Berg}}, \bibinfo {author}
  {\bibfnamefont {N.~H.}\ \bibnamefont {Lindner}},\ and\ \bibinfo {author}
  {\bibfnamefont {Y.}~\bibnamefont {Oreg}},\ }\bibfield  {title} {\bibinfo
  {title} {Superconductivity and fermionic dissipation in quantum hall edges},\
  }\href {https://doi.org/10.1103/PhysRevB.107.L161105} {\bibfield  {journal}
  {\bibinfo  {journal} {Phys. Rev. B}\ }\textbf {\bibinfo {volume} {107}},\
  \bibinfo {pages} {L161105} (\bibinfo {year} {2023})}\BibitemShut {NoStop}%
\bibitem [{\citenamefont {Kurilovich}\ \emph {et~al.}(2023)\citenamefont
  {Kurilovich}, \citenamefont {Raines},\ and\ \citenamefont
  {Glazman}}]{kurilovich2023disorderenabled}%
  \BibitemOpen
  \bibfield  {author} {\bibinfo {author} {\bibfnamefont {V.~D.}\ \bibnamefont
  {Kurilovich}}, \bibinfo {author} {\bibfnamefont {Z.~M.}\ \bibnamefont
  {Raines}},\ and\ \bibinfo {author} {\bibfnamefont {L.~I.}\ \bibnamefont
  {Glazman}},\ }\bibfield  {title} {\bibinfo {title} {Disorder-enabled andreev
  reflection of a quantum hall edge},\ }\href
  {https://doi.org/10.1038/s41467-023-37794-1} {\bibfield  {journal} {\bibinfo
  {journal} {Nat. Commun}\ }\textbf {\bibinfo {volume} {14}},\ \bibinfo {pages}
  {2237} (\bibinfo {year} {2023})}\BibitemShut {NoStop}%
\bibitem [{\citenamefont {Hu}\ \emph {et~al.}(2024)\citenamefont {Hu},
  \citenamefont {Wang},\ and\ \citenamefont {Lian}}]{hu2024resistance}%
  \BibitemOpen
  \bibfield  {author} {\bibinfo {author} {\bibfnamefont {Y.}~\bibnamefont
  {Hu}}, \bibinfo {author} {\bibfnamefont {J.}~\bibnamefont {Wang}},\ and\
  \bibinfo {author} {\bibfnamefont {B.}~\bibnamefont {Lian}},\ }\href@noop {}
  {\bibinfo {title} {Resistance distribution of decoherent quantum
  hall-superconductor edges}} (\bibinfo {year} {2024}),\ \Eprint
  {https://arxiv.org/abs/2405.17550} {arXiv:2405.17550 [cond-mat]} \BibitemShut
  {NoStop}%
\bibitem [{\citenamefont {Nava}\ \emph {et~al.}(2024)\citenamefont {Nava},
  \citenamefont {Egger}, \citenamefont {Hassler},\ and\ \citenamefont
  {Giuliano}}]{nava2024nonabeliana}%
  \BibitemOpen
  \bibfield  {author} {\bibinfo {author} {\bibfnamefont {A.}~\bibnamefont
  {Nava}}, \bibinfo {author} {\bibfnamefont {R.}~\bibnamefont {Egger}},
  \bibinfo {author} {\bibfnamefont {F.}~\bibnamefont {Hassler}},\ and\ \bibinfo
  {author} {\bibfnamefont {D.}~\bibnamefont {Giuliano}},\ }\bibfield  {title}
  {\bibinfo {title} {Non-abelian anyon statistics through ac conductance of a
  majorana interferometer},\ }\href
  {https://doi.org/10.1103/PhysRevLett.133.146604} {\bibfield  {journal}
  {\bibinfo  {journal} {Phys. Rev. Lett.}\ }\textbf {\bibinfo {volume} {133}},\
  \bibinfo {pages} {146604} (\bibinfo {year} {2024})}\BibitemShut {NoStop}%
\bibitem [{\citenamefont {Bondarev}\ \emph {et~al.}(2025)\citenamefont
  {Bondarev}, \citenamefont {Zhang},\ and\ \citenamefont
  {Baranger}}]{bondarev2025}%
  \BibitemOpen
  \bibfield  {author} {\bibinfo {author} {\bibfnamefont {A.}~\bibnamefont
  {Bondarev}}, \bibinfo {author} {\bibfnamefont {G.}~\bibnamefont {Zhang}},\
  and\ \bibinfo {author} {\bibfnamefont {H.~U.}\ \bibnamefont {Baranger}},\
  }\bibfield  {title} {\bibinfo {title} {Transparent graphene-superconductor
  interfaces: Quantum hall and zero field regimes},\ }\href
  {https://doi.org/10.1103/gjcm-qfgz} {\bibfield  {journal} {\bibinfo
  {journal} {Phys. Rev. B}\ }\textbf {\bibinfo {volume} {111}},\ \bibinfo
  {pages} {235444} (\bibinfo {year} {2025})}\BibitemShut {NoStop}%
\bibitem [{\citenamefont {Katayama}\ \emph {et~al.}(2025)\citenamefont
  {Katayama}, \citenamefont {Schnyder}, \citenamefont {Asano},\ and\
  \citenamefont {Ikegaya}}]{katayama2025noisetocurrent}%
  \BibitemOpen
  \bibfield  {author} {\bibinfo {author} {\bibfnamefont {L.}~\bibnamefont
  {Katayama}}, \bibinfo {author} {\bibfnamefont {A.~P.}\ \bibnamefont
  {Schnyder}}, \bibinfo {author} {\bibfnamefont {Y.}~\bibnamefont {Asano}},\
  and\ \bibinfo {author} {\bibfnamefont {S.}~\bibnamefont {Ikegaya}},\
  }\bibfield  {title} {\bibinfo {title} {Giant noise-to-current ratio as a
  signature of dispersing majorana edge modes},\ }\href
  {https://doi.org/10.1103/4qpb-5syl} {\bibfield  {journal} {\bibinfo
  {journal} {Phys. Rev. B}\ }\textbf {\bibinfo {volume} {112}},\ \bibinfo
  {pages} {054501} (\bibinfo {year} {2025})}\BibitemShut {NoStop}%
\bibitem [{\citenamefont {Chang}\ \emph {et~al.}(2013)\citenamefont {Chang},
  \citenamefont {Zhang}, \citenamefont {Feng}, \citenamefont {Shen},
  \citenamefont {Zhang}, \citenamefont {Guo}, \citenamefont {Li}, \citenamefont
  {Ou}, \citenamefont {Wei}, \citenamefont {Wang}, \citenamefont {Ji},
  \citenamefont {Feng}, \citenamefont {Ji}, \citenamefont {Chen}, \citenamefont
  {Jia}, \citenamefont {Dai}, \citenamefont {Fang}, \citenamefont {Zhang},
  \citenamefont {He}, \citenamefont {Wang}, \citenamefont {Lu}, \citenamefont
  {Ma},\ and\ \citenamefont {Xue}}]{chang2013experimental}%
  \BibitemOpen
  \bibfield  {author} {\bibinfo {author} {\bibfnamefont {C.-Z.}\ \bibnamefont
  {Chang}}, \bibinfo {author} {\bibfnamefont {J.}~\bibnamefont {Zhang}},
  \bibinfo {author} {\bibfnamefont {X.}~\bibnamefont {Feng}}, \bibinfo {author}
  {\bibfnamefont {J.}~\bibnamefont {Shen}}, \bibinfo {author} {\bibfnamefont
  {Z.}~\bibnamefont {Zhang}}, \bibinfo {author} {\bibfnamefont
  {M.}~\bibnamefont {Guo}}, \bibinfo {author} {\bibfnamefont {K.}~\bibnamefont
  {Li}}, \bibinfo {author} {\bibfnamefont {Y.}~\bibnamefont {Ou}}, \bibinfo
  {author} {\bibfnamefont {P.}~\bibnamefont {Wei}}, \bibinfo {author}
  {\bibfnamefont {L.-L.}\ \bibnamefont {Wang}}, \bibinfo {author}
  {\bibfnamefont {Z.-Q.}\ \bibnamefont {Ji}}, \bibinfo {author} {\bibfnamefont
  {Y.}~\bibnamefont {Feng}}, \bibinfo {author} {\bibfnamefont {S.}~\bibnamefont
  {Ji}}, \bibinfo {author} {\bibfnamefont {X.}~\bibnamefont {Chen}}, \bibinfo
  {author} {\bibfnamefont {J.}~\bibnamefont {Jia}}, \bibinfo {author}
  {\bibfnamefont {X.}~\bibnamefont {Dai}}, \bibinfo {author} {\bibfnamefont
  {Z.}~\bibnamefont {Fang}}, \bibinfo {author} {\bibfnamefont {S.-C.}\
  \bibnamefont {Zhang}}, \bibinfo {author} {\bibfnamefont {K.}~\bibnamefont
  {He}}, \bibinfo {author} {\bibfnamefont {Y.}~\bibnamefont {Wang}}, \bibinfo
  {author} {\bibfnamefont {L.}~\bibnamefont {Lu}}, \bibinfo {author}
  {\bibfnamefont {X.-C.}\ \bibnamefont {Ma}},\ and\ \bibinfo {author}
  {\bibfnamefont {Q.-K.}\ \bibnamefont {Xue}},\ }\bibfield  {title} {\bibinfo
  {title} {Experimental observation of the quantum anomalous hall effect in a
  magnetic topological insulator},\ }\href
  {https://doi.org/10.1126/science.1234414} {\bibfield  {journal} {\bibinfo
  {journal} {Science}\ }\textbf {\bibinfo {volume} {340}},\ \bibinfo {pages}
  {167} (\bibinfo {year} {2013})}\BibitemShut {NoStop}%
\bibitem [{\citenamefont {Checkelsky}\ \emph {et~al.}(2014)\citenamefont
  {Checkelsky}, \citenamefont {Yoshimi}, \citenamefont {Tsukazaki},
  \citenamefont {Takahashi}, \citenamefont {Kozuka}, \citenamefont {Falson},
  \citenamefont {Kawasaki},\ and\ \citenamefont {Tokura}}]{Checkelsky_2014}%
  \BibitemOpen
  \bibfield  {author} {\bibinfo {author} {\bibfnamefont {J.~G.}\ \bibnamefont
  {Checkelsky}}, \bibinfo {author} {\bibfnamefont {R.}~\bibnamefont {Yoshimi}},
  \bibinfo {author} {\bibfnamefont {A.}~\bibnamefont {Tsukazaki}}, \bibinfo
  {author} {\bibfnamefont {K.~S.}\ \bibnamefont {Takahashi}}, \bibinfo {author}
  {\bibfnamefont {Y.}~\bibnamefont {Kozuka}}, \bibinfo {author} {\bibfnamefont
  {J.}~\bibnamefont {Falson}}, \bibinfo {author} {\bibfnamefont
  {M.}~\bibnamefont {Kawasaki}},\ and\ \bibinfo {author} {\bibfnamefont
  {Y.}~\bibnamefont {Tokura}},\ }\bibfield  {title} {\bibinfo {title}
  {Trajectory of the anomalous hall effect towards the quantized state in a
  ferromagnetic topological insulator},\ }\href
  {https://doi.org/10.1038/nphys3053} {\bibfield  {journal} {\bibinfo
  {journal} {Nat. Phys.}\ }\textbf {\bibinfo {volume} {10}},\ \bibinfo {pages}
  {731} (\bibinfo {year} {2014})}\BibitemShut {NoStop}%
\bibitem [{\citenamefont {Kou}\ \emph {et~al.}(2014)\citenamefont {Kou},
  \citenamefont {Guo}, \citenamefont {Fan}, \citenamefont {Pan}, \citenamefont
  {Lang}, \citenamefont {Jiang}, \citenamefont {Shao}, \citenamefont {Nie},
  \citenamefont {Murata}, \citenamefont {Tang}, \citenamefont {Wang},
  \citenamefont {He}, \citenamefont {Lee}, \citenamefont {Lee},\ and\
  \citenamefont {Wang}}]{kou2014}%
  \BibitemOpen
  \bibfield  {author} {\bibinfo {author} {\bibfnamefont {X.}~\bibnamefont
  {Kou}}, \bibinfo {author} {\bibfnamefont {S.-T.}\ \bibnamefont {Guo}},
  \bibinfo {author} {\bibfnamefont {Y.}~\bibnamefont {Fan}}, \bibinfo {author}
  {\bibfnamefont {L.}~\bibnamefont {Pan}}, \bibinfo {author} {\bibfnamefont
  {M.}~\bibnamefont {Lang}}, \bibinfo {author} {\bibfnamefont {Y.}~\bibnamefont
  {Jiang}}, \bibinfo {author} {\bibfnamefont {Q.}~\bibnamefont {Shao}},
  \bibinfo {author} {\bibfnamefont {T.}~\bibnamefont {Nie}}, \bibinfo {author}
  {\bibfnamefont {K.}~\bibnamefont {Murata}}, \bibinfo {author} {\bibfnamefont
  {J.}~\bibnamefont {Tang}}, \bibinfo {author} {\bibfnamefont {Y.}~\bibnamefont
  {Wang}}, \bibinfo {author} {\bibfnamefont {L.}~\bibnamefont {He}}, \bibinfo
  {author} {\bibfnamefont {T.-K.}\ \bibnamefont {Lee}}, \bibinfo {author}
  {\bibfnamefont {W.-L.}\ \bibnamefont {Lee}},\ and\ \bibinfo {author}
  {\bibfnamefont {K.~L.}\ \bibnamefont {Wang}},\ }\bibfield  {title} {\bibinfo
  {title} {Scale-invariant quantum anomalous hall effect in magnetic
  topological insulators beyond the two-dimensional limit},\ }\href
  {https://doi.org/10.1103/PhysRevLett.113.137201} {\bibfield  {journal}
  {\bibinfo  {journal} {Phys. Rev. Lett.}\ }\textbf {\bibinfo {volume} {113}},\
  \bibinfo {pages} {137201} (\bibinfo {year} {2014})}\BibitemShut {NoStop}%
\bibitem [{\citenamefont {Mogi}\ \emph {et~al.}(2015)\citenamefont {Mogi},
  \citenamefont {Yoshimi}, \citenamefont {Tsukazaki}, \citenamefont {Yasuda},
  \citenamefont {Kozuka}, \citenamefont {Takahashi}, \citenamefont {Kawasaki},\
  and\ \citenamefont {Tokura}}]{mogi2015}%
  \BibitemOpen
  \bibfield  {author} {\bibinfo {author} {\bibfnamefont {M.}~\bibnamefont
  {Mogi}}, \bibinfo {author} {\bibfnamefont {R.}~\bibnamefont {Yoshimi}},
  \bibinfo {author} {\bibfnamefont {A.}~\bibnamefont {Tsukazaki}}, \bibinfo
  {author} {\bibfnamefont {K.}~\bibnamefont {Yasuda}}, \bibinfo {author}
  {\bibfnamefont {Y.}~\bibnamefont {Kozuka}}, \bibinfo {author} {\bibfnamefont
  {K.~S.}\ \bibnamefont {Takahashi}}, \bibinfo {author} {\bibfnamefont
  {M.}~\bibnamefont {Kawasaki}},\ and\ \bibinfo {author} {\bibfnamefont
  {Y.}~\bibnamefont {Tokura}},\ }\bibfield  {title} {\bibinfo {title} {Magnetic
  modulation doping in topological insulators toward higher-temperature quantum
  anomalous hall effect},\ }\href {https://doi.org/10.1063/1.4935075}
  {\bibfield  {journal} {\bibinfo  {journal} {Appl. Phys. Lett.}\ }\textbf
  {\bibinfo {volume} {107}},\ \bibinfo {pages} {182401} (\bibinfo {year}
  {2015})}\BibitemShut {NoStop}%
\bibitem [{\citenamefont {Deng}\ \emph {et~al.}(2020)\citenamefont {Deng},
  \citenamefont {Yu}, \citenamefont {Shi}, \citenamefont {Guo}, \citenamefont
  {Xu}, \citenamefont {Wang}, \citenamefont {Chen},\ and\ \citenamefont
  {Zhang}}]{deng2020quantum}%
  \BibitemOpen
  \bibfield  {author} {\bibinfo {author} {\bibfnamefont {Y.}~\bibnamefont
  {Deng}}, \bibinfo {author} {\bibfnamefont {Y.}~\bibnamefont {Yu}}, \bibinfo
  {author} {\bibfnamefont {M.~Z.}\ \bibnamefont {Shi}}, \bibinfo {author}
  {\bibfnamefont {Z.}~\bibnamefont {Guo}}, \bibinfo {author} {\bibfnamefont
  {Z.}~\bibnamefont {Xu}}, \bibinfo {author} {\bibfnamefont {J.}~\bibnamefont
  {Wang}}, \bibinfo {author} {\bibfnamefont {X.~H.}\ \bibnamefont {Chen}},\
  and\ \bibinfo {author} {\bibfnamefont {Y.}~\bibnamefont {Zhang}},\ }\bibfield
   {title} {\bibinfo {title} {Quantum anomalous hall effect in intrinsic
  magnetic topological insulator mnbi2te4},\ }\href
  {https://doi.org/10.1126/science.aax8156} {\bibfield  {journal} {\bibinfo
  {journal} {Science}\ }\textbf {\bibinfo {volume} {367}},\ \bibinfo {pages}
  {895} (\bibinfo {year} {2020})}\BibitemShut {NoStop}%
\bibitem [{\citenamefont {Huang}\ and\ \citenamefont {Wang}(2025)}]{huang2025}%
  \BibitemOpen
  \bibfield  {author} {\bibinfo {author} {\bibfnamefont {L.}~\bibnamefont
  {Huang}}\ and\ \bibinfo {author} {\bibfnamefont {J.}~\bibnamefont {Wang}},\
  }\bibfield  {title} {\bibinfo {title} {Edge optical effect as a probe of
  chiral topological superconductors},\ }\href
  {https://doi.org/10.1103/PhysRevB.111.L020501} {\bibfield  {journal}
  {\bibinfo  {journal} {Phys. Rev. B}\ }\textbf {\bibinfo {volume} {111}},\
  \bibinfo {pages} {L020501} (\bibinfo {year} {2025})}\BibitemShut {NoStop}%
\bibitem [{\citenamefont {Andreev}(1965)}]{andreev1965thermal}%
  \BibitemOpen
  \bibfield  {author} {\bibinfo {author} {\bibfnamefont {A.~F.}\ \bibnamefont
  {Andreev}},\ }\bibfield  {title} {\bibinfo {title} {Thermal conductivity of
  the intermediate state of superconductors ii},\ }\href@noop {} {\bibfield
  {journal} {\bibinfo  {journal} {Sov. Phys. JETP}\ }\textbf {\bibinfo {volume}
  {20}},\ \bibinfo {pages} {1490} (\bibinfo {year} {1965})}\BibitemShut
  {NoStop}%
\bibitem [{\citenamefont {Blonder}\ \emph {et~al.}(1982)\citenamefont
  {Blonder}, \citenamefont {Tinkham},\ and\ \citenamefont
  {Klapwijk}}]{blonder1982transition}%
  \BibitemOpen
  \bibfield  {author} {\bibinfo {author} {\bibfnamefont {G.~E.}\ \bibnamefont
  {Blonder}}, \bibinfo {author} {\bibfnamefont {M.}~\bibnamefont {Tinkham}},\
  and\ \bibinfo {author} {\bibfnamefont {T.~M.}\ \bibnamefont {Klapwijk}},\
  }\bibfield  {title} {\bibinfo {title} {Transition from metallic to tunneling
  regimes in superconducting microconstrictions: Excess current, charge
  imbalance, and supercurrent conversion},\ }\href
  {https://doi.org/10.1103/PhysRevB.25.4515} {\bibfield  {journal} {\bibinfo
  {journal} {Phys. Rev. B}\ }\textbf {\bibinfo {volume} {25}},\ \bibinfo
  {pages} {4515} (\bibinfo {year} {1982})}\BibitemShut {NoStop}%
\bibitem [{\citenamefont {Lee}\ \emph {et~al.}(1987)\citenamefont {Lee},
  \citenamefont {Stone},\ and\ \citenamefont {Fukuyama}}]{lee1987universal}%
  \BibitemOpen
  \bibfield  {author} {\bibinfo {author} {\bibfnamefont {P.~A.}\ \bibnamefont
  {Lee}}, \bibinfo {author} {\bibfnamefont {A.~D.}\ \bibnamefont {Stone}},\
  and\ \bibinfo {author} {\bibfnamefont {H.}~\bibnamefont {Fukuyama}},\
  }\bibfield  {title} {\bibinfo {title} {Universal conductance fluctuations in
  metals: Effects of finite temperature, interactions, and magnetic field},\
  }\href {https://doi.org/10.1103/PhysRevB.35.1039} {\bibfield  {journal}
  {\bibinfo  {journal} {Phys. Rev. B}\ }\textbf {\bibinfo {volume} {35}},\
  \bibinfo {pages} {1039} (\bibinfo {year} {1987})}\BibitemShut {NoStop}%
\bibitem [{\citenamefont {Zhang}\ and\ \citenamefont
  {Liu}(2020)}]{zhang2020disordered}%
  \BibitemOpen
  \bibfield  {author} {\bibinfo {author} {\bibfnamefont {J.-X.}\ \bibnamefont
  {Zhang}}\ and\ \bibinfo {author} {\bibfnamefont {C.-X.}\ \bibnamefont
  {Liu}},\ }\bibfield  {title} {\bibinfo {title} {Disordered quantum transport
  in quantum anomalous hall insulator-superconductor junctions},\ }\href
  {https://doi.org/10.1103/PhysRevB.102.144513} {\bibfield  {journal} {\bibinfo
   {journal} {Phys. Rev. B}\ }\textbf {\bibinfo {volume} {102}},\ \bibinfo
  {pages} {144513} (\bibinfo {year} {2020})}\BibitemShut {NoStop}%
\bibitem [{sm()}]{sm}%
  \BibitemOpen
  \href@noop {} {\bibinfo {title} {See {{Supplemental Material}} at url for
  more details, which includes
  refs.~\cite{cui2016exact,Zhang2019topological}}}\BibitemShut {NoStop}%
\bibitem [{\citenamefont {Fisher}(1994)}]{fisher1994cooperpair}%
  \BibitemOpen
  \bibfield  {author} {\bibinfo {author} {\bibfnamefont {M.~P.~A.}\
  \bibnamefont {Fisher}},\ }\bibfield  {title} {\bibinfo {title} {Cooper-pair
  tunneling into a quantum hall fluid},\ }\href
  {https://doi.org/10.1103/PhysRevB.49.14550} {\bibfield  {journal} {\bibinfo
  {journal} {Phys. Rev. B}\ }\textbf {\bibinfo {volume} {49}},\ \bibinfo
  {pages} {14550} (\bibinfo {year} {1994})}\BibitemShut {NoStop}%
\bibitem [{\citenamefont {B{\'e}ri}\ \emph {et~al.}(2009)\citenamefont
  {B{\'e}ri}, \citenamefont {Kupferschmidt}, \citenamefont {Beenakker},\ and\
  \citenamefont {Brouwer}}]{beri2009quantum}%
  \BibitemOpen
  \bibfield  {author} {\bibinfo {author} {\bibfnamefont {B.}~\bibnamefont
  {B{\'e}ri}}, \bibinfo {author} {\bibfnamefont {J.~N.}\ \bibnamefont
  {Kupferschmidt}}, \bibinfo {author} {\bibfnamefont {C.~W.~J.}\ \bibnamefont
  {Beenakker}},\ and\ \bibinfo {author} {\bibfnamefont {P.~W.}\ \bibnamefont
  {Brouwer}},\ }\bibfield  {title} {\bibinfo {title} {Quantum limit of the
  triplet proximity effect in half-metal--superconductor junctions},\ }\href
  {https://doi.org/10.1103/PhysRevB.79.024517} {\bibfield  {journal} {\bibinfo
  {journal} {Phys. Rev. B}\ }\textbf {\bibinfo {volume} {79}},\ \bibinfo
  {pages} {024517} (\bibinfo {year} {2009})}\BibitemShut {NoStop}%
\bibitem [{\citenamefont {Galambos}\ \emph {et~al.}(2022)\citenamefont
  {Galambos}, \citenamefont {Ronetti}, \citenamefont {Het{\'e}nyi},
  \citenamefont {Loss},\ and\ \citenamefont {Klinovaja}}]{galambos2022crossed}%
  \BibitemOpen
  \bibfield  {author} {\bibinfo {author} {\bibfnamefont {T.~H.}\ \bibnamefont
  {Galambos}}, \bibinfo {author} {\bibfnamefont {F.}~\bibnamefont {Ronetti}},
  \bibinfo {author} {\bibfnamefont {B.}~\bibnamefont {Het{\'e}nyi}}, \bibinfo
  {author} {\bibfnamefont {D.}~\bibnamefont {Loss}},\ and\ \bibinfo {author}
  {\bibfnamefont {J.}~\bibnamefont {Klinovaja}},\ }\bibfield  {title} {\bibinfo
  {title} {Crossed andreev reflection in spin-polarized chiral edge states due
  to the meissner effect},\ }\href
  {https://doi.org/10.1103/PhysRevB.106.075410} {\bibfield  {journal} {\bibinfo
   {journal} {Phys. Rev. B}\ }\textbf {\bibinfo {volume} {106}},\ \bibinfo
  {pages} {075410} (\bibinfo {year} {2022})}\BibitemShut {NoStop}%
\bibitem [{\citenamefont {Kurilovich}\ and\ \citenamefont
  {Glazman}(2023)}]{kurilovich2023criticality}%
  \BibitemOpen
  \bibfield  {author} {\bibinfo {author} {\bibfnamefont {V.~D.}\ \bibnamefont
  {Kurilovich}}\ and\ \bibinfo {author} {\bibfnamefont {L.~I.}\ \bibnamefont
  {Glazman}},\ }\bibfield  {title} {\bibinfo {title} {Criticality in the
  crossed andreev reflection of a quantum hall edge},\ }\href
  {https://doi.org/10.1103/PhysRevX.13.031027} {\bibfield  {journal} {\bibinfo
  {journal} {Phys. Rev. X}\ }\textbf {\bibinfo {volume} {13}},\ \bibinfo
  {pages} {031027} (\bibinfo {year} {2023})}\BibitemShut {NoStop}%
\bibitem [{\citenamefont {Mardia}\ and\ \citenamefont
  {Jupp}(2010)}]{mardia2010directional}%
  \BibitemOpen
  \bibinfo {editor} {\bibfnamefont {K.~V.}\ \bibnamefont {Mardia}}\ and\
  \bibinfo {editor} {\bibfnamefont {P.~E.}\ \bibnamefont {Jupp}},\ eds.,\ \href
  {https://doi.org/10.1002/9780470316979} {\emph {\bibinfo {title} {Directional
  Statistics}}},\ Wiley Series in Probability and Statistics\ (\bibinfo
  {publisher} {J. Wiley},\ \bibinfo {address} {Chichester New York},\ \bibinfo
  {year} {2010})\BibitemShut {NoStop}%
\bibitem [{\citenamefont {Kawada}\ and\ \citenamefont
  {It{\^o}}(1940)}]{kawada1940probability}%
  \BibitemOpen
  \bibfield  {author} {\bibinfo {author} {\bibfnamefont {Y.}~\bibnamefont
  {Kawada}}\ and\ \bibinfo {author} {\bibfnamefont {K.}~\bibnamefont
  {It{\^o}}},\ }\bibfield  {title} {\bibinfo {title} {On the probability
  distribution on a compact group. i},\ }\href
  {https://doi.org/10.11429/ppmsj1919.22.12_977} {\bibfield  {journal}
  {\bibinfo  {journal} {Proc. Phys.-Math. Soc. Japan. 3rd Series}\ }\textbf
  {\bibinfo {volume} {22}},\ \bibinfo {pages} {977} (\bibinfo {year}
  {1940})}\BibitemShut {NoStop}%
\bibitem [{\citenamefont {Feller}(1991)}]{feller1991introduction}%
  \BibitemOpen
  \bibfield  {author} {\bibinfo {author} {\bibfnamefont {W.}~\bibnamefont
  {Feller}},\ }\href@noop {} {\emph {\bibinfo {title} {An introduction to
  probability theory and its applications, Volume 2}}},\ Vol.~\bibinfo {volume}
  {81}\ (\bibinfo  {publisher} {John Wiley \& Sons},\ \bibinfo {year}
  {1991})\BibitemShut {NoStop}%
\bibitem [{\citenamefont {Thouless}\ and\ \citenamefont
  {Kirkpatrick}(1981)}]{thouless1981conductivity}%
  \BibitemOpen
  \bibfield  {author} {\bibinfo {author} {\bibfnamefont {D.~J.}\ \bibnamefont
  {Thouless}}\ and\ \bibinfo {author} {\bibfnamefont {S.}~\bibnamefont
  {Kirkpatrick}},\ }\bibfield  {title} {\bibinfo {title} {Conductivity of the
  disordered linear chain},\ }\href
  {https://doi.org/10.1088/0022-3719/14/3/007} {\bibfield  {journal} {\bibinfo
  {journal} {J. Phys. C: Solid State Phys.}\ }\textbf {\bibinfo {volume}
  {14}},\ \bibinfo {pages} {235} (\bibinfo {year} {1981})}\BibitemShut
  {NoStop}%
\bibitem [{\citenamefont {Sancho}\ \emph {et~al.}(1984)\citenamefont {Sancho},
  \citenamefont {Sancho},\ and\ \citenamefont {Rubio}}]{sancho1984quick}%
  \BibitemOpen
  \bibfield  {author} {\bibinfo {author} {\bibfnamefont {M.~P.~L.}\
  \bibnamefont {Sancho}}, \bibinfo {author} {\bibfnamefont {J.~M.~L.}\
  \bibnamefont {Sancho}},\ and\ \bibinfo {author} {\bibfnamefont
  {J.}~\bibnamefont {Rubio}},\ }\bibfield  {title} {\bibinfo {title} {Quick
  iterative scheme for the calculation of transfer matrices: Application to mo
  (100)},\ }\href {https://doi.org/10.1088/0305-4608/14/5/016} {\bibfield
  {journal} {\bibinfo  {journal} {J. Phys. F: Met. Phys.}\ }\textbf {\bibinfo
  {volume} {14}},\ \bibinfo {pages} {1205} (\bibinfo {year}
  {1984})}\BibitemShut {NoStop}%
\bibitem [{\citenamefont {Sancho}\ \emph {et~al.}(1985)\citenamefont {Sancho},
  \citenamefont {Sancho}, \citenamefont {Sancho},\ and\ \citenamefont
  {Rubio}}]{sancho1985highly}%
  \BibitemOpen
  \bibfield  {author} {\bibinfo {author} {\bibfnamefont {M.~P.~L.}\
  \bibnamefont {Sancho}}, \bibinfo {author} {\bibfnamefont {J.~M.~L.}\
  \bibnamefont {Sancho}}, \bibinfo {author} {\bibfnamefont {J.~M.~L.}\
  \bibnamefont {Sancho}},\ and\ \bibinfo {author} {\bibfnamefont
  {J.}~\bibnamefont {Rubio}},\ }\bibfield  {title} {\bibinfo {title} {Highly
  convergent schemes for the calculation of bulk and surface green functions},\
  }\href {https://doi.org/10.1088/0305-4608/15/4/009} {\bibfield  {journal}
  {\bibinfo  {journal} {J. Phys. F: Met. Phys.}\ }\textbf {\bibinfo {volume}
  {15}},\ \bibinfo {pages} {851} (\bibinfo {year} {1985})}\BibitemShut
  {NoStop}%
\bibitem [{\citenamefont {MacKinnon}(1985)}]{mackinnon1985calculation}%
  \BibitemOpen
  \bibfield  {author} {\bibinfo {author} {\bibfnamefont {A.}~\bibnamefont
  {MacKinnon}},\ }\bibfield  {title} {\bibinfo {title} {The calculation of
  transport properties and density of states of disordered solids},\ }\href
  {https://doi.org/10.1007/BF01328846} {\bibfield  {journal} {\bibinfo
  {journal} {Z. Physik B: Condens. Matter}\ }\textbf {\bibinfo {volume} {59}},\
  \bibinfo {pages} {385} (\bibinfo {year} {1985})}\BibitemShut {NoStop}%
\bibitem [{\citenamefont {Datta}(2017)}]{datta2017lessons}%
  \BibitemOpen
  \bibfield  {author} {\bibinfo {author} {\bibfnamefont {S.}~\bibnamefont
  {Datta}},\ }\href {https://doi.org/10.1142/8029} {\emph {\bibinfo {title}
  {Lessons from Nanoelectronics: A New Perspective on Transport}}},\ \bibinfo
  {edition} {second edition}\ ed.,\ \bibinfo {series} {Lessons from
  Nanoscience: A Lecture Notes Series}\ No.\ \bibinfo {number} {vol. 5}\
  (\bibinfo  {publisher} {World Scientific},\ \bibinfo {address} {New Jersey},\
  \bibinfo {year} {2017})\BibitemShut {NoStop}%
\bibitem [{\citenamefont {Sun}\ and\ \citenamefont
  {Xie}(2009)}]{sun2009quantum}%
  \BibitemOpen
  \bibfield  {author} {\bibinfo {author} {\bibfnamefont {Q.-f.}\ \bibnamefont
  {Sun}}\ and\ \bibinfo {author} {\bibfnamefont {X.~C.}\ \bibnamefont {Xie}},\
  }\bibfield  {title} {\bibinfo {title} {Quantum transport through a graphene
  nanoribbon--superconductor junction},\ }\href
  {https://doi.org/10.1088/0953-8984/21/34/344204} {\bibfield  {journal}
  {\bibinfo  {journal} {J. Phys.: Condens. Matter}\ }\textbf {\bibinfo {volume}
  {21}},\ \bibinfo {pages} {344204} (\bibinfo {year} {2009})}\BibitemShut
  {NoStop}%
\bibitem [{\citenamefont {Michelsen}\ \emph {et~al.}(2023)\citenamefont
  {Michelsen}, \citenamefont {Recher}, \citenamefont {Braunecker},\ and\
  \citenamefont {Schmidt}}]{michelsen2023supercurrentenabled}%
  \BibitemOpen
  \bibfield  {author} {\bibinfo {author} {\bibfnamefont {A.~B.}\ \bibnamefont
  {Michelsen}}, \bibinfo {author} {\bibfnamefont {P.}~\bibnamefont {Recher}},
  \bibinfo {author} {\bibfnamefont {B.}~\bibnamefont {Braunecker}},\ and\
  \bibinfo {author} {\bibfnamefont {T.~L.}\ \bibnamefont {Schmidt}},\
  }\bibfield  {title} {\bibinfo {title} {Supercurrent-enabled andreev
  reflection in a chiral quantum hall edge state},\ }\href
  {https://doi.org/10.1103/PhysRevResearch.5.013066} {\bibfield  {journal}
  {\bibinfo  {journal} {Phys. Rev. Research}\ }\textbf {\bibinfo {volume}
  {5}},\ \bibinfo {pages} {013066} (\bibinfo {year} {2023})}\BibitemShut
  {NoStop}%
\bibitem [{\citenamefont {Buttiker}(1988)}]{buttiker1988symmetry}%
  \BibitemOpen
  \bibfield  {author} {\bibinfo {author} {\bibfnamefont {M.}~\bibnamefont
  {Buttiker}},\ }\bibfield  {title} {\bibinfo {title} {Symmetry of electrical
  conduction},\ }\href {https://doi.org/10.1147/rd.323.0317} {\bibfield
  {journal} {\bibinfo  {journal} {IBM J. Res. Dev.}\ }\textbf {\bibinfo
  {volume} {32}},\ \bibinfo {pages} {317} (\bibinfo {year} {1988})}\BibitemShut
  {NoStop}%
\bibitem [{\citenamefont {Datta}(1995)}]{datta1995electronic}%
  \BibitemOpen
  \bibfield  {author} {\bibinfo {author} {\bibfnamefont {S.}~\bibnamefont
  {Datta}},\ }\href {https://doi.org/10.1017/CBO9780511805776} {\emph {\bibinfo
  {title} {Electronic Transport in Mesoscopic Systems}}},\ \bibinfo {edition}
  {1st}\ ed.\ (\bibinfo  {publisher} {Cambridge University Press},\ \bibinfo
  {year} {1995})\BibitemShut {NoStop}%
\bibitem [{\citenamefont {{Entin-Wohlman}}\ \emph {et~al.}(2008)\citenamefont
  {{Entin-Wohlman}}, \citenamefont {Imry},\ and\ \citenamefont
  {Aharony}}]{entin-wohlman2008conductance}%
  \BibitemOpen
  \bibfield  {author} {\bibinfo {author} {\bibfnamefont {O.}~\bibnamefont
  {{Entin-Wohlman}}}, \bibinfo {author} {\bibfnamefont {Y.}~\bibnamefont
  {Imry}},\ and\ \bibinfo {author} {\bibfnamefont {A.}~\bibnamefont
  {Aharony}},\ }\bibfield  {title} {\bibinfo {title} {Conductance of
  superconducting-normal hybrid structures},\ }\href
  {https://doi.org/10.1103/PhysRevB.78.224510} {\bibfield  {journal} {\bibinfo
  {journal} {Phys. Rev. B}\ }\textbf {\bibinfo {volume} {78}},\ \bibinfo
  {pages} {224510} (\bibinfo {year} {2008})}\BibitemShut {NoStop}%
\bibitem [{\citenamefont {Cui}\ \emph {et~al.}(2016)\citenamefont {Cui},
  \citenamefont {Yu}, \citenamefont {Iommelli},\ and\ \citenamefont
  {Kong}}]{cui2016exact}%
  \BibitemOpen
  \bibfield  {author} {\bibinfo {author} {\bibfnamefont {G.}~\bibnamefont
  {Cui}}, \bibinfo {author} {\bibfnamefont {X.}~\bibnamefont {Yu}}, \bibinfo
  {author} {\bibfnamefont {S.}~\bibnamefont {Iommelli}},\ and\ \bibinfo
  {author} {\bibfnamefont {L.}~\bibnamefont {Kong}},\ }\bibfield  {title}
  {\bibinfo {title} {Exact distribution for the product of two correlated
  gaussian random variables},\ }\href
  {https://doi.org/10.1109/LSP.2016.2614539} {\bibfield  {journal} {\bibinfo
  {journal} {IEEE Signal Process. Lett.}\ }\textbf {\bibinfo {volume} {23}},\
  \bibinfo {pages} {1662} (\bibinfo {year} {2016})}\BibitemShut {NoStop}%
\bibitem [{\citenamefont {Zhang}\ \emph {et~al.}(2019)\citenamefont {Zhang},
  \citenamefont {Shi}, \citenamefont {Zhu}, \citenamefont {Xing}, \citenamefont
  {Zhang},\ and\ \citenamefont {Wang}}]{Zhang2019topological}%
  \BibitemOpen
  \bibfield  {author} {\bibinfo {author} {\bibfnamefont {D.}~\bibnamefont
  {Zhang}}, \bibinfo {author} {\bibfnamefont {M.}~\bibnamefont {Shi}}, \bibinfo
  {author} {\bibfnamefont {T.}~\bibnamefont {Zhu}}, \bibinfo {author}
  {\bibfnamefont {D.}~\bibnamefont {Xing}}, \bibinfo {author} {\bibfnamefont
  {H.}~\bibnamefont {Zhang}},\ and\ \bibinfo {author} {\bibfnamefont
  {J.}~\bibnamefont {Wang}},\ }\bibfield  {title} {\bibinfo {title}
  {Topological axion states in the magnetic insulator
  $\text{MnBi}_2\text{Te}_4$ with the quantized magnetoelectric effect},\
  }\href {https://doi.org/10.1103/PhysRevLett.122.206401} {\bibfield  {journal}
  {\bibinfo  {journal} {Phys. Rev. Lett.}\ }\textbf {\bibinfo {volume} {122}},\
  \bibinfo {pages} {206401} (\bibinfo {year} {2019})}\BibitemShut {NoStop}%
\end{thebibliography}
\end{document}


\title{Supplementary Materials for ``Revealing superconducting chiral edge modes via resistance distributions''}
\author{Linghao Huang}
\affiliation{State Key Laboratory of Surface Physics and Department of Physics, Fudan University, Shanghai 200433, China} 
\affiliation{Shanghai Research Center for Quantum Sciences, Shanghai 201315, China}
\author{Dongheng Qian}
\affiliation{State Key Laboratory of Surface Physics and Department of Physics, Fudan University, Shanghai 200433, China} 
\affiliation{Shanghai Research Center for Quantum Sciences, Shanghai 201315, China}
\author{Jing Wang}
\thanks{Contact author: wjingphys@fudan.edu.cn}
\affiliation{State Key Laboratory of Surface Physics and Department of Physics, Fudan University, Shanghai 200433, China}
\affiliation{Shanghai Research Center for Quantum Sciences, Shanghai 201315, China}
\affiliation{Institute for Nanoelectronic Devices and Quantum Computing, Fudan University, Shanghai 200433, China}
\affiliation{Hefei National Laboratory, Hefei 230088, China}

\maketitle

\tableofcontents

\newpage

\section{Numerical Results for Edge-mode Scattering Problem}\label{appscatter}
\begin{figure}[htbp]
  \begin{center}
  \includegraphics[width=5.5in,clip=true]{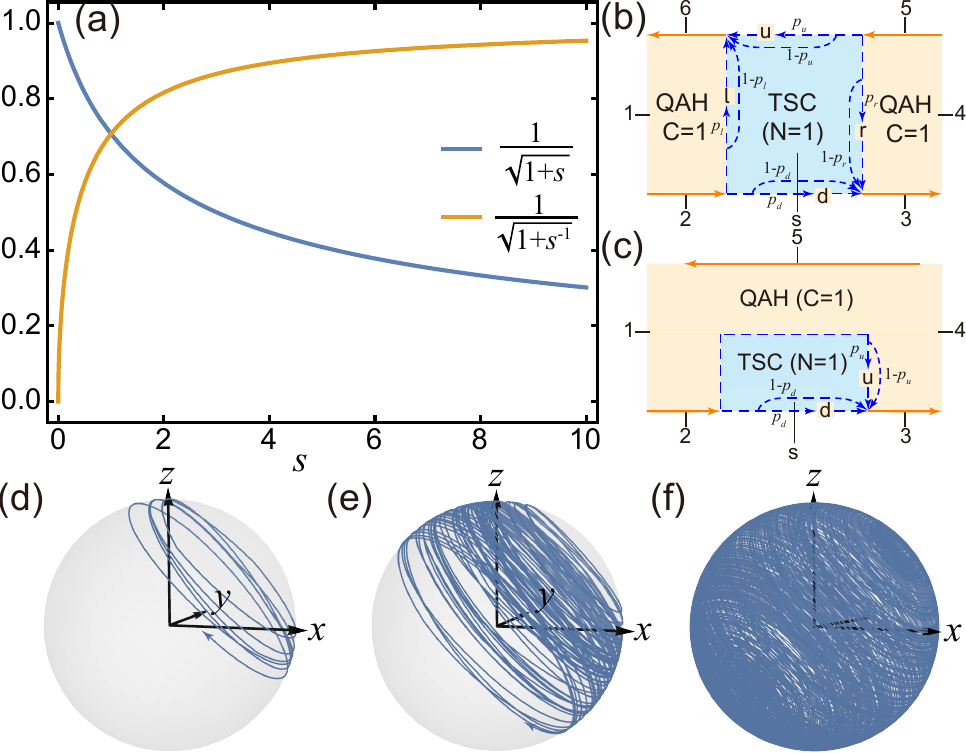}
  \end{center}
  \caption{(a) The $x$ (blue line) and $z$ (orange line) components of $\mathbf{n}_\ell$. Here $s=\frac{\mu_\ell^2}{\varepsilon^2} \left(\frac{v_\ell^2}{\Delta_\ell^2} -1\right)$. Since $v_\ell>\Delta_\ell$, $s$ is always positive. (b) Schematic diagrams for QAH-TSC-QAH junction. There are 7 leads (1-6 and s) attached to the junction, and four floating leads ($l$, $r$, $u$ and $d$) shown at the boundaries of the TSC region. $p_l$ and $1-p_l$ are the probability of the edge modes entering and not entering the floating lead $l$, respectively, with similar meanings for $p_r$, $p_u$, and $p_d$. (c) Schematic diagrams for QAH-TSC junction. There are 6 leads attached to the junction, and two floating leads ($u$ and $d$) at the boundaries of the TSC region. (d)-(f) The evolution of the state.}
  \label{figS1}
\end{figure}

This section provides some numerical results for Section II in the main text. As we pointed out in the main text, the rotational axis always lies in the $xoz$ plane. We plot the $x$ and $z$ components of $\mathbf{n}_\ell$ in Fig.~\ref{figS1}(a). When the parameters in effective Hamiltonian fluctuate slightly due to disorder, $\mathbf{n}_\ell$ changes only slightly around a certain direction, especially when $s=\frac{\mu_\ell^2}{\varepsilon^2} \left(\frac{v_\ell^2}{\Delta_\ell^2} -1\right)$ is large. Therefore, the total transport process can be interpreted as the evolution of a point at north-pole under multiple rotation operations, shown in Figs. \ref{figS1}(d-f), where the parameters are: $\varepsilon=1,~L_\ell=1$, $\mu_\ell$ is uniformly distributed over 0.45 to 0.55 ($U[0.45,0.55]$), $v_\ell\sim U[0.95,1.05],~\Delta_\ell \sim U[0.45,0.55]$, and $L$ is 20, 200, 2000 for (d), (e), (f), respectively. As shown in the figures, when $L$ is small [Fig.~\ref{figS1}(d)], the trajectory can be approximately described as the path of a point rotating around a fixed axis. As $L$ grows, the region of motion for the point gradually disperses [Fig.~\ref{figS1}(e)] until it covers the entire spherical surface [Fig.~\ref{figS1}(f)].

\section{Nonlocal Resistances and Generalized Landauer-B\"uttiker Formula}\label{appLB}
The nonlocal resistances measured in the QAH-SC heterojunction can be calculated using the generalized Landauer-B\"uttiker formula \cite{entin-wohlman2008conductance}. In the configuration of the QAH-SC-QAH  junction shown in Fig. 1(a) of the main text, when leads 2 to 6 are voltage probes, the multi-terminal formula is:
\begin{equation}\label{LB_QAH-SC-QAH}
\begin{aligned}
&I_1=\frac{e^2}{h}(V_1-V_6), \quad I_2=\frac{e^2}{h}(V_2-V_1), \quad
 I_3=\frac{e^2}{h}[(V_3-V_s)+ T_d (V_s-V_2)+ R_d (V_s-V_5)] \\
&I_4=\frac{e^2}{h}(V_4-V_3), \quad I_5=\frac{e^2}{h}(V_5-V_4), \quad
 I_6=\frac{e^2}{h}[(V_6-V_s)+ T_u (V_s-V_5)+ R_u (V_s-V_2)] \\
& I_2=I_3=I_5=I_6=0, \quad I_1+I_2+I_3+I_4+I_5+I_6+I_s=0,
\end{aligned}
\end{equation}
where $T_u=T_{u,\text{ee}}-T_{u,\text{eh}}$ is the charge transmission fraction from lead 5 to 6, $T_d=T_{d,\text{ee}}-T_{d,\text{eh}}$ is that from lead 2 to 3, $R_u=R_{u,\text{ee}}-R_{u,\text{eh}}$ is that from lead 2 to 6, and $R_d=R_{d,\text{ee}}-R_{d,\text{eh}}$ is that from lead 5 to 3. At finite temperature, and as explained in Sec. VI of the main text, the quantities $T$ and $R$ above should be understood as the convolutions of the functions $T(E)$ and $R(E)$ with the kernel $-\frac{\partial f(E-\varepsilon)}{\partial E}$. When the SC is in the $N=2$ phase, $R_u=R_d=0$ due to the absence of edge modes along the interface between QAH and SC. Solving the equations with respect to applied currents yields the quantized values of resistances measured on various leads. For example, if $I_s=0$, then $I_1=-I_4=I$, the non-local resistances are (in units of $h/e^2$):
\begin{equation}
  \begin{aligned}
  R_{14} &= \frac{V_1-V_4}{I} = \frac{2-(R_d + R_u + T_d + T_u)}{(R_d - 1)(R_u - 1) - T_d T_u}, 
  & R_{23} &= \frac{V_2-V_3}{I} = - \frac{R_d R_u - (T_d - 1)(T_u - 1)}{(R_d - 1)(R_u - 1) - T_d T_u}, \\
  R_{56} &= \frac{V_5-V_6}{I} = \frac{R_d R_u - (T_d - 1)(T_u - 1)}{(R_d - 1)(R_u - 1) - T_d T_u}, 
  & R_{2s} &= \frac{V_2-V_s}{I} = \frac{1 - (R_d + T_u)}{(R_d - 1)(R_u - 1) - T_d T_u}, \\
  R_{s3} &= \frac{V_s-V_3}{I} = -1+\frac{1-(R_u+T_d)}{(R_d - 1)(R_u - 1) - T_d T_u}, 
  & \cdots &\phantom{=} \quad
  \end{aligned} 
\end{equation}
When the junction is symmetric such that $T_u=T_d=T,~R_u=R_d=R$, the above expressions can be simplified as:
\begin{equation}
  R_{14} = \frac{2}{1+T-R},~ R_{23} = \frac{1+R-T}{1+T-R},~ R_{56} = \frac{1+R-T}{R-T-1},~
  R_{2s} = \frac{1}{1+T-R},~ R_{s3} = \frac{R-T}{1+T-R},~ \cdots.
\end{equation}

If SC is grounded $V_s=0$ and $I_4=0$, then $I_1=-I_s=I$, the non-local resistances are:
\begin{equation}
  \begin{aligned}
    R_{4s} &= \frac{V_4-V_s}{I} = \frac{T_d}{(R_d - 1)(R_u - 1) - T_d T_u},
    & R_{23} &= \frac{V_2-V_3}{I} = \frac{1-(R_d + T_d)}{(R_d - 1)(R_u - 1) - T_d T_u}, \\
    R_{56} &= \frac{V_5-V_6}{I} = \frac{R_u(1-R_d)+T_d(1-T_u)}{(R_d - 1)(R_u - 1) - T_d T_u},
    & R_{s3} &= \frac{V_s-V_3}{I} = -\frac{T_d}{(R_d - 1)(R_u - 1) - T_d T_u}, \\
    R_{2s} &= \frac{V_2-V_s}{I} = \frac{1-R_d}{(R_d - 1)(R_u - 1) - T_d T_u},
    & \cdots &\phantom{=} \quad
\end{aligned}
\end{equation}
And when $T_u=T_d=T,~R_u=R_d=R$, the above expressions can be simplified as:
\begin{equation}
    R_{4s} = \frac{T}{(R-1)^2-T^2},~ R_{23} = \frac{1}{1+T-R},~ R_{56} = \frac{T-R}{1+T-R},~ R_{s3} = \frac{T}{T^2-(R-1)^2},~ R_{2s} = \frac{1-R}{(R-1)^2-T^2},~ \cdots.
\end{equation}

We point out that although the main text focuses on the behavior of $T$, the behavior of $R$ is similar, because both of them are governed by the edge modes transports.

In the configuration of the QAH-SC junction shown in the shown in Fig. 1(f) of the main text, there is no reflection process due to the chiral nature of edge modes. The multi-terminal formula is:
\begin{equation}\label{LB_QAH-SC}
\begin{aligned}
&I_1=\frac{e^2}{h}(V_1-V_5), \quad I_2=\frac{e^2}{h}(V_2-V_1), \quad
 I_3=\frac{e^2}{h}[(V_3-V_s)+ T (V_s-V_2)] \\
&I_4=\frac{e^2}{h}(V_4-V_3), \quad I_5=\frac{e^2}{h}(V_5-V_4), \quad
 I_2=I_3=I_5=0, \quad I_1+I_2+I_3+I_4+I_5+I_s=0.
\end{aligned}
\end{equation}
where $T=T_{ee}-T_{eh}$ is the charge transmission fraction from lead 2 to 3. If $V_s=0$ while $I_4=0$, then $I_1=-I_s=I$, the non-local resistances are:
\begin{equation}
  R_{4s} = \frac{V_4-V_s}{I} = \frac{T}{1-T}, \quad
  R_{2s} = \frac{V_2-V_s}{I} = \frac{1}{1-T}, \quad
  R_{s3} = \frac{V_s-V_3}{I} = \frac{T}{T-1}, \quad
  \cdots.
\end{equation}

\section{More Analytical Result for Decoherence Effect}\label{appflead}
In this section, we use the floating lead method to phenomenologically describe the decoherence effect under the generalized Landauer-B\"uttiker formalism, in application to QAH-SC-QAH junction and QAH-SC junction. From the same treatment in the main text, we first consider only one floating lead at the edge and use a parameter $p$ to de
scribe the decoherence probability.
For the QAH-SC-QAH junction shown in Fig.~\ref{figS1}(b), part of Eq.~(\ref{LB_QAH-SC-QAH}) is modified:
\begin{equation}
  \begin{aligned}
    &I_d = \frac{e^2}{h}[(V_d-V_s) + p_d T_d^{(1)} (V_s-V_2)] = 0 , \quad
    I_r = \frac{e^2}{h}[(V_r-V_s) + p_r R_r^{(1)} (V_s-V_5)] = 0 , \\
    &I_3 = \frac{e^2}{h}[(V_3-V_s)+ p_d T_d^{(2)} (V_s-V_d) + (1-p_d) T_d^{(1,2)} (V_s-V_2) + 
    p_r R_r^{(2)} (V_s-V_r) + (1-p_r) R_r^{(1,2)} (V_s-V_5)], \\
    &I_u = \frac{e^2}{h}[(V_u-V_s) + p_u T_u^{(1)} (V_s-V_5)] = 0 , \quad
    I_l = \frac{e^2}{h}[(V_l-V_s) + p_l R_l^{(1)} (V_s-V_2)] = 0 , \\
    &I_6 = \frac{e^2}{h}[(V_6-V_s)+ p_u T_u^{(2)} (V_s-V_u) + (1-p_u) T_u^{(1,2)} (V_s-V_5) + 
    p_l R_l^{(2)} (V_s-V_l) + (1-p_l) R_l^{(1,2)} (V_s-V_2)], \\
  \end{aligned}
\end{equation}
where $p_d$ and $1-p_d$ are the probability of the edge modes entering and not entering the floating lead $d$, respectively. $T_d^{(1)}$ is the charge transmission fraction from lead 2 to lead $d$, $T_d^{(2)}$ is from lead $d$ to lead 3, $T_d^{(1,2)}$ is from lead 2 to lead 3, which is the original coherent process. Other symbols have similar interpretations. Therefore:
\begin{equation}
  I_3 = \frac{e^2}{h}[(V_3-V_s)+ T_d^{\text{tot}} (V_s-V_2) + R_r^{\text{tot}} (V_s-V_5)], \qquad
  I_6 = \frac{e^2}{h}[(V_6-V_s)+ T_u^{\text{tot}} (V_s-V_5) + R_l^{\text{tot}} (V_s-V_2)],
\end{equation}
where the effective charge transmission fractions are:
\begin{equation}\label{N2onelead}
  \begin{aligned}
  &T_d^{\text{tot}}= (1-p_d) T_d^{(1,2)} + p_d T_d^{(1)} T_d^{(2)}, \qquad
  R_r^{\text{tot}}= (1-p_r) R_r^{(1,2)} + p_r R_r^{(1)} R_r^{(2)}, \\
  &T_u^{\text{tot}}= (1-p_u) T_u^{(1,2)} + p_u T_u^{(1)} T_u^{(2)}, \qquad
  R_l^{\text{tot}}= (1-p_l) R_l^{(1,2)} + p_l R_l^{(1)} R_l^{(2)}.
  \end{aligned}
\end{equation}
Consequently, one can include the decoherence effect by replacing the charge transmission fractions with the corresponding effective ones.

For the QAH-SC junction shown in Fig.~\ref{figS1}(c), we first consider $N=2$ case, where the the upper route is absent, part of Eq.~(\ref{LB_QAH-SC}) is modified:
\begin{equation}
    I_d = \frac{e^2}{h}[(V_d-V_s) + p_d T_d^{(1)} (V_s-V_2)] = 0, \qquad
    I_3 = \frac{e^2}{h}[(V_3-V_s)+ p_d T_d^{(2)} (V_s-V_d) + (1-p_d) T_d^{(1,2)} (V_s-V_2)].
\end{equation}
Eliminating $V_d$, $T_d^{\text{tot}}$ is the same as the QAH-SC-QAH junction case, i.e., Eq.~(\ref{N2onelead}).
$N=1$ case is slightly complicated: if $p_d$ and $p_u$ represent the probabilities of edge mode entering the floating lead $d$ and $u$, respectively, the coherent component is $\min{(1-p_d,1-p_u)}$, and the remaining part $|p_u-p_d|$ is the incoherent component of the transport process directly from lead 2 to 3. Part of Eq.~(\ref{LB_QAH-SC}) should be modified as:
\begin{equation}
  \begin{aligned}
    &I_d = \frac{e^2}{h}[(V_d-V_s) + p_d T_d^{(1)} (V_s-V_2)] = 0 , \quad
     I_u = \frac{e^2}{h}[(V_u-V_s) + p_u T_u^{(1)} (V_s-V_2)] = 0 , \\
    &I_3 = \frac{e^2}{h}\left[(V_3-V_s)+ \left(p_d T_d^{(1)} T_d^{(2)} + p_u T_u^{(1)} T_u^{(2)} + |p_u-p_d| T_{\text{inco}} + \min{(1-p_d,1-p_u)} T_{\text{co}} \right) (V_s-V_2)\right].
  \end{aligned}
\end{equation}
Therefore:
\begin{equation}\label{N1onelead}
  T_d^{\text{tot}} = p_d T_d^{(1)} T_d^{(2)} + p_u T_u^{(1)} T_u^{(2)} + |p_u-p_d| T_{\text{inco}} + \min{(1-p_d,1-p_u)} T_{\text{co}}.
\end{equation}
Note that the transport process $T_d^{(1)},~T_d^{(2)},~T_u^{(1)},~T_u^{(2)}$ and $T_{\text{inco}}$ corresponding to single Majorana mode transmission from one lead to another, which are analogous to, e.g., the charge transmission process $T_d$ in QAH-SC-QAH junction with $N=1$ without decoherence effect. Thus they follow normal distribution when taking into account of the disorder effect.   

\begin{figure}[b]
  \begin{center}
  \includegraphics[width=6.8in,clip=true]{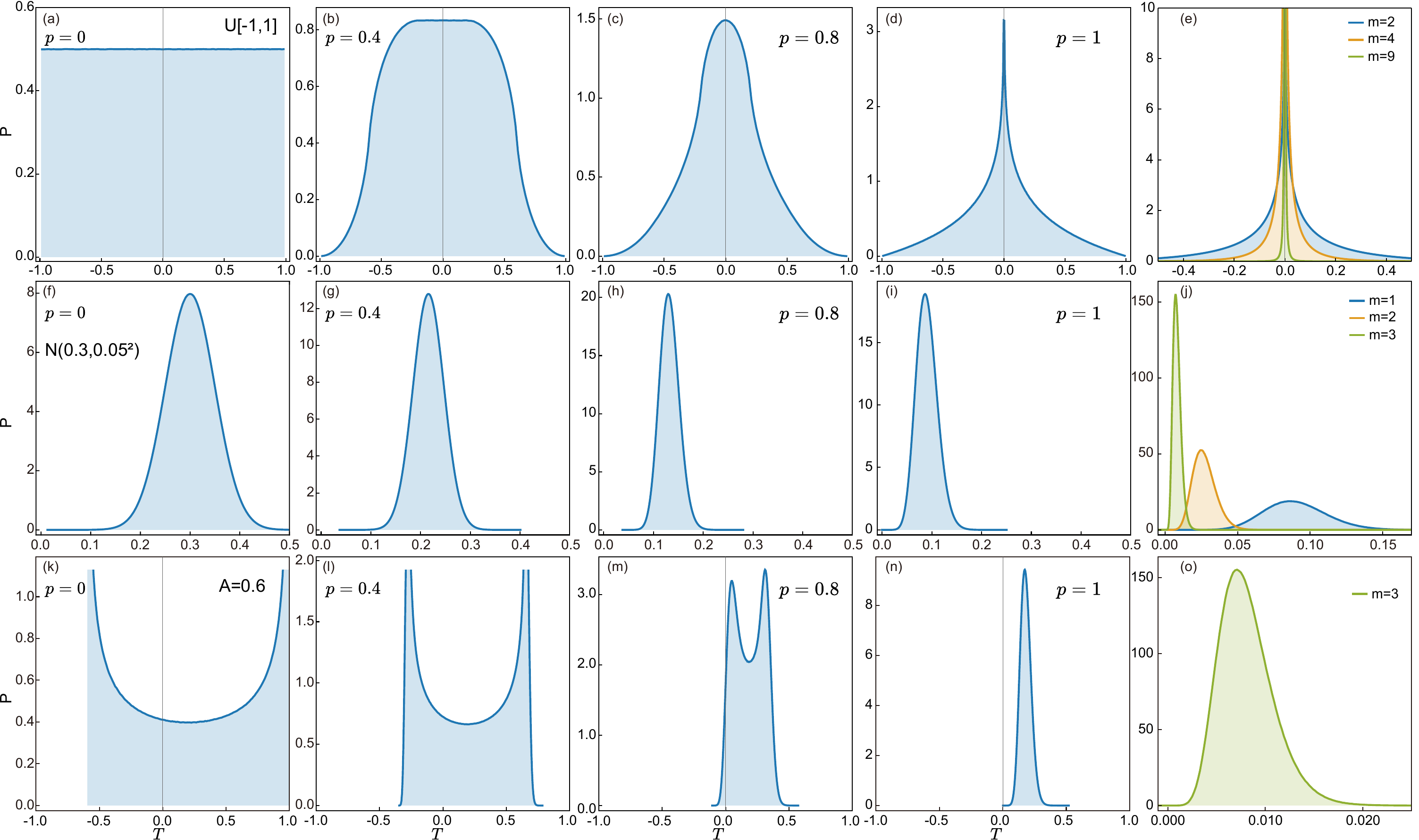}
  \end{center}
  \caption{Decoherence effect on the probability distribution of transmission fraction. We generate $10^7$ random numbers for the charge transmission fraction of each subsegment, then calculate effective charge transmission fraction and obtain its distribution numerically. (a-e) $N=2$ case, using Eq.~(\ref{N2onelead}). The charge transmission fraction for each segment satisfies $U[-1,1]$. In (a-d) there is one floating lead, and in (e) there are multiple floating leads. (f-j) QAH-SC-QAH junction with $N=1$, using Eq.~(\ref{N2onelead}). The charge transmission fraction for each segment satisfies $\mathcal{N}(0.3,0.05^2)$. In (f-i) there is one floating lead, and in (j) there are multiple floating leads. (o) is horizontally scaled version of (j) when there are three floating leads. (k-n) QAH-SC junction with $N=1$, using Eq.~(\ref{N1onelead}). We set $p_u=p_d=p$, $T_d^{(1)},~T_d^{(2)},~T_u^{(1)},~T_u^{(2)},~T_{\text{inco}}$ follow $\mathcal{N}(0.3,0.05^2)$, and $T_{\text{co}}$ follows an arcsine distribution with $A=0.6$.}
  \label{figS2}
\end{figure}

In the following, we focus on the long junction case, where each charge transmission fraction converges to the form of its large $L$ limit. Without loss of generality, we take $T_d^{\text{tot}}$ as an example. For $N=2$ case, $T_d^{(1,2)},~T_d^{(1)}$ and $T_d^{(2)}$ in Eq.~(\ref{N2onelead}) all follow $U[-1,1]$, one can obtain the analytical result of $T_d^{\text{tot}}$. When $p\leq0.5$:
\begin{equation}
    f(T)=
\begin{cases}
    -\frac{1}{2 (p-1)}, & | T| \leq 1-2 p \\ 
    \frac{1}{4 (p-1) p} \left[|T| +(p+| T| -1) \log \left(\left| \frac{p}{p+| T| -1}\right| \right)-1\right], & 2 p-1 < |T| \leq 1 \\
    0, & \text{otherwise}
\end{cases}.
\end{equation}
Apparently, when $p=0$, $f(T)$ reduces to constant $1/2$ over $[-1,1]$. For $p>0.5$:
\begin{equation}
f(T)=
\begin{cases}
 \frac{1}{4 (p-1) p}\left[(|T| -p+1) \log \left(\frac{| T| -p+1}{p}\right)+(| T| +p-1) \log \left(\frac{p}{| p+| T| -1| }\right)+2 (p-1)\right], & | T| <2 p-1 \\
 \frac{1}{4 (p-1) p}\left[|T| +(p+| T| -1) \log \left(\left| \frac{p}{p+| T| -1}\right| \right)-1\right], & 2 p-1 \leq |T| \leq 1 \\
0, & \text{otherwise}
\end{cases},
\end{equation}
and when $p=1$, $f(T)$ becomes $-\frac{1}{2}\log |T|$ over $[-1,1]$. We plot this probability density function in Figs.~\ref{figS2}(a-d), which show that increasing $p$ causes the distribution of $T$ to become more concentrated around zero, and when $p=1$, $f(T)$ becomes logarithmic divergence.
For $N=1$ case, the distribution of $T_d^{\text{tot}}$ is far more complicated, while one can find analytical result for $T_d^{(1)}T_d^{(2)}$, which is product two independent and identically distributed Gaussian random variables. The probability density function is \cite{cui2016exact}:
\begin{equation}\label{twoG}
  f(T=T_d^{(1)}T_d^{(2)}) =e^{-\frac{\mu^2}{\sigma^2}}\sum_{n=0}^{\infty}\sum_{m=0}^{2n}\binom{2n}{m}\frac{\mu^{2n}T^{2n-m}|T|^{m-n}}{\pi(2n)!\sigma^{4n+2}}K_{m-n}\left(\frac{|T|}{\sigma^2}\right),
\end{equation}
where $K_v$ denotes the modified Bessel function of the second kind and order $v$.
The final effective transmission fraction will be a mixture of the above distribution and normal and (or) arcsine distribution, depending on the concrete configuration. We show the results for QAH-SC-QAH junction with $N=1$ case in Fig.~\ref{figS2}(f-i), and QAH-SC junction with $N=1$ case in Fig.~\ref{figS2}(k-n), where, for simplicity, we assume $p_d=p_u=p$ and the same distribution for $T_d^{(1)},~T_d^{(2)},~T_u^{(1)},~T_u^{(2)}$, and $T_{\text{inco}}$. One can find that when $p$ is small, the distributions are similar to that without decoherence effect. When $p\rightarrow1$, $f(T)$ in both cases converge to Eq.~(\ref{twoG}).

For more than one floating leads, the resulting multi-terminal formula is more complicated, while for $N=2$ case one can obtain analytical result for the distribution of $T^\text{tot}$ when charge transmission fraction for each section satisfies $U[-1,1]$:
\begin{equation}
    f(T=T^{\text{tot}}) = \frac{(-1)^{m}}{2 (m)!}[\log (|T|)]^{m}.
\end{equation}
In Fig.~\ref{figS2}(e,j) we plot the results for $N=2$, QAH-SC-QAH junction with $N=1$, respectively (the result for QAH-SC junction with $N=1$ is similar to that for QAH-SC-QAH junction with $N=1$, and is not shown here). When more floating leads are introduced, $f(T^{\text{tot}})$ in $N=2$ case still peaks at zero value, but the divergence rate increases, while $f(T^{\text{tot}})$ in $N=1$ case remains bell-shaped like, but becomes more concentrated with a decrease of in the peak value of $T^{\text{tot}}$.

\section{Technical Details of Numerical Calculations}\label{appeffH}

In the numerical calculations presented in the main text, we consider the QAH state in a magnetic topological insulator thin film with ferromagnetic order. The two-dimensional effective Hamiltonian is $\mathcal{H}_0=\sum_{\mathbf{k}}\psi^{\dag}_{\mathbf{k}}H_0(\mathbf{k})\psi_{\mathbf{k}}$, where $\psi_{\mathbf{k}}=(c^t_{\mathbf{k}\uparrow}, c^t_{\mathbf{k}\downarrow},c^b_{\mathbf{k}\uparrow}, c^b_{\mathbf{k}\downarrow})^T$ and $H_0(\mathbf{k})=k_y\sigma_x\tau_z-k_x\sigma_y\tau_z+m(\mathbf{k})\tau_x+\lambda\sigma_z$. Here, the superscripts $t$ and $b$ denote the top and bottom surface states, respectively. $\sigma_i$ and $\tau_i$ are Pauli matrices for spin and layer, respectively. $\lambda$ is the exchange field. $m(\mathbf{k})=m_0+m_1(k_x^2+k_y^2)$ represents the hybridization between the top and bottom surface states. This model accurately describes the QAH state in Cr-doped (Bi,Sb)$_2$Te$_3$~\cite{chang2013experimental} and odd-layer MnBi$_2$Te$_4$~\cite{Zhang2019topological}. The chiral TSC is induced by superconducting proximity of the QAH state \cite{wang2015chiral}, with the effective BdG Hamiltonian $\mathcal{H}_{\mathrm{bulk}}=\sum_{\mathbf{k}}\Psi^{\dag}_{\mathbf{k}}H_{\mathrm{bulk}}(\mathbf{k})\Psi_{\mathbf{k}}/2$, where $\Psi_{\mathbf{k}}=[(c^t_{\mathbf{k}\uparrow}, c^t_{\mathbf{k}\downarrow}, c^b_{\mathbf{k}\uparrow}, c^b_{\mathbf{k}\downarrow}), (c^{t\dag}_{-{\mathbf{k}}\uparrow}, c^{t\dag}_{-{\mathbf{k}}\downarrow}, c^{b\dag}_{-{\mathbf{k}}\uparrow}, c^{b\dag}_{-{\mathbf{k}}\downarrow})]^T$ and
\begin{equation}\label{QAH+sc}
  H_{\mathrm{bulk}}(\mathbf{k}) =
    \begin{pmatrix}
      H_0(\mathbf{k})-\mu & \Delta(\mathbf{k})\\
      \Delta(\mathbf{k})^\dag & -H_0^*(-{\mathbf{k}})+\mu
    \end{pmatrix}, \qquad
  \Delta(\mathbf{k}) =
  \begin{pmatrix}
    i\Delta_t\sigma_y & 0\\
    0 & i\Delta_b\sigma_y
  \end{pmatrix}.
\end{equation}
Here, $\mu$ is the chemical potential, and $\Delta_t$ and $\Delta_b$ are the pairing gap functions on the top and bottom surface states, respectively. The topological properties of this system were well studied in Ref.~\cite{wang2015chiral}, revealing three TSC phases with BdG Chern numbers $N=0,1,2$. 

We set $v = 1$, $\mu = 0.2$, $m_0 = 1$, $m_1 = 1.5$, and $\lambda = 5$ for the QAH, $\Delta_t = 1$, $\Delta_b = 0$, $\lambda = 1.1$ for the $N=1$ TSC, $\lambda = 3$ for the $N=2$ TSC. To ensure the edge modes along the two boundaries do not overlap, the width of the junction is $30a$ with $a\equiv1$ as the lattice constant. The disorder effect in the TSC region is considered by making $\mu$ a random field, which follows a uniform distribution $\mu\in U[-W_{\text{dis}}/2,W_{\text{dis}}/2]$, with $W_{\text{dis}}$ the strength of the disorder potential. We have verified that using a Gaussian-distributed disorder potential does not alter the conclusions of this study. Since the interface between quantum Hall or QAH insulator and superconductor is highly inhomogeneous, we consider spatial-uncorrelated disorder, i.e., every lattice site has a random $\mu$ independent of each other. In the main text, we set $W_{\text{dis}}=0.2$, and $\varepsilon=0.63$ for QAH-SC-QAH junction with $N=2$ CBEM, $\varepsilon=0.3$ for QAH-SC junction with $N=1$ CBEM. By calculating a sufficient number of different disorder configurations, we obtain the distribution of transmission. We use $1\times10^4$ configurations when plotting the distribution and $500$ configurations when plotting the transmission (reflection)-length relation.

In the calculation of transmission fraction, we adopt the recursive Green's function method~\cite{thouless1981conductivity,sancho1984quick,sancho1985highly,mackinnon1985calculation,datta2017lessons}. The effect of the semi-infinite lead on the center region (TSC region) is manifested in the form of self-energy. The center region is divided into columns, and using the recursive algorithm, one can obtain the Green's function of the center region. At zero temperature, the normal ($T_{\text{ee}}$) and anomalous ($T_{\text{eh}}$) transmission probabilities from lead $i$ to $j$ are calculated by \cite{sun2009quantum}:
\begin{equation}
 T_{{j,\text{e}}\leftarrow {i,\text{e}}} = \text{Tr}\left([\Gamma_j]_{\text{ee}} [G_c]_{\text{ee}} [\Gamma_i]_{\text{ee}} [G_c^\dag]_{\text{ee}}\right), \qquad
 T_{{j,\text{h}}\leftarrow {i,\text{e}}} = \text{Tr}\left([\Gamma_j]_{\text{hh}} [G_c]_{\text{he}} [\Gamma_i]_{\text{ee}} [G_c^\dag]_{\text{eh}}\right),
\end{equation}
respectively, where $\Gamma_j = i \left(\Sigma_j-\Sigma_j^\dag\right)$, $G_c$ is the Green's function of the center region, $\Sigma_j$ is the self-energy for lead $j$, and the subscripts $\text{e}$ ($\text{h}$) denote the electron (hole) sector of the matrix.

\begin{figure}[t]
  \begin{center}
  \includegraphics[width=4.8in,clip=true]{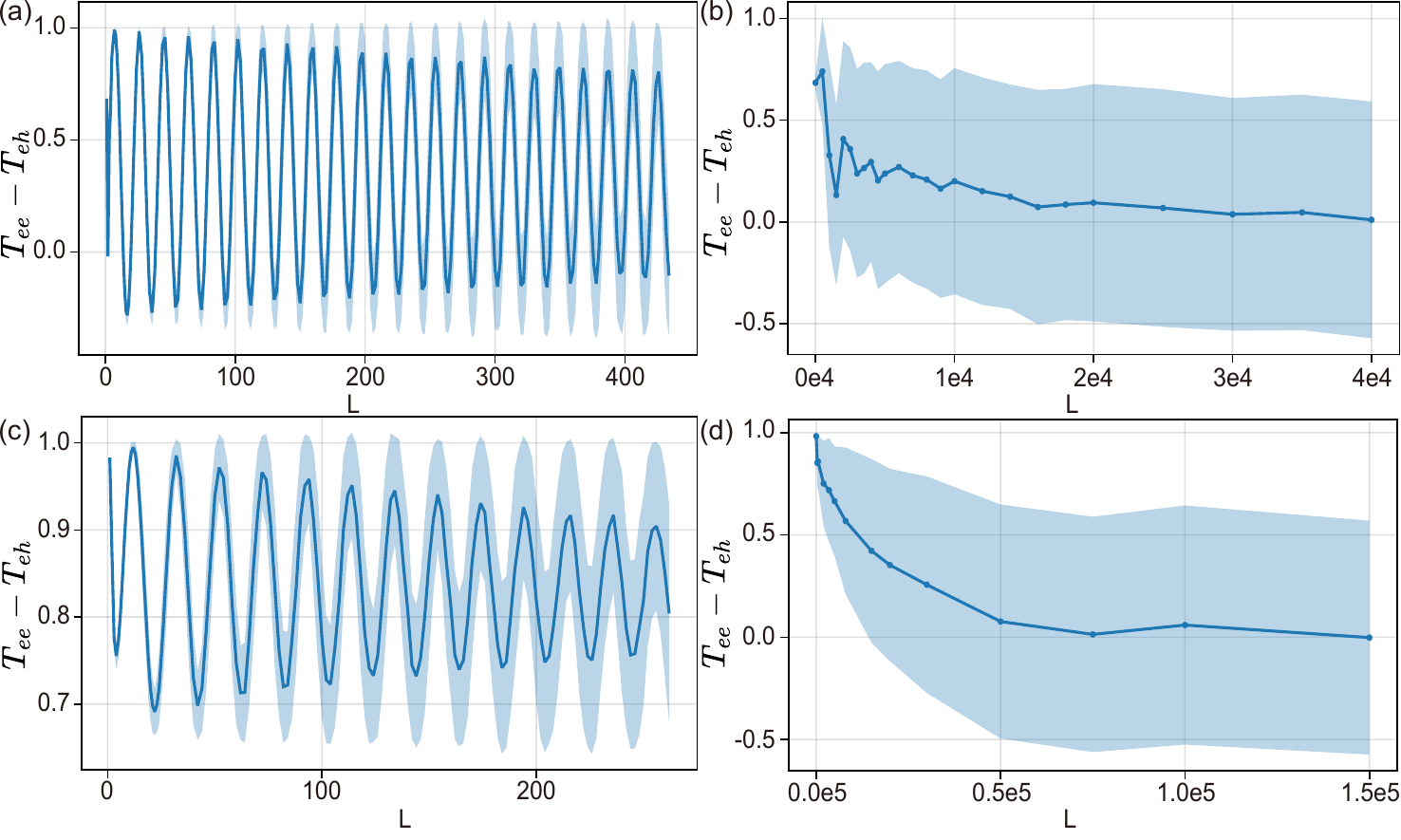}
  \end{center}
  \caption{The relation of charge transmission and the length of the TSC region for the QAH-SC-QAH junction with $N=2$. (a)(b) $\varepsilon=0.63,~W_{\text{dis}}=0.1$ case. (c)(d) $\varepsilon=0.4,~W_{\text{dis}}=0.2$ case. The blue shading represents one standard deviation.}
  \label{figSn1}
\end{figure}
\begin{figure}[htbp]
  \begin{center}
  \includegraphics[width=6.0in,clip=true]{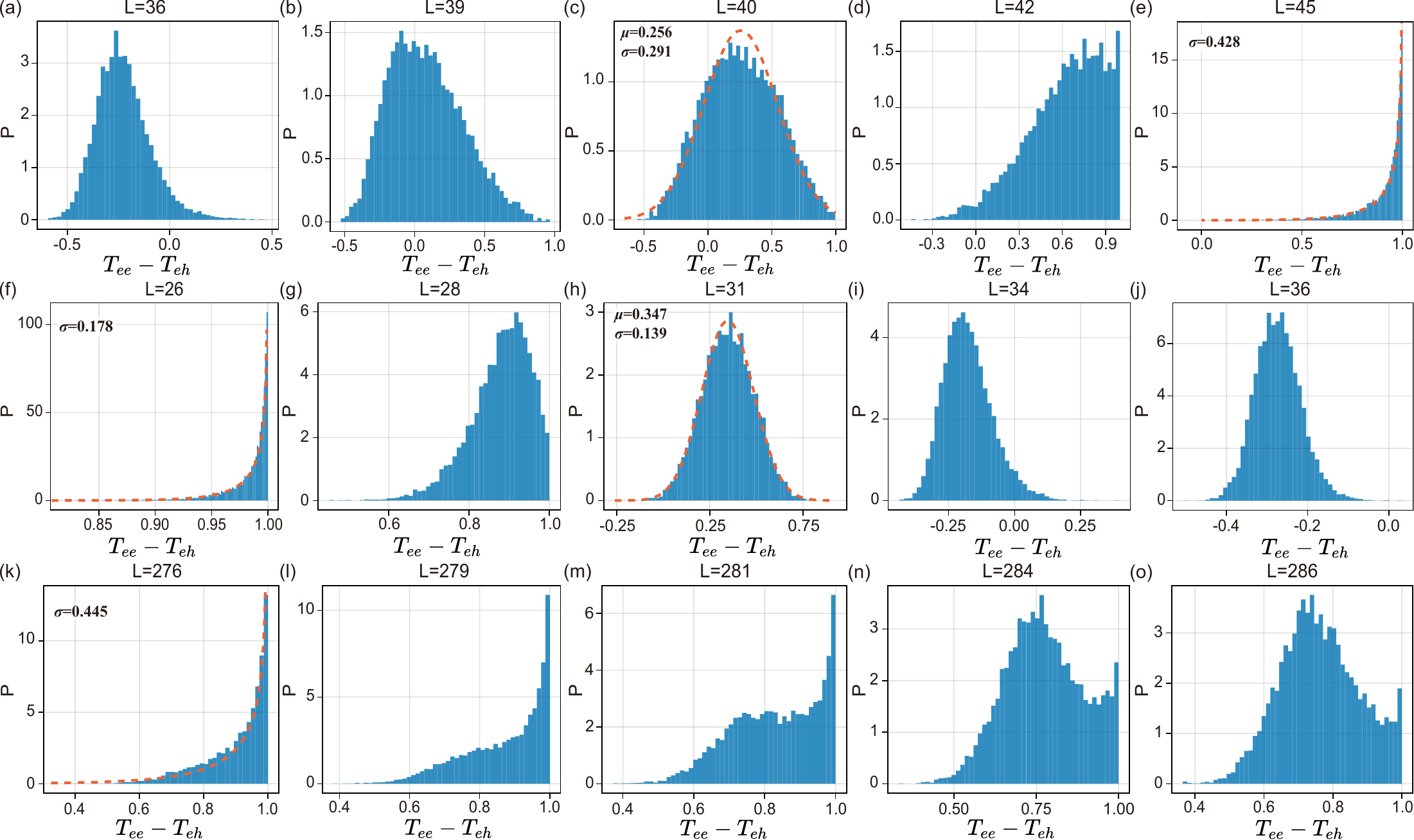}
  \end{center}
  \caption{The distribution of charge transmission for the QAH-SC-QAH junction with $N=2$. (a-e) $\varepsilon=0.63,~W_{\text{dis}}=0.2$ case. (f-j) $\varepsilon=0.63,~W_{\text{dis}}=0.1$ case. (k-o) $\varepsilon=0.4,~W_{\text{dis}}=0.2$ case. The orange dashed lines in some figures are the fitting curves when the distribution is close to normal or chi-squared distribution.}
  \label{figSn2}
\end{figure}

\section{Numerical Results of Two-dimensional Simulation}
\subsection{Numerical results for QAH-SC-QAH junction}\label{appqahtscqah}
In addition to Fig. 2 and Fig. 3 in the main text, we present more numerical results for the QAH-SC-QAH junction.
Fig.~\ref{figSn1} shows the $T-L$ relation for the $N=2$ case under different incident electron energies and disorder potential strengths from the main text. These results are consistent with Fig. 2 in the main text, with only the oscillation amplitude and decay length differing due to different parameters.
The distributions of $T$ at different $L$ are shown in Fig.~\ref{figSn2}. These results further support the conclusions made in the main text. As $L$ increases, the fluctuation of $T$ increases. Since $T$ is always bounded within $[-1,1]$, the resulting distribution deviates from the normal or chi-squared distribution when $L$ is large. 

\begin{figure}[t]
  \begin{center}
  \includegraphics[width=5.0in,clip=true]{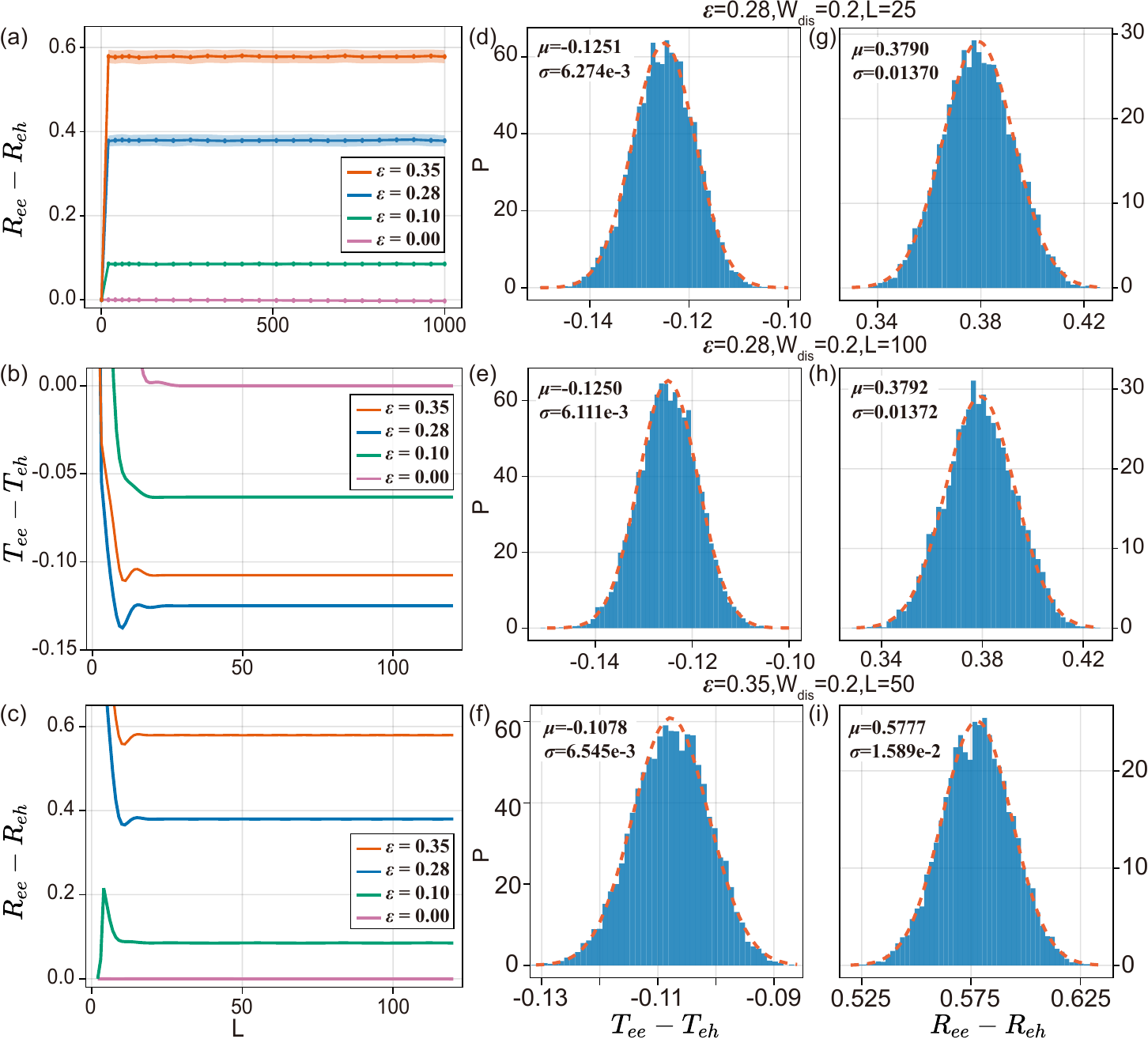}
  \end{center}
  \caption{The distribution of charge transmission (reflection) for the QAH-SC-QAH junction with $N=1$. (a) $R-L$ relation. (b) and (c) are $T-L$ and $R-L$ relations without disorder. (d-f) The distribution of $T$ at different $L$ or $\varepsilon$, and that of $R$ is shown in (g-i). The orange dashed lines are the fitting Gaussian curves.}
  \label{figSn3}
\end{figure}

For the $N=1$ case, the $R-L$ relation is shown in Fig.~\ref{figSn3}(a), which exhibits similar behavior to $T$ shown in the main text. As a supplement, the $T-L$ and $R-L$ relations in the clean case are shown in Fig.~\ref{figSn3}(b) and Fig.~\ref{figSn3}(c). The distributions of $T$ and $R$ are always Gaussian as long as $L$ is large enough to avoid overlap of wavefunctions for edge modes on opposite sides, regardless of different parameters. This is confirmed by Fig.~\ref{figSn3}(d-i).

\begin{figure}[htbp]
  \begin{center}
  \includegraphics[width=4.5in,clip=true]{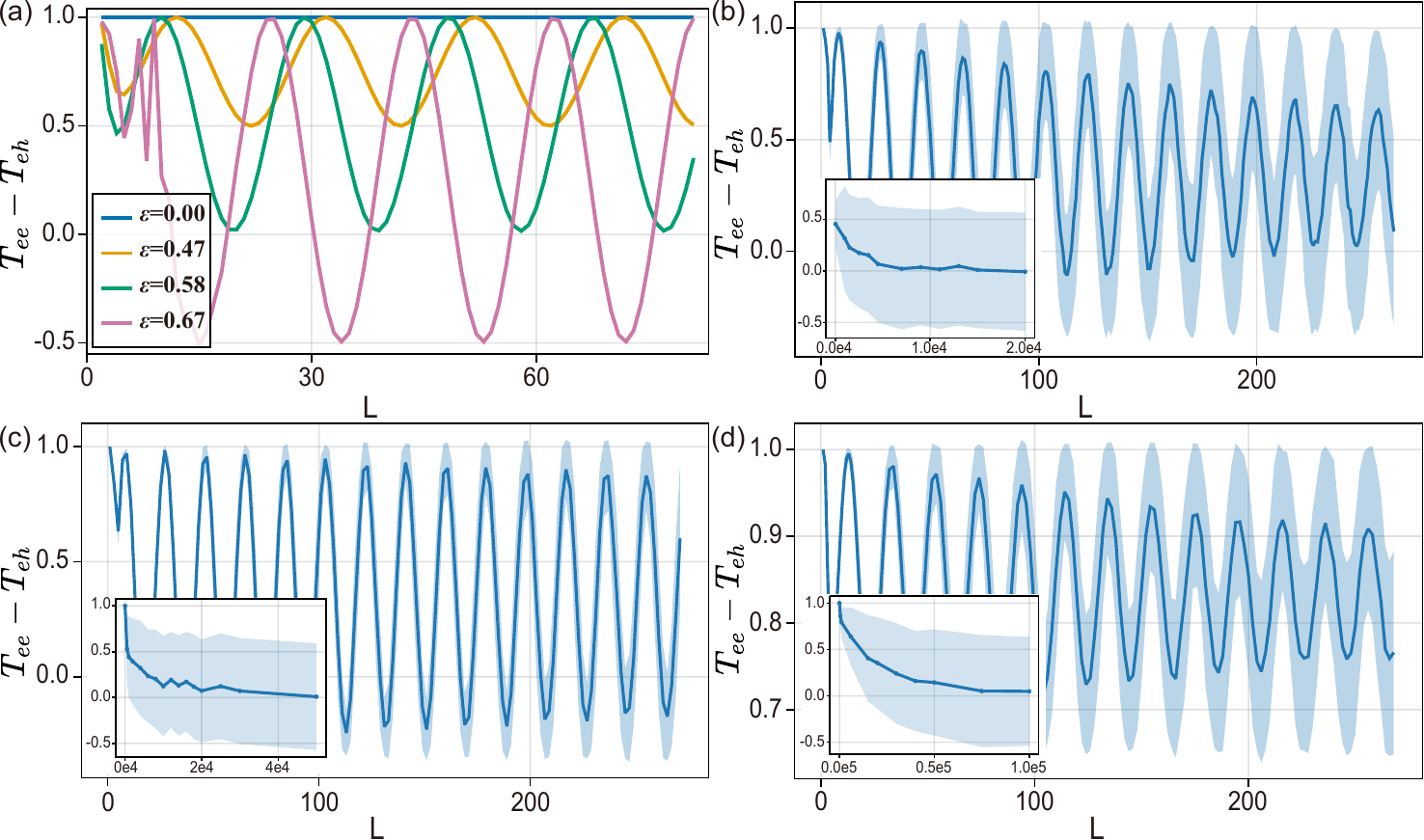}
  \end{center}
  \caption{The relationship between charge transmission and the length of the TSC region for the QAH-SC junction with $N=2$. (a) Disorder-free case. (b) $\varepsilon=0.63,~W_{\text{dis}}=0.2$ case. (c) $\varepsilon=0.63,~W_{\text{dis}}=0.1$ case. (d) $\varepsilon=0.4,~W_{\text{dis}}=0.2$ case. Insets in (b-d) show the corresponding large $L$ cases. The blue shading represents one standard deviation.}
  \label{figSn4}
\end{figure}

\begin{figure}[htbp]
  \begin{center}
  \includegraphics[width=6.5in,clip=true]{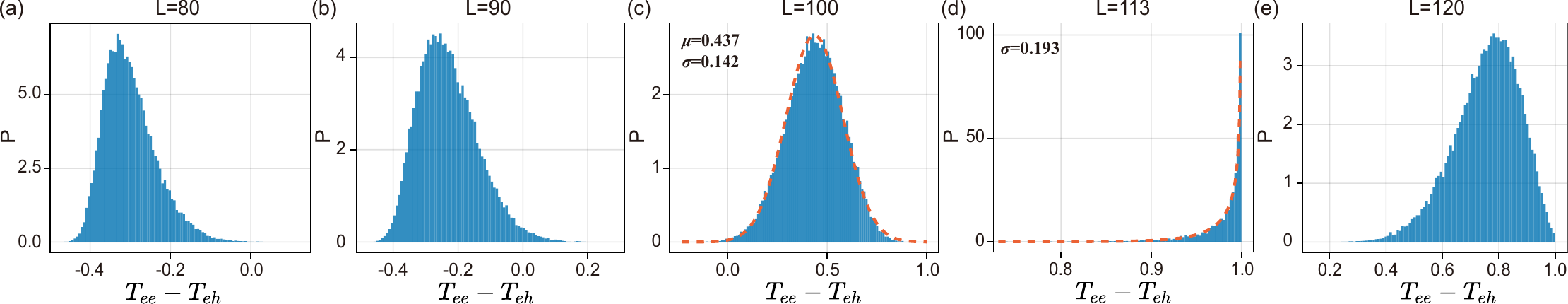}
  \end{center}
  \caption{The distribution of charge transmission for the QAH-SC junction with $N=1$. The parameters are the same as in Fig. 3(c,d) in the main text. The orange dashed lines in some figures are the fitting curves when the distribution is close to the normal or chi-squared distribution.}
  \label{figSn5}
\end{figure}
\begin{figure}[htbp]
  \begin{center}
  \includegraphics[width=4.6in,clip=true]{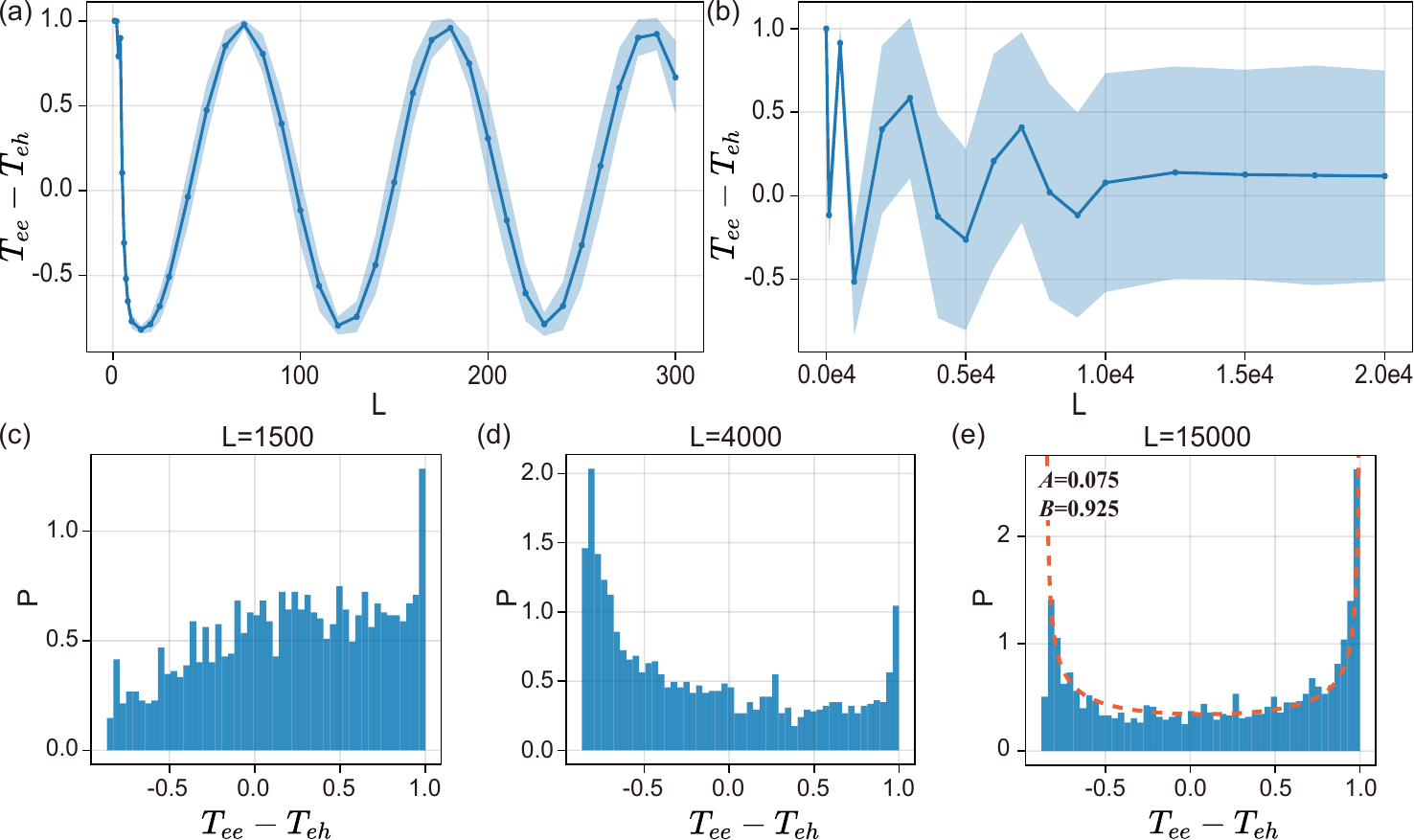}
  \end{center}
  \caption{The case of $\varepsilon=0.18,~W_{\text{dis}}=0.3$ for the QAH-SC junction with $N=1$. (a)(b) The relationship between charge transmission and the length of the TSC region. The blue shading represents one standard deviation. (c-e) The distribution of charge transmission. The orange dashed lines in (e) are the fitting curves.}
  \label{figSn6}
\end{figure}

\subsection{Numerical results for QAH-SC junction}\label{appqahtsc}
In addition to Fig. 3 in the main text, we present further numerical results for the QAH-SC junction.
For the $N=2$ case, the $T-L$ relationship is shown in Fig.~\ref{figSn4}. This is nearly identical to the QAH-SC-QAH junction with $N=2$, as the transmission processes from lead 2 to 3 are exactly the same. The corresponding distribution of $T$ is also the same, thus not shown here.
Fig.~\ref{figSn5} shows the distributions of $T$ at small $L$ with the same parameter setup as in Fig. 3(c,d) in the main text. The results support the conclusion of Case (i) in the main text.
Another parameter setup for the $N=1$ case is shown in Fig.~\ref{figSn6}. Only the oscillation amplitude and decay length differ from Fig. 3(c,d) in the main text. As $L$ increases, the distribution deviates from a normal or chi-squared distribution, gradually approaching a generalized arcsine distribution.

\subsection{Numerical results for $N=0$ cases}\label{appnsc}

\begin{figure}[t]
  \begin{center}
  \includegraphics[width=5.7in,clip=true]{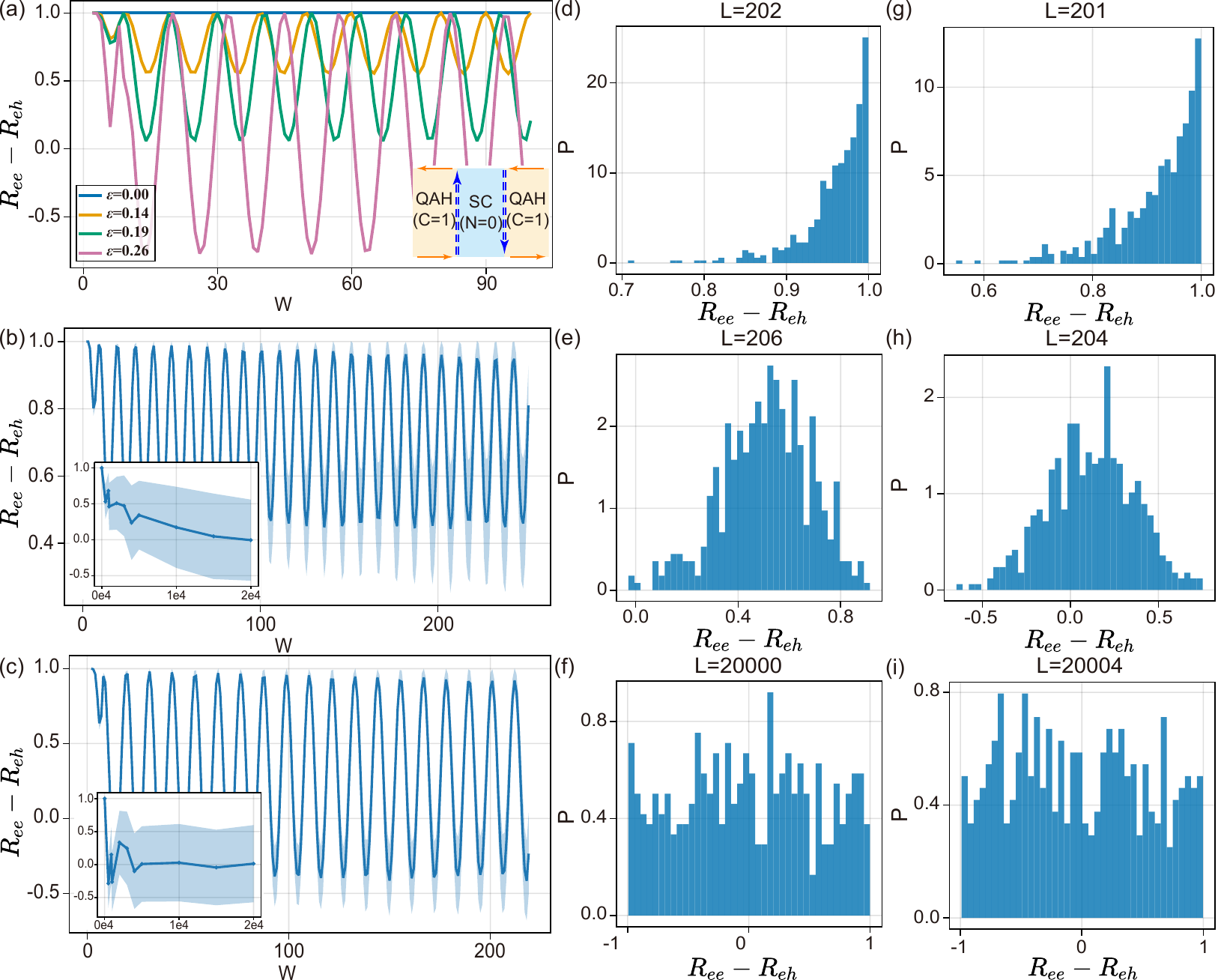}
  \end{center}
  \caption{Charge reflection for the QAH-NSC-QAH junction. (a-c) Relationship between charge reflection and the width of the junction. (d-i) Distribution of charge reflection. (a) Disorder-free case. Inset shows the schematic diagram of the device. (b) and (d-f) $\varepsilon=0.15,~W_{\text{dis}}=0.2$ case. (c) and (g-i) $\varepsilon=0.23,~W_{\text{dis}}=0.2$ case. The blue shading represents one standard deviation.}
  \label{figSn7}
\end{figure}

\begin{figure}[htbp]
  \begin{center}
  \includegraphics[width=5.5in,clip=true]{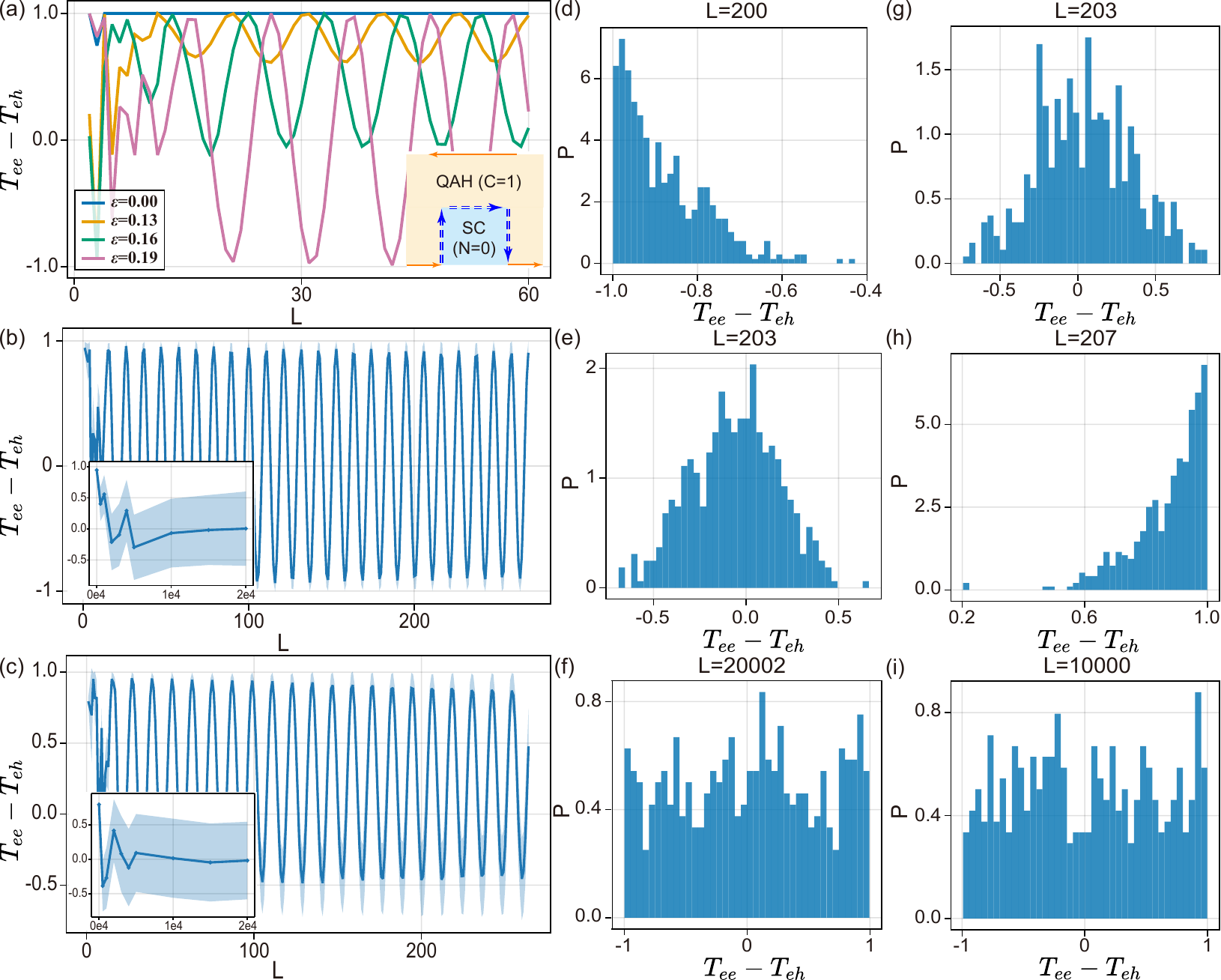}
  \end{center}
  \caption{Charge transmission for the QAH-NSC junction. (a-c) Relationship between charge transmission and the length of the junction. (d-i) Distribution of charge transmission. (a) Disorder-free case. Inset shows the schematic diagram of the device. (b) and (d-f) $\varepsilon=0.19,~W_{\text{dis}}=0.2$ case. (c) and (g-i) $\varepsilon=0.25,~W_{\text{dis}}=0.2$ case. The blue shading represents one standard deviation.}
  \label{figSn8}
\end{figure}

When the SC region of the junction is topologically trivial, i.e., $N=0$, two chiral Bogoliubov
edge modes emerge along the interface between SC and QAH, thus the edge modes transport characteristics is analogous to the $N=2$ case, where those modes emerge along the interface between SC and vacuum. 
In QAH-SC-QAH junction, $T$ is always zero due to the absence of edge modes at the SC/vacuum interface. While one can observe the oscillation of charge reflection fraction $R$ with the width $W$ of the junction, as Andreev conversion occurs at the SC/QAH interface, see Fig.~\ref{figSn7}(a).
In QAH-SC junction, the edge modes are shown in Fig.~\ref{figSn8}(a), allowing investigation of the charge transmission fraction, similar to the $N=2$ case.

We provide numerical calculations for both cases. The effective Hamiltonian for the QAH is:
\begin{equation}
  \mathcal{H}_{\text{QAH}}=\sum_{\mathbf{k}}c^\dag_\mathbf{k}\left[\bm{\zeta}(\mathbf{k})\cdot\bm{\sigma}-\mu_{h}\right]c_{\mathbf{k}},
\end{equation}
and the BdG Hamiltonian for the NSC is:
\begin{equation}
  \mathcal{H}_{\text{SC}}=\sum_{\mathbf{k}}\Psi^\dagger_{\mathbf{k}}\left[\epsilon(\mathbf{k})-\mu_{s}\right]\Psi_{\mathbf{k}}+(\Delta \Psi_{\mathbf{k}}^T i\sigma_y \Psi_{-\mathbf{k}}+\mathrm{H.c.}).
\end{equation}
where $\Psi_{\mathbf{k}}=(c_{\mathbf{k}\uparrow},c_{\mathbf{k}\downarrow})^T$, $\bm{\zeta}(\mathbf{k})=[M-B(\cos k_xa+\cos k_ya), A\sin k_x a, A\sin k_y a ]$, $\bm{\sigma}=(\sigma_x,\sigma_y,\sigma_z)$ are the Pauli matrices, $\epsilon(\mathbf{k})=B(2-\cos k_xa-\cos k_ya)$ is the kinetic energy, $a$ is the lattice constant, $\Delta$ is the pairing amplitude, and $\mu_h$ and $\mu_s$ are the chemical potentials of the QAH and the NSC, respectively \cite{lian2016edgestateinduced}.
The parameters are chosen as: $a=0.8$, $B=1.5625$, $M=2.625$, $A=1.25$, $\Delta=0.3$, $\mu_h=0.2$, and $\mu_s=0.5$. The number of disorder configurations is 500.

The numerical results are shown in Figs.~\ref{figSn7} and \ref{figSn8}. The main tendencies of charge reflection (transmission) are quite similar to those for the $N=2$ case.